\documentclass[12pt,a4paper]{article}
\usepackage[dvips]{epsfig}

\begin{document}
\author{\normalsize\bf Yu.A. Markov$\!\,$\thanks{e-mail:markov@icc.ru}
$\,$and M.A. Markova$^*$}
\title{Nonlinear dynamics of soft fermion\\
excitations in hot QCD plasma I:\\
soft-quark\,--\,soft-gluon scattering}
\date{\it Institute for System Dynamics\\
and Control Theory Siberian Branch\\
of Academy of Sciences of Russia,\\
P.O. Box 1233, 664033 Irkutsk, Russia}

\thispagestyle{empty}
\maketitle{}


\def\theequation{\arabic{section}.\arabic{equation}}

\[
{\bf Abstract}
\]
Within the framework of the hard thermal loop effective theory we
derive a system of Boltzmann-like kinetic equations taking into
account the simplest processes of nonlinear interaction of soft fermionic
and bosonic QCD plasma excitations: elastic scattering of soft-(anti)quark
excitations off soft-gluon and soft-quark excitations,
pair production of soft quark-antiquark excitations,
annihilation into two soft-gluon excitations. The matrix
elements of these processes to leading order in the coupling constant $g$
are obtained. The iterative method of calculation of the matrix
elements for the higher processes of soft-mode interactions is proposed.
The most general expression for the emitted radiant power induced by the
effective currents and effective sources in a quark-gluon plasma (QGP) taking
into account an existence of fermion sector of plasma excitations is defined.
The explicit form of the linearized Boltzmann equation accounting for
scattering of color(less) plasminos off color(less) plasmons is
written out.


\newpage

\section{Introduction}
\setcounter{equation}{0}

In our previous papers \cite{markov1, markov2} on the basis of
a purely gauge sector of the Blaizot-Iancu equations
\cite{blaizot} complemented by the Wong equation \cite{wong} we
have studied in great detail the nonlinear dynamics of soft boson
excitations in hot QCD plasma. In this and two accompanying papers
we would like to enlarge the analysis carried out in
Refs.\,\cite{markov1, markov2} on the fermion sector
of soft plasma excitations. For solving this problem
it is necessary to use the whole system of the Blaizot-Iancu
equations. It is well known that hot QCD plasma including massless quarks
(and antiquarks) possesses doubled dispersion relation for soft
fermion plasma excitations \cite{klimov}. The first branch describes
normal particle excitations (in the subsequent discussion designated by
a symbol $``+"$)
with relation between chirality and helicity at zero temperature.
The second branch is purely collective excitation
(designated further by symbol $``-"$), where usual relation
between chirality and helicity is flipped \cite{pisarski}.
It has been called a {\it plasmino} \cite{braaten1} by analogy with
the plasmon (longitudinal) mode of gluons having also a purely collective
character.

Here, in the first part we have restricted ourselves to study of
the processes connected with interaction of the soft fermion and
boson excitations among themselves without exchange of energy with
hard thermal (or external) partons. We consider the processes of
nonlinear interaction of soft excitations within the framework of
the kinetic approach by the determination of corresponding set
of coupled Boltzmann-like kinetic equations describing some
non-linear aspects of the dynamics of the soft collective
excitations in high-temperature QGP. The construction of such
equations is generally straightforward at a conceptual level, but
it is tedious and intricate in practice. By virtue of complexity
and awkwardness of the kinetic equations for soft plasma modes
taking into account simultaneously the fermion and boson degrees
of freedom, we will consider in details only the simplest
processes of the nonlinear interaction:
soft-(anti)quark\,--\,soft-gluon and
soft-(anti)quark\,--\,soft-quark elastic scattering, production
of soft quark-antiquark pair by fusion of two soft gluons, and
annihilation of soft quark-antiquark pair into two soft-gluon
excitations. At least for weakly excited system corresponding to
the level of thermal fluctuations at the soft momentum scale these
processes are the basic those of the nonlinear interaction of soft
excitations in the medium. In deriving the kinetic equations for
soft fermion and boson modes we use oversimplified approach.
Proceeding from the semiclassical character of the problem, we
initially assume that the structure of the collision terms is
determined by Fermi's golden rules and thus we focus our efforts
exclusively on calculating the scattering probabilities. We
calculate these probabilities in a direct fashion by simple
extracting all relevant contributions in scattering processes
$2\rightarrow 2$ enumerated above. We don't give an explicit proof
of a gauge invariance of the scattering probabilities obtained.
This can be performed in spirit of our previous paper
\cite{markov_PRD} by using the effective Ward identities for hot
gauge theories \cite{braaten2, taylor}.

It should be particularly noted that isolation and study of the nonlinear
dynamics of soft-mode interactions only in QGP presented
in this work have sufficiently artificial character and bear rather limited
sense. Really, by the efforts of a number of authors (see, for example
\cite{braaten3, kobes}) it was shown that the dominant
interactions in QGP are generally those between the soft modes of
interest and the hard particles in the thermal bath. Moreover, the
general transport or equilibration properties of the plasma are in
fact controlled by the hard particles (those with typical momenta
of order $T$, the temperature of the system). In our following
papers the interaction processes of soft and hard modes
of different statistics among themselves and also radiation
processes of soft modes induced by collisions of hard thermal
particles will be considered in details.
The construction of the effective theory for such processes has shown
that it is more convenient to consider separately a theory for
simpler (in the conceptual plan) interaction
processes of the soft modes. Matrix elements of
interaction processes of soft and hard excitations include, as
constitutive blocks, the expressions defining matrix elements of
self-interaction of soft excitations provided that external momenta
are put on relevant mass-shells. This significantly facilitates general
analysis and classification of individual terms in very complicated and tangled
expressions for matrix elements of the interaction processes
between the soft excitations and the hard particles including
simultaneously the soft and hard modes of Fermi and Bose statistics.

Incidentally there are circumstances at which the kinetic equations
for the soft quasi-particles may be relevant independently: for
instance, when we are interested in the thermalization of such
soft modes or in their damping rates. Besides at studying the
processes with soft-gluon excitations only we have shown \cite{markov1,
markov2} that there exists a certain level of intensity of
plasma excitations  above which the nonlinear interaction processes of
soft boson excitations among themselves start to dominate over the
interaction processes of soft modes with hard thermal particles,
thus defining in particular qualitatively different mechanism for the
damping of a soft gluons. In other words, for large values of
soft-gluon occupation number (i.e. when a subsystem of soft boson
excitations is far from equilibrium) dynamics of hot QCD plasma
is defined by scattering processes of soft
quasiparticles off each other. In the second part
\cite{markov_II} we will show that and for the case of fermion
sector of plasma excitations in QGP there exists a critical
threshold above which the interaction processes of soft fermion
and boson excitations among themselves start to play dominating
role in general dynamics of the system.

In the light of discussed above a study of soft-excitations dynamics
only can represent independent interest, however this
is beyond the scope of our work. Our ultimate goal is the study of new
(not considered earlier in the literature) possible mechanisms of
energy losses of high-energy particles propagating through hot QCD
medium associated with an existence in medium soft-(anti)quark
excitations on a level with soft-gluon excitations. This is a main
subject of the following papers. Research of the energy
losses of energetic partons in QGP at present is of great
interest with respect to {\it jet quenching} phenomenon
\cite{gyulassy, kovner}. The basic purpose of present work is
development of the general technique for calculation of relevant
probabilities and in particular, deriving the semiclassical
formula for the emitted radiant power in QGP taking into account
fermion sector of plasma excitations (section 8). This formula
(with the further updating) is basic formula for defining energy
losses of energetic parton (quark or gluon) travelling through
quark-gluon plasma.

The derivation of the Boltzmann equation describing evolution of
the number density of fermion quasiparticles in QGP was also
considered by Ni\'egawa \cite{niegawa}. Within the framework of
the Keldysh-Schwinger formalism the kinetic equations for normal
and abnormal modes here, are defined from the requirement of the
absence of the large contributions due to pinch singularities of
the perturbative scheme proposed in Ref.\,\cite{niegawa}. The
formalism suggested by Ni\'egawa is quite rigorous and possesses
great generality at least for quasiuniform systems near
equilibrium. Unfortunately in this paper particular expressions
for collision terms wasn't given. In this sense, our work having
more phenomenological character supplements the paper
\cite{niegawa} with examples of concrete calculation of the
collision terms within the HTL effective theory.

The paper is organized as follows. In section 2, preliminary
comments with regard to deriving the system of the Boltzmann
equations describing the processes of nonlinear interaction of
soft fermion and boson excitations in QGP are given. In section 3,
the system of nonlinear integral equations for gauge potential
$A^a_{\mu}$ and quark wave function $\psi^i_{\alpha}$ is written
out. On the basis of perturbative solutions of these equations the
notions of the effective currents and sources playing a key
position in our research are given. In section 4, an algorithm of
the successive calculation of the effective amplitudes defining
matrix elements of the processes we are interested is presented.
In sections 5, 6 and 7, the details of calculations of the
probabilities for the simplest processes of interaction:
soft-(anti)quark\,--\,soft-gluon and
soft-(anti)quark\,--\,soft-quark elastic scattering, production
of soft quark-antiquark pair by fusion of two soft gluons, and
annihilation of soft quark-antiquark pair into two soft-gluon
excitations are presented. Section 8 is devoted to deriving the
most general formula for the emitted radiant power in QGP. It was
shown how this formula within the framework of Tsytovich
correspondence principle allows by a direct way to calculate the
probabilities of higher order processes of nonlinear interaction
of soft plasma modes. In section 9, the explicit form of
linearized Boltzmann equation for colorless plasminos is written
out and some speculations concerning the extension of the
formalism developed to colorcharged plasminos are given. In
Conclusion some features of dynamics of soft excitations is
briefly discussed and connection with more traditional approaches
based on imaginary- and real-time field theories is traced.
Finally, in Appendixes summary of various properties of
HTL-amplitudes used in the text and the explicit form of the
conjugate effective amplitudes are given.

\section{Preliminaries}
\setcounter{equation}{0}

In this paper we consider a construction of effective kinetic
theory for colorless soft-quark and soft-gluon plasma excitations propagating
in a hot quark-gluon plasma. We assume that localized number densities of
soft-quark $n^{\pm}({\bf q},x) \equiv (n^{\pm\,ij}_{\bf q})$,
soft-antiquark $\bar{n}^{\pm}({\bf q},x) \equiv (\bar{n}^{\pm\,ij}_{\bf q})$,
and soft-gluon $N^{t,\,l}({\bf k}, x)\equiv(N_{\bf k}^{t,\,l\,ab})$ excitations
are diagonal in a color space
\[
n^{\pm\,ij}_{\bf q}=\delta^{ij} n^{\pm}_{\bf q},\quad
\bar{n}^{\pm\,ij}_{\bf q}=\delta^{ij}\bar{n}^{\pm}_{\bf q},\quad
N_{\bf k}^{t,\,l\,ab} = \delta^{ab} N_{\bf k}^{t,\,l} ,
\]
where $i,j=1,2,\ldots,N_c$ and $a,b = 1, \ldots , N_c^2 - 1$ for $SU(N_c)$
gauge group. We consider a change of the number densities of the colorless
soft-(anti)quark excitations as a result of their interactions among themselves
and with the colorless soft-gluon excitations.

The dispersion relations for soft-quark modes $\omega^{\pm}({\bf q})\equiv
\omega_{\bf q}^{\pm}$ and soft-gluon modes $\omega^{t,\,l}({\bf k})\equiv
\omega_{\bf k}^{t,\,l}$ are defined by equations
\begin{equation}
{\rm Re}\,^{\ast}\!\Delta^{\!-1}_{\pm}(q^0,{\bf q}) = 0,\quad
{\rm Re}\,^{\ast}\!\Delta^{\!-1\,t,\,l}(k^0,{\bf k}) = 0 ,
\label{eq:2q}
\end{equation}
where
\[
\,^{\ast}\!\Delta^{\!-1}_{\pm}(q^0,{\bf q})=q^0 \mp |{\bf q}| +
\frac{\omega_0^2}{|{\bf q}|}
\biggl[1 - \biggl(1\mp\frac{|{\bf q}|}{q^0}\biggr)
F\biggl(\frac{q^0}{\vert {\bf q}\vert}\bigg)\bigg],
\]
and
\[
\,^{\ast}\!\Delta^{\!-1\,t}(k^0,{\bf k}) = k^2-\frac{3}{2}\,\omega_{pl}^2
\bigg[\frac{(k^0)^2}{{\bf k}^2} - \frac{k^2}{{\bf k}^2}
F \bigg( \frac{k^0}{\vert {\bf k}\vert}\bigg)\bigg]\;,\;
\]
\[
\,^{\ast}\!\Delta^{\!-1\,l}(k^0,{\bf k}) = k^2\bigg(
1 + \frac{3\,\omega_{pl}^2}{{\bf k}^2}
\bigg[1 - F \bigg( \frac{k^0}{\vert {\bf k}\vert}\bigg)\bigg]\bigg)\;,\;
F (x) \equiv \frac{x}{2} \bigg[ \ln \bigg \vert \frac{1 + x}{1 - x}
\bigg \vert - i \pi
\theta ( 1 - \vert x \vert ) \bigg]
\]
are inverse quark and gluon scalar propagators.
$\omega_0^2= g^2C_FT^2/8$ and $\omega_{pl}^2 = g^2(2N_c+N_f) T^2/18$ are
plasma frequencies squared of the quark and gluon sectors of plasma excitations.
Hereafter, we denote momenta of the soft-quark fields by
$q,\,q^{\prime}\!,\,q_1,\ldots$ and momenta of the
soft-gauge fields by $k,\,k^{\prime}\!,\,k_1,\ldots\,\,$.

We expect a time-space evolution of the scalar functions
$n_{\bf q}^{(f)},\,f=\pm$ and $N_{\bf k}^{(b)},\,b=t,\,l$ to be described by
the self-consistent system of Boltzmann-like equations
\begin{equation}
\frac{\partial n_{\bf q}^{(f)}}{\partial t} +
{\bf v}_{\bf q}^{(f)}\cdot\frac{\partial n_{\bf q}^{(f)}}{\partial {\bf x}} =
- n_{\bf q}^{(f)}\,\Gamma_{\rm d}^{(f)}[n_{\bf q}^{\pm},N_{\bf k}^{t,\,l}] +
(1 - n_{\bf q}^{(f)})\Gamma_{\rm i}^{(f)}[n_{\bf q}^{\pm},N_{\bf k}^{t,\,l}],
\label{eq:2w}
\end{equation}
\begin{equation}
\frac{\partial N_{\bf k}^{(b)}}{\partial t} +
{\bf v}_{\bf k}^{(b)}\cdot\frac{\partial N_{\bf k}^{(b)}}{\partial {\bf x}} =
- N_{\bf k}^{(b)}\,\Gamma_{\rm d}^{(b)}[n_{\bf q}^{\pm},N_{\bf k}^{t,\,l}] +
(1 + N_{\bf k}^{(b)})\Gamma_{\rm i}^{(b)}[n_{\bf q}^{\pm},N_{\bf k}^{t,\,l}],
\label{eq:2e}
\end{equation}
where  ${\bf v}_{\bf q}^{(f)}=\partial\omega_{\bf q}^{(f)}/\partial{\bf q}$
and ${\bf v}_{\bf k}^{(b)}=\partial\omega_{\bf k}^{(b)}/\partial{\bf k}$
are the group velocities of soft fermionic and
bosonic excitations, respectively. The generalized decay rates
$\Gamma_{\rm d}^{(f\!,\,b)}$ and inverse decay rates
$\Gamma_{\rm i}^{(f\!,\,b)}$ are (non-linear) functionals of
the soft-(anti)quark and soft-gluon number densities. Here, for the sake of
brevity we drop dependence on soft-antiquark occupation numbers
$\bar{n}_{\bf q}^{\pm}$ on the right-hand side of Eqs.\,(\ref{eq:2w}) and
(\ref{eq:2e}). The equation for $\bar{n}_{\bf q}^{(f)}$ is obtained from
(\ref{eq:2w}) by replacement
$n_{\bf q}^{(f)}\rightleftharpoons\bar{n}_{\bf q}^{(f)}$.

The decay and regenerating rates can be formally represented in the form of
functional expansion in powers of the soft-(anti)quark and soft-gluon number
densities
\begin{equation}
\Gamma_{\rm d}^{(f\!,\,b)}[n_{\bf q}^{\pm},N_{\bf k}^{t,\,l}] =
\sum_{n = 1}^{\infty}
\Gamma_{\rm d}^{(f\!,\,b)(2n + 1)}[n_{\bf q}^{\pm},N_{\bf k}^{t,\,l}],\;\;
\Gamma_{\rm i}^{(f\!,\,b)}[n_{\bf q}^{\pm},N_{\bf k}^{t,\,l}] =
\sum_{n = 1}^{\infty}
\Gamma_{\rm i}^{(f\!,\,b)(2n + 1)}[n_{\bf q}^{\pm},N_{\bf k}^{t,\,l}],
\label{eq:2r}
\end{equation}
where
$\Gamma_{{\rm d},\,{\rm i}}^{(f\!,\,b)(2n + 1)}
[n_{\bf q}^{\pm},N_{\bf k}^{t,\,l}]$
collect the contributions of the total $n$th power in
$n_{\bf q}^{\pm},\,\bar{n}_{\bf q}^{\pm}$ and $N_{\bf k}^{t,\,l}$.
Unfortunately, the general structure of the expressions for arbitrary $n$
is very cumbersome and therefore we restrict our consideration to the simplest
case for $n=1$.

The fermion decay and regenerating rates in the lowest order of the nonlinear
interaction ($n=1$ in Eq.\,(\ref{eq:2r})) can be formally represented in the
following form:
\begin{equation}
\Gamma_{\rm d}^{(f)}[n_{\bf q}^{\pm},\!N_{\bf k}^{t,\,l}]\!=
\!\!\!\sum\limits_{f_1=\pm}\sum\limits_{\,b_1,\,b_2=t,\,l}\Biggl\{
2\!\int\!\!d{\cal T}_{qg\rightarrow qg}^{(fb_1;\,f_1b_2)}
\,{\it w}_{qg\rightarrow qg}^{(fb_1;\,f_1b_2)}
({\bf q}, {\bf k}_1;\, {\bf q}_1, {\bf k}_2)
\Bigl(1-n_{{\bf q}_1}^{(f_1)}\Bigr)
N_{{\bf k}_1}^{(b_1)}\Bigl(1+N_{{\bf k}_2}^{(b_2)}\Bigr)
\label{eq:2t}
\end{equation}
\[
\hspace{4cm}
+\!\int\!d{\cal T}_{q\bar{q}\rightarrow gg}^{(ff_1;\,b_1b_2)}\,
{\it w}_{q\bar{q}\rightarrow gg}^{(ff_1;\,b_1b_2)}
({\bf q}, {\bf q}_1;\, {\bf k}_1, {\bf k}_2)n_{{\bf q}_1}^{(f_1)}
\Bigl(1+N_{{\bf k}_1}^{(b_1)}\Bigr)
\Bigl(1+N_{{\bf k}_2}^{(b_2)}\Bigr)\!\Biggr\}
\]
\[
\hspace{2cm}
+\sum\limits_{f_1,\,f_2,\,f_3=\pm}\Biggl\{
\int\!d{\cal T}_{qq\rightarrow qq}^{(ff_1;\,f_2f_3)}
\, {\it w}_{qq\rightarrow qq}^{(ff_1;\,f_2f_3)}
({\bf q}, {\bf q}_1;\, {\bf q}_2, {\bf q}_3)n_{{\bf q}_1}^{(f_1)}
\Bigl(1-n_{{\bf q}_2}^{(f_2)})\Bigr)
\Bigl(1-n_{{\bf q}_3}^{(f_3)}\Bigr)
\]
\[
\hspace{4cm}
+\,2
\!\int\!d{\cal T}_{q\bar{q}\rightarrow q\bar{q}}^{(ff_1;\,f_2f_3)} \,
{\it w}_{q\bar{q}\rightarrow q\bar{q}}^{(ff_1;\,f_2f_3)}
({\bf q}, {\bf q}_1;\, {\bf q}_2, {\bf q}_3)\bar{n}_{{\bf q}_1}^{(f_1)}
\Bigl(1-\bar{n}_{{\bf q}_2}^{(f_2)}\Bigr)
\Bigl(1-n_{{\bf q}_3}^{(f_3)}\Bigr)\!\Biggr\},
\]
and in turn,
\begin{equation}
\Gamma_{\rm i}^{(f)}[n_{\bf q}^{\pm},N_{\bf k}^{t,\,l}] =
\!\!\!\sum\limits_{f_1=\pm}\sum\limits_{\,b_1,\,b_2=t,\,l}\Biggl\{
2\!\int\!d{\cal T}_{qg\rightarrow qg}^{(fb_1;\,f_1b_2)}
\,{\it w}_{qg\rightarrow qg}^{(fb_1;\,f_1b_2)}
({\bf q}, {\bf k}_1;\, {\bf q}_1, {\bf k}_2)n_{{\bf q}_1}^{(f_1)}
\Bigl(1+N_{{\bf k}_1}^{(b_1)}\Bigr)
N_{{\bf k}_2}^{(b_2)}
\hspace{0.8cm}
\label{eq:2y}
\end{equation}
\[
\hspace{4.5cm}
+\!\int\!d{\cal T}_{q\bar{q}\rightarrow gg}^{(ff_1;\,b_1b_2)} \,
{\it w}_{q\bar{q}\rightarrow gg}^{(ff_1;\,b_1b_2)}
({\bf q}, {\bf q}_1;\, {\bf k}_1, {\bf k}_2)
\Bigl(1-n_{{\bf q}_1}^{(f_1)}\Bigr)
N_{{\bf k}_1}^{(b_1)}N_{{\bf k}_2}^{(b_2)}\!\Biggr\}
\]
\[
\hspace{2.2cm}
+\sum\limits_{f_1,\,f_2,\,f_3=\pm}\Biggl\{
\int\!d{\cal T}_{qq\rightarrow qq}^{(ff_1;\,f_2f_3)}
\, {\it w}_{qq\rightarrow qq}^{(ff_1;\,f_2f_3)}
({\bf q}, {\bf q}_1;\, {\bf q}_2, {\bf q}_3)
\Bigl(1-n_{{\bf q}_1}^{(f_1)}\Bigr)
n_{{\bf q}_2}^{(f_2)}n_{{\bf q}_3}^{(f_3)}
\]
\[
\hspace{4.1cm}
+\,2
\!\int\!d{\cal T}_{q\bar{q}\rightarrow q\bar{q}}^{(ff_1;\,f_2f_3)}\,
{\it w}_{q\bar{q}\rightarrow q\bar{q}}^{(ff_1;\,f_2f_3)}
({\bf q}, {\bf q}_1;\, {\bf q}_2, {\bf q}_3)
\Bigl(1-\bar{n}_{{\bf q}_1}^{(f_1)}\Bigr)
\bar{n}_{{\bf q}_2}^{(f_2)}n_{{\bf q}_3}^{(f_3)}\Biggr\}.
\]
The functions ${\it w}_{qg\rightarrow qg}^{(fb_1;\,f_1b_2)},\,
{\it w}_{q\bar{q}\rightarrow gg}^{(ff_1;\,b_1b_2)}$ are
probabilities for `elastic' scattering of soft-quark excitations off
soft-gluon excitations and annihilation of soft quark-antiquark pair into two
soft-gluon excitations, and ${\it w}_{qq\rightarrow qq}^{(ff_1;\,f_2f_3)},
\,{\it w}_{q\bar{q}\rightarrow q\bar{q}}^{(ff_1;\,f_2f_3)}$ are
probabilities for `elastic' scattering of soft-quark excitations off
soft-quark and soft-antiquark excitations, respectively. The phase-space
measures for these processes are
\begin{equation}
\int\!d{\cal T}_{qg\rightarrow qg}^{(fb_1;\,f_1b_2)}\equiv\!
\int\!\!\frac{d{\bf q}_1}{(2\pi)^3}\!\int\!\!\frac{d{\bf k}_1}{(2\pi)^3}\!
\int\!\!\frac{d{\bf k}_2}{(2 \pi)^3}\,
(2\pi)^4\delta({\bf q}+{\bf k}_1-{\bf q}_1-{\bf k}_2)
\,\delta(\omega_{\bf q}^{(f)}+\omega_{{\bf k}_1}^{(b_1)}-
\omega_{{\bf q}_1}^{(f_1)}-\omega_{{\bf k}_2}^{(b_2)}),
\label{eq:2u}
\end{equation}
\[
\int\!d{\cal T}_{q\bar{q}\rightarrow gg}^{(ff_1;\,b_1b_2)}\!\equiv\!
\int\!\!\frac{d{\bf q}_1}{(2\pi)^3}\!\int\!\!\frac{d{\bf k}_1}{(2\pi)^3}\!
\int\!\!\frac{d{\bf k}_2}{(2 \pi)^3}\,
(2\pi)^4\delta({\bf q}+{\bf q}_1-{\bf k}_1-{\bf k}_2)
\,\delta(\omega_{\bf q}^{(f)}+\omega_{{\bf q}_1}^{(f_1)}-
\omega_{{\bf k}_1}^{(b_1)}-\omega_{{\bf k}_2}^{(b_2)}),
\hspace{0.5cm}
\]
\[
\int\!d{\cal T}_{qq\rightarrow qq}^{(ff_1;\,f_2f_3)}\!=\!
\int\!d{\cal T}_{q\bar{q}\rightarrow q\bar{q}}^{(ff_1;\,f_2f_3)}\!\equiv\!
\int\!\!\frac{d{\bf q}_1}{(2\pi)^3}\!\int\!\!\frac{d{\bf q}_1}{(2\pi)^3}\!
\int\!\!\frac{d{\bf q}_2}{(2 \pi)^3}\,
(2\pi)^4\delta({\bf q}+{\bf q}_1-{\bf q}_2-{\bf q}_3)
\]
\[
\times\,
\delta(\omega_{\bf q}^{(f)}+\omega_{{\bf q}_1}^{(f_1)}-
\omega_{{\bf q}_2}^{(f_2)}-\omega_{{\bf q}_3}^{(f_3)}).
\]

For boson sector of the plasma excitations the generalized rates
$\Gamma_{\rm d}^{(b)}$ and $\Gamma_{\rm i}^{(b)}$ to the lowest order in the
nonlinear interactions of soft modes have a similar structure
\begin{equation}
\Gamma_{\rm d}^{(b)}[n_{\bf q}^{\pm},N_{\bf k}^{t,\,l}] =
\!\!\!\sum\limits_{b_1=t,\,l}\sum\limits_{\,f_1,\,f_2=\pm}\Biggl\{
\int\!d{\cal T}_{gq\rightarrow gq}^{(bf_1;\,b_1f_2)}
\, {\it w}_{gq\rightarrow gq}^{(bf_1;\,b_1f_2)}
({\bf k}, {\bf q}_1;\, {\bf k}_1, {\bf q}_2)
\Bigl(1+N_{{\bf k}_1}^{(b_1)}\Bigr)
n_{{\bf q}_1}^{(f_1)}
\Bigl(1-n_{{\bf q}_2}^{(f_2)}\Bigr)
\label{eq:2i}
\end{equation}
\[
\hspace{4.4cm}
+\!\int\!d{\cal T}_{g\bar{q}\rightarrow g\bar{q}}^{(bf_1;\,b_1f_2)}\,
{\it w}_{g\bar{q}\rightarrow g\bar{q}}^{(bf_1;\,b_1f_2)}
({\bf k}, {\bf q}_1;\, {\bf k}_1, {\bf q}_2)
\Bigl(1+N_{{\bf k}_1}^{(b_1)}\Bigr)
\bar{n}_{{\bf q}_1}^{(f_1)}
\Bigl(1-\bar{n}_{{\bf q}_2}^{(f_2)}\Bigr)
\]
\[
\hspace{4.4cm}
+\!\int\!d{\cal T}_{gg\rightarrow q\bar{q}}^{(b\,b_1;\,f_1f_2)}\,
{\it w}_{gg\rightarrow q\bar{q}}^{(b\,b_1;\,f_1f_2)}
({\bf k}, {\bf k}_1;\, {\bf q}_1, {\bf q}_2)N_{{\bf k}_1}^{(b_1)}
\Bigl(1-\bar{n}_{{\bf q}_1}^{(f_1)}\Bigr)
\Bigl(1-n_{{\bf q}_2}^{(f_2)}\Bigr)\!\Biggr\},
\vspace{1cm}
\]
\[
\Gamma_{\rm i}^{(b)}[n_{\bf q}^{\pm},N_{\bf k}^{t,\,l}] =
\!\!\!\sum\limits_{b_1=t,\,l}\sum\limits_{\,f_1,\,f_2=\pm}\Biggl\{
\int\!d{\cal T}_{gq\rightarrow gq}^{(bf_1;\,b_1f_2)}
\, {\it w}_{gq\rightarrow gq}^{(bf_1;\,b_1f_2)}
({\bf k}, {\bf q}_1;\, {\bf k}_1, {\bf q}_2)N_{{\bf k}_1}^{(b_1)}
\Bigl(1-n_{{\bf q}_1}^{(f_1)}\Bigr)
n_{{\bf q}_2}^{(f_2)}
\]
\[
\hspace{4.5cm}
+\!\int\!d{\cal T}_{g\bar{q}\rightarrow g\bar{q}}^{(bf_1;\,b_1f_2)}\,
{\it w}_{g\bar{q}\rightarrow g\bar{q}}^{(bf_1;\,b_1f_2)}
({\bf k}, {\bf q}_1;\, {\bf k}_1, {\bf q}_2)N_{{\bf k}_1}^{(b_1)}
\Bigl(1-\bar{n}_{{\bf q}_1}^{(f_1)}\Bigr)
\bar{n}_{{\bf q}_2}^{(f_2)}
\]
\[
\hspace{5.05cm}
+\!\int\!d{\cal T}_{gg\rightarrow q\bar{q}}^{(b\,b_1;\,f_1f_2)} \,
{\it w}_{gg\rightarrow q\bar{q}}^{(b\,b_1;\,f_1f_2)}
({\bf k}, {\bf k}_1;\, {\bf q}_1, {\bf q}_2)
\Bigl(1+N_{{\bf k}_1}^{(b_1)}\Bigr)
\bar{n}_{{\bf q}_1}^{(f_1)}n_{{\bf q}_2}^{(f_2)}\Biggr\},
\hspace{0.8cm}
\]
where, ${\it w}_{gq\rightarrow gq}^{(bf_1;\,b_1f_2)},
\,{\it w}_{g\bar{q}\rightarrow g\bar{q}}^{(bf_1;\,b_1f_2)}$
are probabilities for `elastic' scattering of soft-gluon excitations off
soft-quark and soft-antiquark excitations, respectively, and
${\it w}_{gg\rightarrow q\bar{q}}^{(bb_1;\,f_1f_2)}$ is the
probability for soft quark-antiquark pair creation by two soft-gluon fusion.
The phase-space measures for these processes are
\begin{equation}
\int\!d{\cal T}_{gq\rightarrow gq}^{(bf_1;\,b_1f_2)}=\!
\int\!d{\cal T}_{g\bar{q}\rightarrow g\bar{q}}^{(bf_1;\,b_1f_2)}
\equiv\!\!
\int\!\!\frac{d{\bf k}_1}{(2\pi)^3}\!\int\!\!\frac{d{\bf q}_1}{(2\pi)^3}\!
\int\!\!\frac{d{\bf q}_2}{(2 \pi)^3}\,
(2\pi)^4\delta({\bf k}+{\bf q}_1-{\bf k}_1-{\bf q}_2)
\label{eq:2o}
\end{equation}
\[
\times
\delta(\omega_{{\bf k}}^{(b)} + \omega_{{\bf q}_1}^{(f_1)}-
\omega_{{\bf k}_1}^{(b_1)} - \omega_{{\bf q}_2}^{(f_2)}),
\]
\[
\int\!d{\cal T}_{gg \rightarrow q\bar{q}}^{(bb_1;\,f_1f_2)}\equiv\!
\int\!\!\frac{d{\bf k}_1}{(2\pi)^3}\!\int\!\!\frac{d{\bf q}_1}{(2\pi)^3}\!
\int\!\!\frac{d{\bf q}_2}{(2 \pi)^3}\,
(2\pi)^4\delta({\bf k}+{\bf k}_1-{\bf q}_1-{\bf q}_2)\,
\delta(\omega_{{\bf k}}^{(b)}+\omega_{{\bf k}_1}^{(b_1)}-
\omega_{{\bf q}_1}^{(f_1)}-\omega_{{\bf q}_2}^{(f_2)}).
\hspace{1.2cm}
\]

We have written out above the most general expressions for the
fermion and boson decay and regenerating rates to the lowest order
in the nonlinear interaction taking into account all possible
channels of transitions from initial two-quasiparticle states to
different types of the final two-quasiparticle states. However it
is clear that not all these processes of scattering, annihilations
and fusions are kinematically permissible\footnote{Here, as
a kinematic permissibility we mean generally speaking not only the
existence of a solution for the system of conservation laws of
energy and momentum corresponding to concrete reaction of the
nonlinear interaction of two quasiparticles, but also selection
over helicity conservation of a quasiparticle system.}. In this
work we do not {\it set} as a purpose to determine all permissible
channels of reactions that require as a first step a study of a
system of equations determining the conservation laws of energy
and momentum of two-quasiparticles interacting system.
Unfortunately at present there not exist general analytic
methods\footnote{In the work of Kadomtsev and Kontorovich
\cite{kadomtsev} the geometrical method of solving the system of
conservation laws for two-to-two scattering processes of
quasiparticles with arbitrary dispersion laws was suggested.
Unfortunately, this approach having sufficiently general character
is restricted only to the case of two-dimensional momenta of
quasiparticles.} permitting in a direct way to define whether or
not $\delta$-functions have supports different from zero in
integration measures (\ref{eq:2u}) and (\ref{eq:2o}) and thus for
solving this problem it should be used numerical methods. In the
remaining part of the paper we restrict our consideration to
calculation of the probabilities of only processes, which
certainly exist (for example, the elastic scattering) leaving the
general case for a separate research.

\section{\bf Effective currents and effective amplitudes}
\setcounter{equation}{0}

We use the metric $g^{\mu \nu} = {\rm diag}(1,-1,-1,-1)$, choose units such
that $c=k_{B}=1$ and note $x=(t,{\bf x}),\,k=(k_0,{\bf k}),\,q=(q_0,{\bf q})$
etc. As was mentioned above, we consider ${\rm SU}(N_c)$ gauge
theory with $N_f$ flavors of massless quarks. The color indices for the
adjoint representation $a,b, \ldots$ run from 1 to $N_c^2-1$, while those for
the fundamental representation $i,j, \ldots$ run from 1 to $N_c$.
The Greek indices $\alpha, \beta, \ldots$ for the spinor representation run
from 1 to 4. In the following discussion we will use notation adopted
in our previous paper \cite{markov_PRD}.

The input equations for the effective theory under consideration are
self-consistent system of the field equations:
the Yang-Mills equation for gauge potential $A_{\mu}^a$ (k) (Eq.\,(4.9) in
Ref.\,\cite{markov_PRD})
\[
\,^{\ast}{\cal D}^{-1}_{\mu\nu}(k) A^{a\nu}(k) =
- j^{A(2)\,a}_{\mu}(A,A)(k) - j^{A(3)\,a}_{\mu}(A,A,A)(k) -
j^{\Psi(0,2)\,a}_{\mu}(\bar{\psi},\psi)(k)
\]
\begin{equation}
-j^{\Psi(1,2)\,a}_{\mu}(A,\bar{\psi},\psi)(k),
\label{eq:3q}
\end{equation}
and the Dirac equation for soft-quark field (Eq.\,(3.3) in \cite{markov_PRD})
supplemented by its Dirac conjugate equation for completeness of the 
description
\begin{equation}
\,^{\ast}\!S^{-1}_{\alpha\beta}(q)\psi^i_{\beta}(q) =
-\,\eta^{(1,1)\,i}_{\alpha}(A,\psi)(q) - \eta^{(2,1)\,i}_{\alpha}(A,A,\psi)(q),
\label{eq:3w}
\end{equation}
\[
\bar{\psi}^i_{\beta}(-q)\,^{\ast}\!S^{-1}_{\beta\alpha}(-q) =
\bar{\eta}^{(1,1)\,i}_{\alpha}(A^{\ast},\bar{\psi})(-q) +
\bar{\eta}^{(2,1)\,i}_{\alpha}(A^{\ast},A^{\ast},\bar{\psi})(-q).
\]
Here, in the last line we take into account the following relations:
$\gamma^0\,^{\ast}\!S^{\dagger}(-q)\gamma^0=-\,^{\ast}\!S(q)$ and
$\bar{\eta}(-q)=\eta^{\dagger}(q)\gamma^0$.
On the right-hand side of Eqs.\,(\ref{eq:3q}), (\ref{eq:3w}) in the expansion 
of induced currents and sources
we keep the terms up to the third order in interacting fields containing  
a relevant information on the two-to-two scattering processes.
The expansion terms of induced currents\footnote{Here, we somewhat
redefine HTL-induced currents
$j^{A(2)}$, $j^{A(3)}$, $j^{\Psi(0,2)}$ and source $\eta^{(1,1)}$ including
bare vertices in their definition (see, below).}
$j^A$ and $j^{\psi}$ on the right-hand side of Eq.\,(\ref{eq:3q}) in the
{\it hard thermal loop} (HTL) approximation have the following structure:
\[
j^{A(2)a}_{\mu}(A,A)(k)=
\frac{1}{2!}\,g\, (T^a)^{bc}\!\!\int\!^\ast\Gamma_{\mu\nu\lambda}(k,-k_1,-k_2)
A^{b\nu}(k_1)A^{c\lambda}(k_2)\,
\delta(k-k_1-k_2) dk_1 dk_2,
\hspace{0.3cm}
\]
\begin{equation}
j^{\Psi(0,2)a}_{\mu}(\bar{\psi},\psi)(k)=
g(t^a)^{ij}\!\!\int\!\!\,^{\ast}\Gamma^{(G)}_{\mu,\,\alpha\beta}(k;q_1,-q_2)
\bar{\psi}^i_{\alpha}(-q_1)\psi^j_{\beta}(q_2)
\,\delta(k + q_1 - q_2) dq_1 dq_2,
\hspace{0.3cm}
\label{eq:3e}
\end{equation}
\[
j^{\Psi(1,2)a}_{\mu}(A,\bar{\psi},\psi)(k)=
g^2\!\!\int\!\delta\Gamma^{(G)ab,\,ij}_{\mu\nu,\,\alpha\beta}(k,-k_1;q_1,-q_2)
A^{b\nu}(k_1)\bar{\psi}^i_{\alpha}(-q_1)\psi^j_{\beta}(q_2)
\hspace{2.5cm}
\]
\[
\times\,\delta(k+q_1-k_1-q_2)dk_1dq_1dq_2,
\]
where $(T^a)^{bc}\equiv -if^{abc}$.
$^{\ast}\Gamma_{\mu\nu\lambda}$ is HTL-resummed three-gluon vertex and
$^{\ast}\Gamma^{(G)}_{\mu,\,\alpha\beta}$ is HTL-resummed vertex between
quark pair and gluon. They represent a sum of bare vertex and
corresponding HTL-correction.
$\delta\Gamma^{(G)ab,\,ij}_{\mu\nu,\,\alpha\beta}$ is an effective
(HTL-induced) vertex between quark pair and two gluons.
The current $j^{A(3)}$ defines a self-action of soft bosonic field
considered in Ref.\,\cite{markov1, markov3}. The explicit form of
HTL-amplitudes in integrands in
Eq.\,(\ref{eq:3e}) can be found in Ref.\,\cite{blaizot}. The superscript $(G)$
points to the fact that in the coordinate representation the time argument of
external gluon leg incoming in the vertex functions
$^{\ast}\Gamma^{(G)}_{\mu,\,\alpha \beta}$ and
$\delta\Gamma^{(G)ab,\,ij}_{\mu\nu,\,\alpha\beta}$ is largest.

Furthermore, the induced sources on the right-hand side of Dirac equation
(\ref{eq:3w}) have the following structure:
\begin{equation}
\eta^{(1,1)i}_{\alpha}(A,\psi)(q) =  g\,(t^a)^{ij}\!\!\int\!
\,^{\ast}\Gamma^{(Q)}_{\mu,\,\alpha\beta}(k_1;q_1,-q)
A^{a\mu}(k_1)\psi^j_{\beta}(q_1)\,\delta(q - q_1 - k_1)dk_1dq_1,
\label{eq:3r}
\end{equation}
\[
\eta^{(2,1)i}_{\alpha}(A,A,\psi)(q) = \frac{1}{2!}\,g^2\!\!\int\!
\delta\Gamma^{(Q)ab,\,ij}_{\mu\nu,\,\alpha\beta}(k_1,k_2;q_1,-q)
A^{a\mu}(k_1)A^{b\nu}(k_2)\psi^j_{\beta}(q_1)
\]
\[
\hspace{1.7cm}
\times\,\delta(q - q_1 - k_1 -k_2)dq_1dk_1dk_2.
\]
Here, $\,^{\ast}\Gamma^{(Q)}_{\mu,\,\alpha\beta}$ and
$\delta\Gamma^{(Q)ab,\,ij}_{\mu\nu,\,\alpha\beta}$ are HTL-resummed vertices
between quark pair and gluon, and quark pair and two gluons, respectively.
The superscript $(Q)$ denotes that these vertices in the coordinate
representation have the largest time argument for external quark incoming leg.

For conjugate induced sources $\bar{\eta}^{(1,1)a}_{\alpha}$ and
$\bar{\eta}^{(2,1)a}_{\alpha}$ by virtue of properties (A.3) and (A.4),
we have from (\ref{eq:3r})
\begin{equation}
\bar{\eta}^{(1,1)i}_{\alpha}(A^{\ast},\bar{\psi})(-q) =
g\,(t^a)^{ji}\!\!\int\!
\,^{\ast}\Gamma^{(Q)}_{\mu,\,\beta\alpha}(-k_1;-q_1,q)
A^{\ast a\mu}(k_1)\bar{\psi}^j_{\beta}(-q_1)\,\delta(q - q_1 - k_1)dk_1dq_1,
\label{eq:3rr}
\end{equation}
\[
\bar{\eta}^{(2,1)i}_{\alpha}(A^{\ast},A^{\ast},\bar{\psi})(-q)
= -\,\frac{1}{2!}\,g^2\!\!\int\!
\delta\Gamma^{(Q)ba,\,ji}_{\mu\nu,\,\beta\alpha}(-q,q_1;k_1,k_2)
A^{\ast a\mu}(k_1)A^{\ast b\nu}(k_2)\bar{\psi}^j_{\beta}(-q_1)
\hspace{0.9cm}
\]
\[
\hspace{1.3cm}
\times\,\delta(q - q_1 - k_1 -k_2)dq_1dk_1dk_2.
\]
The $\delta\Gamma^{(G)ab,\,ij}_{\mu\nu,\,\alpha\beta}$ and
$\delta\Gamma^{(Q)ab,\,ij}_{\mu\nu,\,\alpha\beta}$ vertices don't
exist at tree level and to leading order they arise
entirely from the HTL \cite{braaten2, taylor, blaizot}. In the subsequent
discussion we are needed their color structure, therefore an explicit
form of the vertex functions is given in Appendix A.

Lastly $\,^{\ast}{\cal D}_{\mu\nu}(k)$ and $\,^{\ast}\!S(q)$ are medium
modified gluon (in the covariant gauge) and quark propagators
\[
\,^{\ast}{\cal D}_{\mu\nu}(k) =
- P_{\mu \nu}(k) \,^{\ast}\!\Delta^t(k) -
Q_{\mu \nu}(k) \,^{\ast}\!\Delta^l(k)
+\,\xi D_{\mu\nu}(k)\Delta^{0}(k),
\]
\[
\,^{\ast}\!S(q) = h_{+}(\hat{\mathbf q}) \,^{\ast}\!\triangle_{+}(q) +
h_{-}(\hat{\mathbf q}) \,^{\ast}\!\triangle_{-}(q).
\]
Here in the first line, the Lorentz matrices are defined by
\[
P_{\mu \nu}(k) = g_{\mu \nu} - D_{\mu \nu}(k) - Q_{\mu \nu}(k), \quad
Q_{\mu \nu}(k) = \frac{\bar{u}_{\mu} (k) \bar{u}_{\nu} (k)}{\bar{u}^2(k)},
\quad D_{\mu\nu}(k)=\frac{k_{\mu}k_{\nu}}{k^2},
\]
\[
\Delta^0(k)=\frac{1}{k^2},\quad
\bar{u}_{\mu}(k)=k^2u_{\mu} - k_{\mu}(k\cdot u).
\]
The scalar transverse and longitudinal gluon propagators are
$\,^{\ast}\!\Delta^{t,\,l}(k)=1/(k^2-\delta\Pi^{t,\,l}(k))$, where 
$\delta\Pi^{t,\,l}(k)$
are transverse and longitudinal soft-gluon self-energies, $\xi$ is a gauge
fixing parameter. Let us assume that we are in a rest frame of a heat bath, so that
$u_{\mu}=(1,0,0,0).$ In the soft-quark propagator the matrix functions
$h_{\pm}(\hat{\bf q})=(\gamma^0 \mp \hat{{\bf q}}\cdot{\bf\gamma})/2$ with
$\hat{{\bf q}} \equiv {\bf q}/\vert {\bf q} \vert$ are the spinor projectors
onto eigenstates of helicity and
\begin{equation}
\,^{\ast}\!\triangle_{\pm}(q) =
-\,\frac{1}{q^0\mp [\,\vert{\bf q}\vert + \delta\Sigma_{\pm}(q)]},
\label{eq:3t}
\end{equation}
where $\delta\Sigma_{\pm}(q)$ are scalar quark self-energies for
normal $(+)$ and plasmino $(-)$ modes.

The system of nonlinear integral equations (\ref{eq:3q}), (\ref{eq:3w}) can
be perturbatively solved (at least in the weak-field limit) by the
approximation scheme method. Discarding the nonlinear terms in $A$, $\psi$ and
$\bar{\psi}$ on the right-hand side of Eqs.\,(\ref{eq:3q}), (\ref{eq:3w}),
we obtain in zero approximation
\[
\,^{\ast}{\cal D}^{-1}_{\mu\nu}(k) A^{a\nu}(k) = 0,\quad
\,^{\ast}\!S^{-1}_{\alpha\beta}(q)\psi^i_{\beta}(q) = 0,\quad
\bar{\psi}^i_{\beta}(-q)\,^{\ast}\!S^{-1}_{\beta\alpha}(-q) = 0.
\]
The solutions of these equations denoted by $A^{(0)a}_{\mu}(k)$,
$\psi^{(0)i}_{\alpha}(q)$ and $\bar{\psi}^{(0)i}_{\alpha}(-q)$ are the
solutions for free fields. By iterating Eqs.\,(\ref{eq:3q}), (\ref{eq:3w})
in general case we obtain the series
\begin{equation}
A_{\mu}^a(k)=\sum_{s=1}^{\infty}A_{\mu}^{(s-1)\,a}(k),\quad
\psi^{i}_{\alpha}(q)=\sum_{s=1}^{\infty}\psi^{(s-1)i}_{\alpha}(q),\quad
\bar{\psi}^i_{\alpha}(-q)=
\sum_{s=1}^{\infty}\bar{\psi}^{(s-1)i}_{\alpha}(-q),
\label{eq:3y}
\end{equation}
where $A_{\mu}^{(s-1)\,a}(k)$, $\psi^{(s-1)i}_{\alpha}(q)$ and
$\bar{\psi}^{(s-1)i}_{\alpha}(-q)$ are contributions of $g^{s-1}$ order to
the soft-gluon and soft-quark interacting fields\footnote{Note that by virtue 
of choice of the right-hand sides of Eqs.\,(\ref{eq:3q}) and (\ref{eq:3w}), 
expansions (\ref{eq:3y}) accurate up to $s=3$. For higher $s$ it is necessary 
to add the subsequent terms of the expansions of induced currents
$j^A, \, j^{\psi}$ and sources $\eta, \, \bar{\eta}$ to the right-hand sides
of Eqs.\,(\ref{eq:3q}), (\ref{eq:3w}).}. Substituting expansions (\ref{eq:3y})
into system (\ref{eq:3q}), (\ref{eq:3w}), we obtain iterative solutions 
of higher order in $g$:

\noindent{\it to the first order}
\begin{equation}
A_{\mu}^{(1)a}(k) =
-^{\ast}{\cal D}_{\mu\nu}(k)[\,\tilde{j}^{A(2)a\nu}(A^{(0)},A^{(0)})+
\tilde{j}^{\Psi(0,2)a\nu}(\bar{\psi}^{(0)},\psi^{(0)})],
\label{eq:3u}
\end{equation}
\[
\psi^{(1)i}_{\alpha}(q) = -\,^{\ast}\!S_{\alpha\beta}(q)
\tilde{\eta}^{(1,1)\,i}_{\beta}(A^{(0)},\psi^{(0)}),\quad
\bar{\psi}^{(1)i}_{\alpha}(-q) =
\tilde{\bar{\eta}}^{(1,1)\,i}_{\beta}(A^{\ast(0)},\bar{\psi}^{(0)})
\,^{\ast}\!S_{\beta\alpha}(-q),
\]
{\it to the second order}
\begin{equation}
A_{\mu}^{(2)a}(k) =
-^{\ast}{\cal D}_{\mu\nu}(k)[\,\tilde{j}^{A(3)a\nu}(A^{(0)},A^{(0)},A^{(0)})+
\tilde{j}^{\Psi(1,2)a\nu}(A^{(0)},\bar{\psi}^{(0)}\psi^{(0)})],
\label{eq:3i}
\end{equation}
\[
\psi^{(2)i}_{\alpha}(q) = -\,^{\ast}\!S_{\alpha\beta}(q)
[\,\tilde{\eta}^{(2,1)\,i}_{\beta}(A^{(0)},A^{(0)},\psi^{(0)})+
\tilde{\eta}^{(0,3)\,i}_{\beta}(\bar{\psi}^{(0)},\psi^{(0)},\psi^{(0)})],
\hspace{0.5cm}
\]
\[
\bar{\psi}^{(2)i}_{\alpha}(-q) =
[\,\tilde{\bar{\eta}}^{(2,1)\,i}_{\beta}
(A^{\ast(0)},A^{\ast(0)},\bar{\psi}^{(0)})+
\tilde{\bar{\eta}}^{(0,3)\,i}_{\beta}
(\bar{\psi}^{(0)},\bar{\psi}^{(0)},\psi^{(0)})]
\,^{\ast}\!S_{\beta\alpha}(-q)
\]
and so on. Here, on the right-hand side new
effective\footnote{Hereafter all effective quantities such as
currents, sources, vertices etc. will be denoted by the same
letters as initial quantities with tilde above only.} currents and
sources being functionals of free fields appear. The effective
currents and sources to first order approximation (\ref{eq:3u}) are associated
with initial currents (\ref{eq:3e}) and sources (\ref{eq:3r}), (\ref{eq:3rr})
by a simple way
\[
\tilde{j}^{A(2)a\nu}(A^{(0)},A^{(0)})\equiv j^{A(2)a\nu}(A^{(0)},A^{(0)}),
\quad
\tilde{j}^{\Psi(0,2)a\nu}(\bar{\psi}^{(0)},\psi^{(0)})\equiv
j^{\Psi(0,2)a\nu}(\bar{\psi}^{(0)},\psi^{(0)}),
\]
\[
\tilde{\eta}^{(1,1)\,i}_{\beta}(A^{(0)},\psi^{(0)})\equiv
{\eta}^{(1,1)\,i}_{\beta}(A^{(0)},\psi^{(0)}),\quad\;\;\;
\tilde{\bar{\eta}}^{(1,1)\,i}_{\beta}(A^{\ast(0)},\bar{\psi}^{(0)})\equiv
{\bar{\eta}}^{(1,1)\,i}_{\beta}(A^{\ast(0)},\bar{\psi}^{(0)}).
\]
The explicit form of the effective current $\tilde{j}^{A(3)a\nu}$ to second 
approximation order (\ref{eq:3i}) was obtained in Ref.\,\cite{markov3}. The 
remaining effective currents and sources have the following structures:
\[
\tilde{j}^{\Psi(1,2)a}_{\mu}(A^{(0)},\bar{\psi}^{(0)},\psi^{(0)})=
g^2\!\!\int\!
\!\,^{\ast}\tilde{\Gamma}^{(G)aa_1,\,ij}_{\mu{\mu}_1,\,\alpha\beta}
(k,-k_1;q_1,-q_2)
A^{(0)a_1{\mu}_1}(k_1)\bar{\psi}^{(0)i}_{\alpha}(-q_1)
\psi^{(0)j}_{\beta}(q_2)
\hspace{0.2cm}
\]
\begin{equation}
\times\,\delta(k+q_1-k_1-q_2)dk_1dq_1dq_2,
\label{eq:3o}
\end{equation}
\[
\tilde{\eta}^{(2,1)i}_{\alpha}(A^{(0)},A^{(0)},\psi^{(0)}) =
\frac{g^2}{2!}\!\int\!
\!\,^{\ast}\tilde{\Gamma}^{(Q)a_1a_2,\,ij}_{\mu_1\mu_2,\alpha\beta}
(k_1,k_2;q_1,-q)
A^{(0)a_1\mu_1}(k_1)A^{(0)a_2\mu_2}(k_2)\psi^{(0)j}_{\beta}(q_1)
\hspace{0.5cm}
\]
\begin{equation}
\times\,\delta(q - q_1 - k_1 -k_2)dq_1dk_1dk_2,
\label{eq:3p}
\end{equation}
\[
\tilde{\eta}^{(0,3)\,i}_{\alpha}(\bar{\psi}^{(0)},\psi^{(0)},\psi^{(0)})=
\frac{g^2}{2!}\!\int\!
\!\,^{\ast}\tilde{\Gamma}^{ii_1i_2i_3}_{\alpha\alpha_1\alpha_2\alpha_3}
(q,q_1,-q_2,-q_3)
\bar{\psi}^{(0)i_1}_{\alpha_1}(-q_1)
\psi^{(0)i_2}_{\alpha_2}(q_2)\psi^{(0)i_3}_{\alpha_3}(q_3)
\hspace{0.6cm}
\]
\begin{equation}
\times\,\delta(q + q_1 - q_2 -q_3)dq_1dq_2dq_3,
\label{eq:3a}
\end{equation}
and correspondingly, (Dirac) conjugate effective sources are
\[
\tilde{\bar{\eta}}^{(2,1)i}_{\alpha}
(A^{\ast(0)},A^{\ast(0)},\bar{\psi}^{(0)}) =
\frac{g^2}{2!}\!\!\int\!
\!\,^{\ast}\tilde{\bar{\Gamma}}^{(Q)a_1a_2,\,ij}_{\mu_1\mu_2,\alpha\beta}
(k_1,k_2;q_1,-q)
A^{\ast(0)a_1\mu_1}(k_1)A^{\ast(0)a_2\mu_2}(k_2)\bar{\psi}^{(0)j}_{\beta}(-q_1)
\]
\begin{equation}
\times\,\delta(q - q_1 - k_1 -k_2)dq_1dk_1dk_2,
\label{eq:3s}
\end{equation}
\[
\tilde{\bar{\eta}}^{(0,3)\,i}_{\beta}
(\bar{\psi}^{(0)},\bar{\psi}^{(0)},\psi^{(0)})=
\frac{g^2}{2!}\!\int\!
\!\,^{\ast}\tilde{\bar{\Gamma}}^{ii_1i_2i_3}_{\alpha\alpha_1\alpha_2\alpha_3}
(q,q_1,-q_2,-q_3)
\bar{\psi}^{(0)i_3}_{\alpha_3}(-q_3)
\bar{\psi}^{(0)i_2}_{\alpha_2}(-q_2)
\psi^{(0)i_1}_{\alpha_1}(q_1)
\hspace{1.7cm}
\]
\begin{equation}
\times\,\delta(q + q_1 - q_2 -q_3)dq_1dq_2dq_3.
\label{eq:3d}
\end{equation}
The effective amplitudes entering into the integrand of the second iteration
(\ref{eq:3o})\,--\,(\ref{eq:3d}) (and also entering into higher 
iterations $s>3$)
represent highly nontrivial combinations of the HTL-amplitudes and equilibrium
soft-gluon and soft-quark propagators. These effective amplitudes
can be defined in a direct way starting from the connection between the
effective and initial currents. Thus for example the effective
amplitude in (\ref{eq:3o}) is defined from the relation
\[
\tilde{j}^{\Psi(1,2)a}_{\mu}(A^{(0)},\bar{\psi}^{(0)},\psi^{(0)})(k)=
j^{\Psi(1,2)a}_{\mu}(A^{(0)},\bar{\psi}^{(0)},\psi^{(0)})(k)
\]
\[
+
\,\{
j^{A(2)a\nu}(-^{\ast}{\cal D}
\tilde{j}^{\Psi(0,2)}(\bar{\psi}^{(0)},\psi^{(0)}),A^{(0)})
+j^{A(2)a\nu}(A^{(0)},-^{\ast}{\cal D}
\tilde{j}^{\Psi(0,2)}(\bar{\psi}^{(0)},\psi^{(0)}))
\}\hspace{0.2cm}
\]
\[
+\,\{
j^{\Psi(0,2)a\nu}
(\tilde{\bar{\eta}}^{(1,1)}(A^{\ast(0)},\bar{\psi}^{(0)})\!\,^{\ast}\!S,
\psi^{(0)})
+j^{\Psi(0,2)a\nu}(\bar{\psi}^{(0)},
-\,^{\ast}\!S\tilde{\eta}^{(1,1)}(A^{(0)},\psi^{(0)}))
\}.
\]
However, such a direct approach for determination of the explicit
form of the effective amplitudes is very complicated even in the
second order approximation and as a consequence, ineffective. In
the forthcoming section we will consider somewhat different
approach to the calculation of the effective amplitudes, which
allows us to avoid many intermediate operations and to a certain
extent to automate the calculation procedure. This approach is
extension of the calculating method for purely gluonic effective
amplitudes suggested in Ref.\,\cite{markov1} to the case of an
existence of soft-quark degree of freedom in the system. New
features here, is appearing effective sources (\ref{eq:3a}), (\ref{eq:3d}), 
which don't have counterparts in the expansion of the induced currents and
sources in initial system (\ref{eq:3q}), (\ref{eq:3w}).

\section{\bf Calculation of effective amplitudes}
\setcounter{equation}{0}

The calculating algorithm of the effective amplitudes is based on the
following idea. The total induced current $j^{Aa}_{\mu}+j^{\Psi
a}_{\mu}$, and induced source $\eta_{\alpha}^i$ have two
representations: by means of interacting and free fields that
must be equal each other
\begin{equation}
j^{Aa}_{\mu}[A]+j^{\Psi a}_{\mu}[A,\bar{\psi},\psi]=
\sum_{s=2}^{\infty} j^{A(s)a}_{\mu}\underbrace{(A,\ldots,A)}_{s}+
\sum_{n=0}^{\infty}j^{\Psi(n,\,2)a}_{\mu}(\underbrace{A,\ldots,A}_{n},
\bar{\psi},\psi)=
\label{eq:4q}
\end{equation}
\[
\sum_{s=2}^{\infty}\tilde{j}^{A(s)a}_{\mu}
(\underbrace{A^{(0)},\ldots,A^{(0)}}_{s})
+ \sum_{n=0}^{\infty}\sum_{l=1}^{\infty}
\tilde{j}^{\Psi(n,\,2l)a}_{\mu}(\underbrace{A^{(0)},\ldots,A^{(0)}}_{n},
\underbrace{\bar{\psi}^{(0)},
\psi^{(0)},\ldots,\bar{\psi}^{(0)},\psi^{(0)}}_{2l}),
\]
\begin{equation}
\eta_{\alpha}^i[A,\psi]=
\sum_{s=1}^{\infty}\eta^{(s,\,1)i}_{\alpha}
(\underbrace{A,\ldots,A}_{s},\psi)=
\label{eq:4w}
\end{equation}
\[
\sum_{s=1}^{\infty}\tilde{\eta}^{(s,\,1)i}_{\alpha}
(\underbrace{A^{(0)},\ldots,A^{(0)}}_{s},\psi^{(0)})+
\sum_{n=0}^{\infty}\sum_{l=1}^{\infty}
\tilde{\eta}^{(n,\,2l+1)i}_{\alpha}(\underbrace{A^{(0)},\ldots,A^{(0)}}_{n},
\underbrace{\bar{\psi}^{(0)},
\psi^{(0)},\ldots,\psi^{(0)}}_{2l+1}).
\]
Equation (\ref{eq:4w}) is to be supplemented by a similar equation for 
conjugate source $\bar{\eta}^{\,i}_{\alpha}$. The explicit form of the 
functions $j^{A(s)a}_{\mu}(A,\ldots,A)$ and
$\tilde{j}^{A(s)a}_{\mu}(A^{(0)},\ldots,A^{(0)})$ was defined in
Ref.\,\cite{markov1} (Eqs.\,(3.9) and (7.8)). The original currents
$j^{\Psi(n,\,2)a}_{\mu}(A,\ldots,A,\bar{\psi},\psi)$ and sources
$\eta^{(n,\,1)i}_{\alpha}(A,\ldots,A,\psi)$ are expressed as
\[
j^{\Psi(n,\,2)a}_{\mu}(A,\ldots,A,\bar{\psi},\psi)(k)=
\frac{1}{n!}\,g^{n+1}\!\!\int\!
\,^{\ast}\Gamma^{(G)aa_1\ldots a_n,\,ij}_{\mu\mu_1\ldots\mu_n,\,\alpha\beta}
(k,-k_1,\ldots,-k_n;q_1,-q_2)
\]
\[
\times
A^{a_1\mu_1}(k_1)\ldots A^{a_n\mu_n}(k_n)
\bar{\psi}^i_{\alpha}(-q_1)\psi^j_{\beta}(q_2)
\]
\begin{equation}
\times\delta\!\left(k-\sum_{i=1}^{n}k_i+q_1-q_2\right)\!
dq_1dq_2\!\prod_{i=1}^{n}\!dk_i,\!\quad n=0,1,\ldots\,,
\label{eq:4e}
\end{equation}
\[
\eta^{(s,\,1)i}_{\alpha}(A,\ldots,A,\psi)(q)=
\frac{1}{s!}\,g^{s}\!\!\int\!
\,^{\ast}\Gamma^{(Q)a_1\ldots a_s,\,ij}_{\mu_1\ldots\mu_s,\,\alpha\beta}
(k_1,\ldots,k_s;q_1,-q)
\]
\[
\times
A^{a_1\mu_1}(k_1)\ldots A^{a_s\mu_s}(k_s)
\psi^j_{\beta}(q_1)
\,\delta\!\left(q-q_1-\sum_{i=1}^{s}k_i\right)\!
dq_1\prod_{i=1}^{s}dk_i,\quad s=1,2,\ldots\,.
\]
Here, the HTL-amplitudes
$\,^{\ast}\Gamma^{(G)aa_1\ldots a_n,\,ij}_{\mu\mu_1\ldots\mu_n,\,\alpha\beta}$
for $n=0$ and
$\,^{\ast}\Gamma^{(Q)a_1\ldots a_s,\,ij}_{\mu_1\ldots\mu_s,\,\alpha\beta}$
for $s=1$ represent the sum of bare vertices and corresponding HTL-correction.
For $n>0$, $s>1$ we rename $^\ast\Gamma^{(G)}\equiv\delta\Gamma^{(G)}$,
$^\ast\Gamma^{(Q)}\equiv\delta\Gamma^{(Q)}$.

The interacting fields on the left-hand side of equations (\ref{eq:4q})
and (\ref{eq:4w}) are defined by the expansion
\[
A^{a\mu}(k) = A^{(0)\,a\mu}(k) -\!
\,^{\ast}\tilde{\cal D}^{\mu\nu}(k)\Biggl[\,\sum_{s=2}^{\infty}
\tilde{j}^{A(s)a}_{\nu}(\underbrace{A^{(0)},\ldots,A^{(0)}}_{s})
\]
\begin{equation}
+ \sum_{n=0}^{\infty}\sum_{l=1}^{\infty}
\tilde{j}^{\Psi(n,\,2l)a}_{\mu}(\underbrace{A^{(0)},\ldots,A^{(0)}}_{n},
\underbrace{\bar{\psi}^{(0)},
\psi^{(0)},\ldots,\bar{\psi}^{(0)},\psi^{(0)}}_{2l})\biggr],
\label{eq:4r}
\end{equation}
\[
\psi^i_{\alpha}(q)=
\psi^{(0)i}_{\alpha}(q) -
\,^\ast\!S_{\alpha\beta}(q)
\Biggl[\,
\sum_{s=1}^{\infty}\tilde{\eta}^{(s,\,1)i}_{\beta}
(\underbrace{A^{(0)},\ldots,A^{(0)}}_{s},\psi^{(0)})
\]
\[
+
\sum_{n=0}^{\infty}\sum_{l=1}^{\infty}
\tilde{\eta}^{(n,\,2l+1)a}_{\alpha}
(\underbrace{A^{(0)},\ldots,A^{(0)}}_{n},
\underbrace{\bar{\psi}^{(0)},
\psi^{(0)},\ldots,\psi^{(0)}}_{2l+1})\biggr].
\]
Similar expansion holds for conjugate function $\bar{\psi}_{\alpha}^i(-q)$.
The calculation of purely gluonic effective currents $\tilde{j}^{A(s)a}_{\mu}
(A^{(0)},\ldots,A^{(0)})$ was considered in
Ref.\,\cite{markov1}. The remaining effective currents and
sources have the following structure:
\[
\tilde{j}^{\Psi(n,\,2l)a}_{\mu}(A^{(0)},\ldots,A^{(0)},
\bar{\psi}^{(0)},\psi^{(0)},\ldots,\bar{\psi}^{(0)},\psi^{(0)})(k)=
\]
\[
=
\frac{1}{n!}\,\frac{1}{(l!)^2}\,g^{n+l}\!\!\int\!
\,^{\ast}\tilde{\Gamma}^{(G)aa_1\ldots a_n,\,i_1\ldots i_l j_1\ldots j_l
}_{\mu\mu_1\ldots\mu_n,\,\alpha_1\ldots\alpha_l\beta_1\ldots\beta_l}
(k,-k_1,\ldots,-k_n;q_1,\ldots q_l,-q^{\prime}_1,\ldots-q^{\prime}_l)
\]
\[
\times
A^{(0)a_1\mu_1}(k_1)\ldots A^{(0)a_n\mu_n}(k_n)
\bar{\psi}^{(0)i_1}_{\alpha_1}(-q_1)\ldots
\bar{\psi}^{(0)i_l}_{\alpha_l}(-q_l)
\psi^{(0)j_1}_{\beta_1}(q_1^{\prime})\ldots
\psi^{(0)j_l}_{\beta_l}(q_l^{\prime})
\]
\begin{equation}
\times
\,\delta\!\left(k-\sum_{i=1}^{n}k_i+\sum_{i=1}^{l}(q_i-q_i^{\prime})\right)
\!\prod_{i=1}^{n}\!dk_i\prod_{j=1}^{l}\!dq_jdq_j^{\prime},
\label{eq:4t}
\end{equation}
\[
\tilde{\eta}^{(s,\,1)i}_{\alpha}(A^{(0)},\ldots,A^{(0)},\psi^{(0)})(q)=
\frac{1}{s!}\,g^{s}\!\!\int\!
\,^{\ast}\tilde{\Gamma}^{(Q)a_1\ldots a_s,\,ij}_{\mu_1\ldots\mu_s,
\,\alpha\beta}
(k_1,\ldots,k_s;q_1,-q)
\]
\[
\times
A^{(0)a_1\mu_1}(k_1)\ldots A^{(0)a_s\mu_s}(k_s)
\psi^{(0)j}_{\beta}(q_1)
\,\delta\!\left(q-q_1-\sum_{i=1}^{s}k_i\right)\!
dq_1\prod_{i=1}^{s}dk_i,
\]
etc. The substitution of Eqs.\,(\ref{eq:4r}), (\ref{eq:4t}) into (\ref{eq:4q})
and (\ref{eq:4w}) turns the last equations into identities. Now we functionally
differentiate the right- and left-hand sides of equalities (\ref{eq:4q}) and
(\ref{eq:4w}) with respect to free fields $A^{(0)a\mu}$,
$\psi_{\alpha}^{(0)i}(q)$ and $\bar{\psi}_{\alpha}^{(0)i}(-q)$ considering
Eq.\,(\ref{eq:4t}) for differentiation on the right-hand side and
Eqs.\,(\ref{eq:4e}), (\ref{eq:4r}), (\ref{eq:4t}) for differentiation
on the left-hand side, and set
$A^{(0) a \mu} = \psi_{\alpha}^{(0)i} = \bar{\psi}_{\alpha}^{(0)i} = 0$
after all calculations. The desired effective amplitudes will appear both on
the left-hand side and on the right-hand side. However principal effect here,
is that the effective amplitudes on the left-hand side will have at least 
one less external soft leg than on the right-hand side. This enables us to 
calculate them in a recurrent way. Below we will give some examples.

The second functional derivative of the induced current $j^{\Psi a}_{\mu}$
with respect to $\bar{\psi}^{(0)}$ and $\psi^{(0)}$ yields
\begin{equation}
\left.\frac{\delta^2j_{\mu}^{\Psi a}[A,\bar{\psi},\psi](k)}
{\delta \psi^{(0)\,j}_{\beta}(q_2)\delta \bar{\psi}^{(0)\,i}_{\alpha}(-q_1)}\,
\right|_{A^{(0)}=\psi^{(0)}=\bar{\psi}^{(0)}=0}
=
g\,^{\ast}\tilde{\Gamma}^{(G)a,\,ij}_{\mu,\,\alpha\beta}(k;q_1,-q_2)
\delta(k+q_1-q_2)
\label{eq:4y}
\end{equation}
\[
=g\,(t^a)^{ij}\,^{\ast}\Gamma^{(G)}_{\mu,\,\alpha\beta}(k;q_1,-q_2)
\delta(k+q_1-q_2),
\]
and the second functional derivative of the induced source $\eta_{\alpha}^i$
with respect to $A^{(0)}$ and $\psi^{(0)}$ gives
\begin{equation}
\left.\frac{\delta^2\eta_{\alpha}^i[A,\psi](q)}
{\delta A^{(0)a\mu} \delta \psi^{(0)\,i_1}_{\alpha_1}(q_1)}\,
\right|_{A^{(0)}=\psi^{(0)}=\bar{\psi}^{(0)}=0}
=
g\,^{\ast}\tilde{\Gamma}^{(Q)a,\,ii_1}_{\mu,\,\alpha\alpha_1}(k;q_1,-q)
\delta(q-q_1-k)
\label{eq:4u}
\end{equation}
\[
=g\,(t^a)^{ii_1}\,^{\ast}\Gamma^{(Q)}_{\mu,\,\alpha\alpha_1}(k;q_1,-q)
\delta(q-q_1-k).
\]
Eqs.\,(\ref{eq:4y}) and (\ref{eq:4u}) show that the effective amplitudes
of decay of soft-gluon excitation into pair of soft quark-antiquark excitations
(or inverse process of pair annihilation into soft gluon) coincide with usual
HTL-resummed vertex between quark pair and gluon. This decay process is
kinematically forbidden and hence the $\delta$-functions don't support on
the mass-shell of plasma excitations.

The first nontrivial examples arise in calculation of the next derivative.
It defines the effective amplitudes for the processes with four soft plasma
excitations: the process of nonlinear interaction of two soft bosonic and
two fermionic excitations and the process of four soft fermionic
interaction. For the first process two effective amplitudes are possible. The
first amplitude arises from the variation
\[
\left.\frac{\delta^3\!
\left(j_{\mu}^{A a}[A](k)+j_{\mu}^{\Psi a}[A,\bar{\psi},\psi](k)\right)}
{\delta A^{(0)a_1\mu_1}(k_1)
\delta \psi^{(0)\,j}_{\beta}(q_2)\delta \bar{\psi}^{(0)\,i}_{\alpha}(-q_1)}\,
\right|_{A^{(0)}=\psi^{(0)}=\bar{\psi}^{(0)}=0}=
\]
\[
=\!\int\Biggl\{
\frac{\delta^2 j_{\mu}^{A(2)a}(k)}
{\delta A^{a_1^{\prime}\mu_1^{\prime}}(k_1^{\prime})
\delta A^{a_2^{\prime}\mu_2^{\prime}}(k_2^{\prime})}\,
\frac{\delta A^{a_2^{\prime}\mu_2^{\prime}}(k_2^{\prime})}
{\delta A^{(0)a_1\mu_1}(k_1)}\,
\frac{\delta^2A^{a_1^{\prime}\mu_1^{\prime}}(k_1^{\prime})}
{\delta \psi^{(0)j}_{\beta}(q_2)\delta\bar{\psi}^{(0)i}_{\alpha}(-q_1)}\,
\,dk_1^{\prime}dk_2^{\prime}
\hspace{1cm}
\]
\[
+\,\frac{\delta^3 j_{\mu}^{\Psi(1,2)a}(k)}
{\delta A^{a_1^{\prime}\mu_1^{\prime}}(k_1^{\prime})
\delta \psi^{j_1^{\prime}}_{\beta_1^{\prime}}(q_2^{\prime})
\delta\bar{\psi}^{i_1^{\prime}}_{\alpha_1^{\prime}}(-q_1^{\prime})}\,
\frac{\delta A^{a_1^{\prime}\mu_1^{\prime}}(k_1^{\prime})}
{\delta A^{(0)a_1\mu_1}(k_1)}\,
\frac{\delta \psi^{j_1^{\prime}}_{\beta_1^{\prime}}(q_2^{\prime})}
{\delta \psi^{(0)j}_{\beta}(q_2)}\,
\frac{\delta\bar{\psi}^{i_1^{\prime}}_{\alpha_1^{\prime}}(-q_1^{\prime})}
{\delta\bar{\psi}^{(0)\,i}_{\alpha}(-q_1)}\,\,
dk_1^{\prime}dq_1^{\prime}dq_2^{\prime}
\hspace{0.2cm}
\]
\[
+\,\frac{\delta^2 j_{\mu}^{\Psi(0,2)a}(k)}
{\delta \psi^{j_1^{\prime}}_{\beta_1^{\prime}}(q_2^{\prime})
\delta\bar{\psi}^{i_1^{\prime}}_{\alpha_1^{\prime}}(-q_1^{\prime})}\,
\frac{\delta^2\psi^{j_1^{\prime}}_{\beta_1^{\prime}}(q_2^{\prime})}
{\delta A^{(0)a_1\mu_1}(k_1)\delta \psi^{(0)j}_{\beta}(q_2)}\,
\frac{\delta\bar{\psi}^{i_1^{\prime}}_{\alpha_1^{\prime}}(-q_1^{\prime})}
{\delta\bar{\psi}^{(0)i}_{\alpha}(-q_1)}\,\,
dq_1^{\prime}dq_2^{\prime}
\hspace{2.9cm}
\]
\[
\hspace{0.2cm}
+\,\frac{\delta^2 j_{\mu}^{\Psi(0,2)a}(k)}
{\delta \psi^{j_1^{\prime}}_{\beta_1^{\prime}}(q_2^{\prime})
\delta\bar{\psi}^{i_1^{\prime}}_{\alpha_1^{\prime}}(-q_1^{\prime})}\,
\frac{\delta \psi^{j_1^{\prime}}_{\beta_1^{\prime}}(q_2^{\prime})}
{\delta \psi^{(0)j}_{\beta}(q_2)}\,
\frac{\delta^2\bar{\psi}^{i_1^{\prime}}_{\alpha_1^{\prime}}(-q_1^{\prime})}
{\delta A^{(0)a_1\mu_1}(k_1)\delta\bar{\psi}^{(0)i}_{\alpha}(-q_1)}\,\,
dq_1^{\prime}dq_2^{\prime}
\Biggl\}\Bigg|_{A^{(0)}=\psi^{(0)}=\bar{\psi}^{(0)}=0}.
\]
Here, on the right-hand side we keep the terms different from zero
only. Taking into account Eqs.\,(\ref{eq:3e}), (\ref{eq:4r}),
(\ref{eq:4y}) and (\ref{eq:4u}), from the last expression we find
the effective amplitude determining effective current
(\ref{eq:3o})
\begin{equation}
\!\,^{\ast}\tilde{\Gamma}^{(G)aa_1,\,ij}_{\mu{\mu}_1,\,\alpha\beta}
(k,-k_1;q_1,-q_2)=
\delta{\Gamma}^{(G)aa_1,\,ij}_{\mu{\mu}_1,\,\alpha\beta}(k,-k_1;q_1,-q_2)
\label{eq:4i}
\end{equation}
\[
-\,
[t^{a},t^{a_1}]^{ij}
\,^\ast\Gamma_{\mu\nu\mu_1}(k,-k+k_1,-k_1)
\,^{\ast}{\cal D}^{\nu\nu^{\prime}}(k-k_1)
\,^{\ast}\Gamma^{(G)}_{\nu^{\prime},\,\alpha\beta}(-q_1+q_2;q_1,-q_2)
\]
\[
-\,(t^{a}t^{a_1})^{ij}
\,^{\ast}\Gamma^{(G)}_{\mu,\,\alpha\gamma}(k;q_1,-k-q_1)
\,^{\ast}S_{\gamma\gamma^{\prime}}(k+q_1)
\,^{\ast}\Gamma^{(Q)}_{\mu_1,\,\gamma^{\prime}\beta}(k_1;q_2,-k_1-q_2)
\hspace{0.65cm}
\]
\[
+\,(t^{a_1}t^{a})^{ij}
\,^{\ast}\Gamma^{(Q)}_{\mu_1,\,\alpha\gamma}(k_1;-q_1,q_1-k_1)
\,^{\ast}S_{\gamma\gamma^{\prime}}(k_1-q_1)
\,^{\ast}\Gamma^{(G)}_{\mu,\,\gamma^{\prime}\beta}(k;-k+q_2,-q_2),
\]
where $[\,,]$ denotes a commutator.
The second similar effective amplitude results from variation
\[
\left.\frac{\delta^3\eta^i_{\alpha}[A,\psi](q)}
{\delta A^{(0)a_1\mu_1}(k_1)\delta A^{(0)a_2\mu_2}(k_2)
\delta \psi^{(0)\,j}_{\beta}(q_1)}\,
\right|_{A^{(0)}=\psi^{(0)}=\bar{\psi}^{(0)}=0}.
\]
The calculations analogous to previous ones lead to the effective
amplitude determining effective source (\ref{eq:3p})
\begin{equation}
\!\,^{\ast}\tilde{\Gamma}^{(Q)a_1a_2,\,ij}_{\mu_1\mu_2,\,\alpha\beta}
(k_1,k_2;q_1,-q)=
\delta{\Gamma}^{(Q)a_1a_2,\,ij}_{\mu_1\mu_2,\,\alpha\beta}(k_1,k_2;q_1,-q)
\label{eq:4o}
\end{equation}
\[
+\,[t^{a_1},t^{a_2}]^{ij}
\,^{\ast}\Gamma^{(Q)}_{\nu,\,\alpha\beta}(-q+q_1;q,-q_1)
\,^{\ast}{\cal D}^{\nu\nu^{\prime}}(k_1+k_2)
\,^\ast\Gamma_{\nu^{\prime}\mu_1\mu_2}(k_1+k_2,-k_1,-k_2)
\]
\[
-\,(t^{a_1}t^{a_2})^{ij}
\,^{\ast}\Gamma^{(Q)}_{\mu_1,\,\alpha\gamma}(-k_1;q,-q+k_1)
\,^{\ast}\!S_{\gamma\gamma^{\prime}}(q-k_1)
\,^{\ast}\Gamma^{(Q)}_{\mu_2,\,\gamma^{\prime}\beta}(-k_2;q_1+k_2,-q_1)
\hspace{0.3cm}
\]
\[
-\,(t^{a_2}t^{a_1})^{ij}
\,^{\ast}\Gamma^{(Q)}_{\mu_2,\,\alpha\gamma}(-k_2;q,-q+k_2)
\,^{\ast}\!S_{\gamma\gamma^{\prime}}(q-k_2)
\,^{\ast}\Gamma^{(Q)}_{\mu_1,\,\gamma^{\prime}\beta}(-k_1;q_1+k_1,-q_1).
\hspace{0.2cm}
\]
From the definition of this effective amplitude it follows symmetry property
with respect to permutation of soft-gluon legs useful in the subsequent
discussion
\begin{equation}
\!\,^{\ast}\tilde{\Gamma}^{(Q)a_1a_2,\,ij}_{\mu_1\mu_2,\,\alpha\beta}
(k_1,k_2;q_1,-q)=
\!\,^{\ast}\tilde{\Gamma}^{(Q)a_2a_1,\,ij}_{\mu_2\mu_1,\,\alpha\beta}
(k_2,k_1;q_1,-q).
\label{eq:4p}
\end{equation}

The effective amplitude of four soft-quark interaction follows from the
variation
\[
\left.\frac{\delta^3\eta^i_{\alpha}[A,\psi](q)}
{\delta \psi^{(0)i_3}_{\alpha_3}(q_3)
\delta \bar{\psi}^{(0)i_2}_{\alpha_2}(-q_2)
\delta \psi^{(0)i_1}_{\alpha_1}(q_1)}\,
\right|_{A^{(0)}=\psi^{(0)}=\bar{\psi}^{(0)}=0}
\]
\[
=\!\int\Biggl\{
-\,\frac{\delta^2 \eta_{\alpha}^{(1,1)i}(q)}
{\delta A^{a_1^{\prime}\mu_1^{\prime}}(k_1^{\prime})
\delta \psi^{i_1^{\prime}}_{\alpha_1^{\prime}}(q_1^{\prime})}\,
\frac{\delta \psi^{i_1^{\prime}}_{\alpha_1^{\prime}}(q_1^{\prime})}
{\delta \psi^{(0)i_3}_{\alpha_3}(q_3)}\,
\frac{\delta^2A^{a_1^{\prime}\mu_1^{\prime}}(k_1^{\prime})}
{\delta \psi^{(0)i_1}_{\alpha_1}(q_1)
\delta\bar{\psi}^{(0)i_2}_{\alpha_2}(-q_2)}\,
\,dk_1^{\prime}dq_1^{\prime}
\hspace{3cm}
\]
\[
\hspace{0.5cm}
+\,\frac{\delta^2\eta_{\alpha}^{(1,1)i}(q)}
{\delta A^{a_1^{\prime}\mu_1^{\prime}}(k_1^{\prime})
\delta \psi^{i_1^{\prime}}_{\alpha_1^{\prime}}(q_1^{\prime})}\,
\frac{\delta^2A^{a_1^{\prime}\mu_1^{\prime}}(k_1^{\prime})}
{\delta \psi^{(0)i_3}_{\alpha_1}(q_3)
\delta\bar{\psi}^{(0)i_2}_{\alpha_2}(-q_2)}\,
\frac{\delta \psi^{i_1^{\prime}}_{\alpha_1^{\prime}}(q_1^{\prime})}
{\delta \psi^{(0)i_1}_{\alpha_1}(q_1)}\,
\,dk_1^{\prime}dq_1^{\prime}
\Biggl\}\Bigg|_{A^{(0)}=\psi^{(0)}=\bar{\psi}^{(0)}=0}.
\]
We kept again the terms different from zero only on the right-hand side.
Taking into account Eqs.\,(\ref{eq:3r}), (\ref{eq:4t}) and (\ref{eq:4u}), we
easily derive the explicit form of the effective amplitude in integrand of
effective source (\ref{eq:3a})
\begin{equation}
\!\,^{\ast}\tilde{\Gamma}^{ii_1i_2i_3}_{\alpha\alpha_1\alpha_2\alpha_3}
(q,q_1,-q_2,-q_3)=
\label{eq:4a}
\end{equation}
\[
=(t^a)^{ii_2}(t^a)^{i_1i_3}
\,^{\ast}\Gamma^{(Q)\mu}_{\alpha\alpha_2}(q-q_2;q_2,-q)
\,^{\ast}{\cal D}_{\mu\mu^{\prime}}(q-q_2)
\,^{\ast}\Gamma^{(G)\mu^{\prime}}_{\alpha_1\alpha_3}(-q_1+q_3;q_1,-q_3)
\]
\[
\hspace{0.2cm}
+\,(t^a)^{ii_3}(t^a)^{i_1i_2}
\,^{\ast}\Gamma^{(Q)\mu}_{\alpha\alpha_3}(q-q_3;q_3,-q)
\,^{\ast}{\cal D}_{\mu\mu^{\prime}}(q-q_3)
\,^{\ast}\Gamma^{(G)\mu^{\prime}}_{\alpha_1\alpha_2}(-q_1+q_2;q_1,-q_2).
\]
Here, the following property of antisymmetry with respect to permutation of two
last soft-quark legs holds
\[
\!\,^{\ast}\tilde{\Gamma}^{ii_1i_2i_3}_{\alpha\alpha_1\alpha_2\alpha_3}
(q,q_1,-q_2,-q_3)=
-\,
\!\,^{\ast}\tilde{\Gamma}^{ii_1i_3i_2}_{\alpha\alpha_1\alpha_3\alpha_2}
(q,q_1,-q_3,-q_2).
\]
There is no such a property in general case for permutation of two
first soft-quark legs by virtue of inequality of the
HTL-amplitudes with different time ordering of external legs:
$\,^{\ast}\Gamma^{(Q)\mu}_{\alpha\beta}(q-q_1;q_1,-q) \ne
\,^{\ast}\Gamma^{(G)\mu}_{\alpha\beta}(q-q_1;q_1,-q)$.

For convenience of subsequent references in Appendix B we give the
explicit form of the effective amplitudes for conjugate effective
current
$\tilde{j}^{\ast\Psi(1,2)a}_{\mu}(A^{\ast(0)},\bar{\psi}^{(0)},\psi^{(0)})$
and Dirac conjugate effective sources
$\tilde{\bar{\eta}}^{(2,1)i}_{\alpha}
(A^{\ast(0)},A^{\ast(0)},\bar{\psi}^{(0)})$ and
$\tilde{\bar{\eta}}^{(0,3)\,i}_{\beta}
(\bar{\psi}^{(0)},\bar{\psi}^{(0)},\psi^{(0)})$.

\section{\bf Boltzmann equation for elastic scattering of soft-quark
excitations off soft-gluon excitations}
\setcounter{equation}{0}

The initial equation for deriving the Boltzmann equation describing a change of
the number densities of the soft fermion excitations owing to their scattering
off the soft boson excitations is equation (Eq.\,(4.4) in
Ref.\,\cite{markov_PRD})
\begin{equation}
{\rm Sp} \bigg( \frac{\partial}{\partial q^{\mu}} [\,-\!\not\!q +
\delta \Sigma^{\rm H}(q)]
\frac{\partial \Upsilon^{ji}(q,x)}{\partial x_{\mu}} \bigg) =
\label{eq:5q}
\end{equation}
\[
=\,ig\!\int\!dq^{\prime} dq_1 dk_1 \Bigl\{
\langle \bar{\psi}^i_{\alpha}(-q)A^{a\mu}(k_1)[\,^{\ast}
\Gamma^{(Q)\,a}_{\mu}(k_1;q_1,-q^{\prime})
\psi(q_1)]^j_{\alpha} \rangle
\,\delta(q^{\prime} - q_1 - k_1)
\hspace{1cm}
\]
\[
\hspace{2.5cm}
-\,\langle [\bar{\psi}(-q_1)
\,^{\ast}\Gamma^{(Q)\,a}_{\mu}(-k_1;-q_1,q)]^i_{\alpha}
A^{\ast\,a\mu}(k_1)
\psi^j_{\alpha}(q^{\prime})\rangle
\,\delta(q - q_1 - k_1)\Bigr\}
\]
\[
+\,ig^2\!\!\int\!dq^{\prime}dq_1dk_1dk_2\Big\{
\langle A^{a\mu}(k_1) A^{b\nu}(k_2)\bar{\psi}^i_{\alpha}(-q)
[\delta\Gamma^{(Q)\,ab}_{\mu\nu}(-k_1,-k_2;q^{\prime},-q_1)
\psi(q_1)]^j_{\alpha}\rangle
\hspace{0.3cm}
\]
\[
\times\,
\delta(q^{\prime} - q_1 - k_1 - k_2)
\]
\[
\hspace{0.5cm}
-\,\langle A^{\ast\,a\mu}(k_1) A^{\ast\,b\nu}(k_2)[\bar{\psi}(-q_1)
\delta\Gamma^{(Q)\,ba}_{\mu\nu}(k_2,k_1;-q,q_1)]^i_{\alpha}
\psi^j_{\alpha}(q^{\prime})\rangle
\,\delta(q - q_1 - k_1 - k_2)\Big\}.
\]
Here, $\not\!\!q = \gamma^{\mu}q_{\mu}$, the Dirac trace is denoted by
${\rm Sp}$, and $\delta\Sigma^{\rm H}(q)$ is ``Hermitian'' part of the quark
self-energy. The angle brackets denote an average with respect to any density
matrix.

Using above-obtained expressions (\ref{eq:3u}) and (\ref{eq:3i}) for
gauge potential and quark wave functions by a simple search we extract all
sixth-order correlators of free fields responsible for the scattering processes
of soft fermion excitations off the soft boson excitations to the lowest order
in the coupling constant. As a guide rule for choice of the relevant terms a
simple fact will be used. When we make decoupling of the sixth-order
correlators in the terms of the pairs (in order to define the product
$n^{(f_1)}_{{\bf q}_1} N_{{\bf k}_1}^{(b_1)} N_{{\bf k}_2}^{(b_2)}$) we keep
the terms, which correctly reproduce relevant momentum-energy conservation
laws. These conservation laws encoded in $\delta$-functions of the first two
measures of integration in Eq.\,(\ref{eq:2u}). Besides the number of appearing
terms can be considerably cut if we
note that it is necessary to keep only such terms in intermediate expressions,
which contain the quark propagators $^{\ast}S(-q)$ and $^{\ast}S(q^{\prime})$.
These propagators give later on the terms proportional to
$Z_{\pm}({\bf q})\delta (q^0 - \omega_{\bf q}^{\pm})$,
i.e., the factors take into account the existence of soft-quark excitations
with wave vector ${\bf q}$ and energies $\omega_{\bf q}^{\pm}$ in spite
of the fact that the number densities of soft-quark excitations
$n_{\bf q}^{\pm}$ are explicitly absent on the right-hand side.

Let us consider the first term on the right-hand side of Eq.\,(\ref{eq:5q})
containing the third-order correlation function of interacting fields
\begin{equation}
ig\!\int\!dq^{\prime} dq_1 dk_1\,
\langle \bar{\psi}^i_{\alpha}(-q)A^{a\mu}(k_1)\,^{\ast}
\Gamma^{(Q)a,\,jj^{\prime}}_{\mu,\,\alpha\beta}(k_1;q_1,-q^{\prime})
\psi^{j^{\prime}}_{\beta}(q_1)\rangle
\,\delta(q^{\prime} - q_1 - k_1).
\label{eq:5w}
\end{equation}
For this term there exist two substitutions resulting in the
sixth-order correlation function of free fields with conservation laws
(\ref{eq:2u}). The first substitution is
\[
\bar{\psi}^i_{\alpha}(-q)\rightarrow
\tilde{\bar{\eta}}^{(2,1)i}_{\gamma}
(A^{\ast(0)},A^{\ast(0)},\bar{\psi}^{(0)})(-q)\,^{\ast}\!S_{\gamma\alpha}(-q),
\quad
\psi^{j^{\prime}}_{\beta}(q_1)\rightarrow \psi^{(0)j^{\prime}}_{\beta}(q_1),
\]
\[
A^{a\mu}(k_1)\rightarrow
-\,^{\ast}{\cal D}^{\mu\nu}(k_1)
\tilde{j}^{A(2)a}_{\nu}(A^{(0)},A^{(0)})(k_1)
\]
and the second one is
\[
\bar{\psi}^i_{\alpha}(-q)\rightarrow
\tilde{\bar{\eta}}^{(2,1)i}_{\gamma}
(A^{\ast(0)},A^{\ast(0)},\bar{\psi}^{(0)})\,^{\ast}\!S_{\gamma\alpha}(-q),
\quad
A^{a\mu}(k_1)\rightarrow A^{(0)a\mu}(k_1),
\]
\[
\psi^{j^{\prime}}_{\beta}(q_1)\rightarrow
-\,^{\ast}\!S_{\beta\rho}(q_1)
\tilde{\eta}^{(1,1)j^{\prime}}_{\rho}
(A^{(0)},\psi^{(0)})(q_1).
\]
Here, the product of the correlation functions of free gauge and quark fields
appears in the following form:
\[
\langle A^{\ast(0)a_1\mu_1}(k_1)A^{\ast(0)a_2\mu_2}(k_2)
A^{(0)a_3\mu_3}(k_3)A^{(0)a_4\mu_4}(k_4)\rangle
\langle \bar{\psi}^{(0)i}_{\alpha}(-q)\psi^{(0)j}_{\beta}(q_1)\rangle
\]
\begin{equation}
\simeq\{\,\delta^{a_1a_4}I^{\mu_1\mu_4}(k_1)\delta(k_1-k_4)\,
\delta^{a_2a_3}I^{\mu_2\mu_3}(k_2)\delta(k_2-k_3)
\label{eq:5e}
\end{equation}
\[
+\, \delta^{a_1a_3}I^{\mu_1\mu_3}(k_1)\delta(k_1-k_3)\,
\delta^{a_2a_4}I^{\mu_2\mu_4}(k_2)\delta(k_2-k_4)\}\,
\delta^{ij} \Upsilon_{\beta\alpha}(q_1)\delta(q_1-q),
\]
where on the right-hand side we use definitions of the correlation functions
$I^{\mu\nu}(k)$ and $\Upsilon_{\alpha\beta}(q)$ in the equilibrium
\cite{markov_PRD}. Simple algebraical transformations enables us to lead
term (\ref{eq:5w}) to the following form:
\[
\frac{ig^4}{2!}\!\int\! dq_1 dk_1 dk_2\,
\!\,^{\ast}\tilde{\bar{\Gamma}}^{(Q)a_1a_2,\,il}_{\mu_1\mu_2,\,\alpha\beta}\!
(k_1,k_2;q_1,-q)
\,^{\ast}\!S_{\alpha\alpha^{\prime}}(-q)
\]
\begin{equation}
\times\Bigl\{[t^{a_1},t^{a_2}]^{jl}
\,^{\ast}\Gamma^{(Q)}_{\nu,\,\alpha^{\prime}\beta^{\prime}}(q-q_1;q_1,-q)
\,^{\ast}{\cal D}^{\nu\nu^{\prime}}(k_1+k_2)
\,^\ast\Gamma_{\nu^{\prime}\mu_1^{\prime}\mu_2^{\prime}}(k_1+k_2,-k_1,-k_2)
\label{eq:5r}
\end{equation}
\[
-\,(t^{a_1}t^{a_2})^{jl}
\,^{\ast}\Gamma^{(Q)}_{\mu_1^{\prime},\,\alpha^{\prime}\gamma}(-k_1;q,-q+k_1)
\,^{\ast}\!S_{\gamma\gamma^{\prime}}(q-k_1)
\,^{\ast}\Gamma^{(Q)}_{\mu_2^{\prime},\,\gamma^{\prime}\beta^{\prime}}
(-k_2;q_1+k_2,-q_1)
\hspace{0.2cm}
\]
\[
-\,(t^{a_2}t^{a_1})^{jl}
\,^{\ast}\Gamma^{(Q)}_{\mu_2^{\prime},\,\alpha^{\prime}\gamma}(-k_2;q,-q+k_2)
\,^{\ast}\!S_{\gamma\gamma^{\prime}}(q-k_2)
\,^{\ast}\Gamma^{(Q)}_{\mu_1^{\prime},\,\gamma^{\prime}\beta^{\prime}}
(-k_1;q_1+k_1,-q_1)
\Bigr\}
\]
\[
\times\,\Upsilon_{\beta^{\prime}\!\beta}(q_1)I^{\mu_1\mu_1^{\prime}}(k_1)
I^{\mu_2\mu_2^{\prime}}(k_2)
\delta(q-q_1-k_1-k_2).
\]
In deriving Eq.\,(\ref{eq:5r}) we have used property (B.4) for the effective
amplitude $\tilde{\bar{\Gamma}}^{(Q)a_1a_2,\,il}_{\mu_1\mu_2,\,\alpha\beta}$.

Now we turn to the third term on the right-hand side of Eq.\,(\ref{eq:5q})
containing the fourth-order correlation function of interacting fields
\begin{equation}
ig^2 \int\!dq^{\prime}dq_1dk_1dk_2\,
\langle A^{a\mu}(k_1) A^{b\nu}(k_2)\bar{\psi}^i_{\alpha}(-q)
\delta\Gamma^{(Q)ab,\,jj^{\prime}}_{\mu\nu,\,\alpha\beta}(-k_1,-k_2;q^{\prime},-q_1)
\psi^{j^{\prime}}_{\beta}(q_1)\rangle
\label{eq:5t}
\end{equation}
\[
\times\,
\delta(q^{\prime} - q_1 - k_1 - k_2).
\]
Here, there exists a unique replacement, which leads to relevant conservation
laws (\ref{eq:2u})
\[
\bar{\psi}^i_{\alpha}(-q)\rightarrow
\tilde{\bar{\eta}}^{(2,1)i}_{\gamma}
(A^{\ast(0)},A^{\ast(0)},\bar{\psi}^{(0)})\,^{\ast}\!S_{\gamma\alpha}(-q),
\quad
\psi^{j^{\prime}}_{\beta}(q_1)\rightarrow
\psi^{(0)j^{\prime}}_{\beta}(q_1),
\]
\[
A^{a\mu}(k_1)\rightarrow A^{(0)a\mu}(k_1),\quad
A^{b\nu}(k_2)\rightarrow A^{(0)b\nu}(k_2).
\]
This replacement with decomposition (\ref{eq:5e}) results equation
(\ref{eq:5t}) in the following form:
\[
\frac{ig^4}{2!}\!\int\! dq_1 dk_1 dk_2\,
\!\,^{\ast}\tilde{\bar{\Gamma}}^{(Q)a_1a_2,\,il}_{\mu_1\mu_2,\,\alpha\beta}\!
(k_1,k_2;q_1,-q)
\,^{\ast}\!S_{\alpha\alpha^{\prime}}(-q)
\,\delta{\Gamma}^{(Q)a_1a_2,\,jl}_{\mu_1^{\prime}\mu_2^{\prime},
\,\alpha^{\prime}\beta^{\prime}}\!
(k_1,k_2;q_1,-q)
\]
\[
\times\,\Upsilon_{\beta^{\prime}\!\beta}(q_1)I^{\mu_1\mu_1^{\prime}}(k_1)
I^{\mu_2\mu_2^{\prime}}(k_2)
\delta(q-q_1-k_1-k_2).
\]
If we now add the last expression to Eq.\,(\ref{eq:5r}) and take into account
the definition of effective amplitude (\ref{eq:4o}), then we get
\begin{equation}
\frac{ig^4}{2!}\!\int\! dq_1 dk_1 dk_2\,
\!\,^{\ast}\tilde{\bar{\Gamma}}^{(Q)a_1a_2,\,il}_{\mu_1\mu_2,\,\alpha\beta}\!
(k_1,k_2;q_1,-q)
\,^{\ast}\!S_{\alpha\alpha^{\prime}}(-q)
\,\!\,^{\ast}\tilde{\Gamma}^{(Q)a_1a_2,\,jl}_{\mu_1^{\prime}\mu_2^{\prime},
\,\alpha^{\prime}\beta^{\prime}}\!
(k_1,k_2;q_1,-q)
\label{eq:5y}
\end{equation}
\[
\times\,\Upsilon_{\beta^{\prime}\!\beta}(q_1)I^{\mu_1\mu_1^{\prime}}(k_1)
I^{\mu_2\mu_2^{\prime}}(k_2)
\delta(q-q_1-k_1-k_2).
\]
The remaining second and fourth terms on the right-hand side of
Eq.\,(\ref{eq:5q}) after similar reasoning give in a sum expression
complex conjugate to that of (\ref{eq:5y}).
To consider the simplest interaction process of plasminos with plasmons,
it is necessary to fulfill the following replacements:
\begin{equation}
I^{\mu_1\mu_1^{\prime}}(k_1)\rightarrow
Q^{\mu_1\mu_1^{\prime}}(k_1)I^l(k_1),\quad
I^{\mu_2\mu_2^{\prime}}(k_2)\rightarrow
Q^{\mu_2\mu_2^{\prime}}(k_2)I^l(k_2),
\label{eq:5u}
\end{equation}
\[
\Upsilon_{\beta^{\prime}\!\beta}(q_1)\rightarrow
(h_{-}({\hat{\bf q}}_1))_{\beta^{\prime}\!\beta}
\Upsilon^{-}({\bf q}_1)\delta(q_1^0-\omega^{-}_{{\bf q}_1}) +
(h_{+}({\hat{\bf q}}_1))_{\beta^{\prime}\!\beta}
\bar{\Upsilon}^{-}(-{\bf q}_1)\delta(q_1^0+\omega^{-}_{{\bf q}_1}),
\]
\[
\,^{\ast}\!S_{\alpha\alpha^{\prime}}(-q)\rightarrow
- \,(h_{-}({\hat{\bf q}}))_{\alpha\alpha^{\prime}}
(\,^{\ast}\!\Delta_{-}(q))^{\ast}.
\]
We have determined the function
$\Upsilon_{\beta^{\prime}\!\beta}(q_1)$ in the {\it quasiparticle
approximation}. The spectral density $\Upsilon^{-}({\bf q}_1)$
describes the plasmino branch of the soft quark excitations and
$\bar{\Upsilon}^{-}(-{\bf q}_1)$ describes antiplasmino branch. To
take into account weakly inhomogeneous and non-stationary in the
medium it is sufficient within an accepted accuracy to replace
equilibrium spectral densities $I^l(k_i)$, $\Upsilon^{-}({\bf
q}_1)$ and $\bar{\Upsilon}^{-}(-{\bf q}_1)$ by off-equilibrium
ones in the Wigner form (Eq.\,(3.6) in Ref.\,\cite{markov_PRD}):
\[
I^l(k_i)\rightarrow I^l_{k_i}\equiv I^l(k_i,x),\; \;i=1,2,
\]
\[
\Upsilon^{-}({\bf q}_1)\rightarrow \Upsilon_{{\bf q}_1}^{-} \equiv
\Upsilon^{-}({\bf q}_1,x),\;
\bar{\Upsilon}^{-}(-{\bf q}_1)\rightarrow  \bar{\Upsilon}_{-{\bf q}_1}^{-}
\equiv \bar{\Upsilon}^{-}(-{\bf q}_1,x)
\]
slowly depending on $x=(t,{\bf x})$.

After replacements (\ref{eq:5u}) we perform a trivial integration in $dq_1^0$.
For antiplasmino part of the function $\Upsilon_{\beta^{\prime}\!\beta}(q_1)$
we make a replacement of variable in integrand (\ref{eq:5y}):
${\bf q}_1 \rightarrow -{\bf q}_1$
($\omega_{{\bf q}_1}^{-}\rightarrow \omega_{{\bf q}_1}^{-})$. Finally instead
of (\ref{eq:5y}) for this special case of nonlinear interaction of soft modes
in QGP we obtain
\[
\frac{g^4}{2!}\,(i\,^{\ast}\!\Delta_{-}(q))^{\ast}\!\!
\int\! d{\bf q}_1 dk_1 dk_2\,
\frac{1}{\bar{u}^2(k_1)\bar{u}^2(k_2)}\,
\]
\[
\times\Bigl[\,\bar{u}^{\mu_1}(k_1)\bar{u}^{\mu_2}(k_2)
\!\,^{\ast}\tilde{\bar{\Gamma}}^{(Q)a_1a_2,\,il}_{\mu_1\mu_2,\,\alpha\beta}\!
(k_1,k_2;q_1,-q)
(h_{-}({\hat{\bf q}}))_{\alpha\alpha^{\prime}}
\]
\[
\hspace{0.4cm}
\times
\bar{u}^{\mu_1^{\prime}}(k_1)\bar{u}^{\mu_2^{\prime}}(k_2)\,
\!\,^{\ast}\tilde{\Gamma}^{(Q)a_1a_2,\,jl}_{\mu_1^{\prime}\mu_2^{\prime},
\,\alpha^{\prime}\beta^{\prime}}\!
(k_1,k_2;q_1,-q)(h_{-}({\hat{\bf q}}_1))_{\beta^{\prime}\!\beta}
\]
\begin{equation}
\times\,
\Upsilon_{{\bf q}_1}^{-}I_{k_1}^lI_{k_2}^l\,
\delta(q-q_1-k_1-k_2)
\label{eq:5i}
\end{equation}
\[
\hspace{0.6cm}
+\,
\bar{u}^{\mu_1}(k_1)\bar{u}^{\mu_2}(k_2)
\!\,^{\ast}\tilde{\bar{\Gamma}}^{(Q)a_1a_2,\,il}_{\mu_1\mu_2,\,\alpha\beta}\!
(k_1,k_2;-q_1,-q)
(h_{-}({\hat{\bf q}}))_{\alpha\alpha^{\prime}}
\]
\[
\hspace{0.8cm}
\times\,
\bar{u}^{\mu_1^{\prime}}(k_1)\bar{u}^{\mu_2^{\prime}}(k_2)
\,\!\,^{\ast}\tilde{\Gamma}^{(Q)a_1a_2,\,jl}_{\mu_1^{\prime}\mu_2^{\prime},
\,\alpha^{\prime}\beta^{\prime}}\!
(k_1,k_2;-q_1,-q)(h_{-}({\hat{\bf q}}_1))_{\beta^{\prime}\!\beta}
\]
\[
\hspace{1.4cm}
\times\,
\bar{\Upsilon}_{{\bf q}_1}^{-}I_{k_1}^lI_{k_2}^l\,
\delta(q+q_1-k_1-k_2)\Bigr]_{q^0_1=\omega^{-}_{{\bf q}_1}}.
\]
The expression obtained enables us to define the probability of the
plasmino-plasmon scattering
$w_{qg \rightarrow qg}^{(-l;\,-l)}(q, q_1;k_1,k_2)$ and
the probability of pair annihilation into two plasmons
$w_{q\bar{q}\rightarrow gg}^{(--;\,ll)}(q, q_1;k_1, k_2)$.
At first we express projectors
$h_{-}({\hat{\bf q}})$ and $h_{-}({\hat{\bf q}_1})$ in terms of
simultaneous eigenspinors of chirality and helicity
\begin{equation}
(h_{-}({\hat{\bf q}}))_{\alpha\alpha^{\prime}}=
\sum\limits_{\lambda=\pm}v_{\alpha}(\hat{\bf q},\lambda)
\bar{v}_{\alpha^{\prime}}(\hat{\bf q},\lambda),\quad
(h_{-}({\hat{\bf q}}_1))_{\beta^{\prime}\!\beta}=
\sum\limits_{\lambda_1=\pm}v_{\beta^{\prime}}(\hat{\bf q}_1,\lambda_1)
\bar{v}_{\beta}(\hat{\bf q}_1,\lambda_1).
\label{eq:5o}
\end{equation}
Furthermore, we take also the boson spectral functions $I^l_{k_1}$ and
$I^l_{k_2}$ in the form of the quasiparticle approximation
\begin{equation}
I^l_{k_i} = I_{{\bf k}_i}^l \delta (k^0_i - \omega_{{\bf k}_i}^l)
+ I_{-{\bf k}_i}^l \delta (k^0_i + \omega_{{\bf k}_i}^l), \quad
I_{{\bf k}_i}^l\equiv I^l({\bf k}_i,x),\;\;i=1,2.
\label{eq:5p}
\end{equation}
For the first term in integrand of (\ref{eq:5i}) in product
$I^l_{k_1} I^l_{k_2}$ it is necessary to keep only `crossed' terms
\[
I_{{\bf k}_1}^l \delta (k^0_1 - \omega_{{\bf k}_1}^l)
I_{-{\bf k}_2}^l \delta (k^0_2 + \omega_{{\bf k}_2}^l) +
I_{-{\bf k}_1}^l \delta (k^0_1 + \omega_{{\bf k}_1}^l)
I_{{\bf k}_2}^l \delta (k^0_2 - \omega_{{\bf k}_2}^l).
\]
Employing properties (\ref{eq:4p}) and (B.4) it can be shown that
the second term here with replacement $k_1 \rightleftharpoons k_2$
gives a contribution, which is equal to the first term.

For the second term in integrand of (\ref{eq:5i}) in product
$I^l_{k_1} I^l_{k_2}$ we are needed only to keep one term of a `direct'
product
\[
I_{{\bf k}_1}^l \delta (k^0_1 - \omega_{{\bf k}_1}^l)
I_{{\bf k}_2}^l \delta (k^0_2 - \omega_{{\bf k}_2}^l).
\]
The $\delta$-functions give us a possibility to perform integration
in $dk^0_1 dk^0_2$. Now we turn to (anti)plasmino and plasmon
number densities accordingly setting
\begin{equation}
n_{\mathbf q}^{-} = (2\pi)^3\,
2{\rm Z}_{-}^{-1}({\bf q}) \Upsilon_{\bf q}^{-},\;\,
1 - \bar{n}_{\bf q}^{-} = (2\pi)^3\,
2{\rm Z}_{-}^{-1}({\bf q}) \bar{\Upsilon}_{\bf q}^{-},\;\,
N_{{\bf k}}^l = -\,(2\pi)^3\,2\omega_{{\bf k}}^l {\rm Z}_l^{-1}({\bf k})
I_{{\bf k}}^l,
\label{eq:5a}
\end{equation}
where ${\rm Z}_{-}({\bf q})$ and ${\rm Z}_l({\bf p})$ are residues
of the HTL-resummed quark and gluon propagators at plasmino and plasmon
poles, respectively.

As was stated above, a presence of the second and fourth terms
gives us an expression complex conjugate to that of (\ref{eq:5i}). It is not
difficult to see that in practice this comes to a simple replacement of
the factor $(i^{\ast}\Delta_{-}(q))^{\ast}$ in Eq.\,(\ref{eq:5i}) by
\[
(i\,^{\ast}\!\Delta_{-}(q))^{\ast} +
i\,^{\ast}\!\Delta_{-}(q)\simeq-\,2\pi{\rm Z}_{-}({\bf q})
\delta (q^0 - \omega_{\bf q}^{-}).
\]

Taking into account the above-mentioned reasoning, we introduce the following
matrix element of plasmino-plasmon elastic scattering
\begin{equation}
{\rm T}^{a_1a_2,\,ii_1}_{\lambda\lambda_1}(-{\bf k}_1,{\bf k}_2;{\bf q}_1,-{\bf q})
\equiv g^2
\left(\frac{{\rm Z}_{-}({\bf q})}{2}\right)^{\!1/2}\!
\left(\frac{{\rm Z}_{-}({\bf q}_1)}{2}\right)^{\!1/2}\!
\left(\frac{{\rm Z}_l({\bf k}_1)}{2\omega_{{\bf k}_1}^l}\right)^{\!1/2}\!
\left(\frac{{\rm Z}_l({\bf k}_2)}{2\omega_{{\bf k}_2}^l}\right)^{\!1/2}
\label{eq:5s}
\end{equation}
\[
\times\,
\Biggl(\frac{\bar{u}^{\mu_1}(k_1)}{\sqrt{\bar{u}^2(k_1)}}\Biggr)
\Biggl(\frac{\bar{u}^{\mu_1}(k_2)}{\sqrt{\bar{u}^2(k_2)}}\Biggr)
\Bigl[\,\bar{v}_{\beta}(\hat{\bf q}_1,\lambda_1)
\!\,^{\ast}\tilde{\bar{\Gamma}}^{(Q)a_1a_2,\,ii_1}_{\mu_1\mu_2,\,\alpha\beta}\!
(-k_1,k_2;q_1,-q) v_{\alpha}(\hat{\bf q},\lambda)\Bigr]_{\,{\rm on-shell}}.
\]
The probability of plasmino-plasmon scattering is defined as
\begin{equation}
\delta^{ji} {\it w}_{qg\rightarrow qg}^{(-\,l;\,-\,l)}
({\bf q}, {\bf k}_1; {\bf q}_1, {\bf k}_2)=\!\!\!
\sum\limits_{\lambda\!,\,\lambda_1=\,\pm}\!\!
{\rm T}^{a_1a_2,\,ii_1}_{\lambda\lambda_1}(-{\bf k}_1,{\bf k}_2;{\bf q}_1,-{\bf q})
\,({\rm T}^{a_1a_2,\,ji_1}_{\lambda\lambda_1}
(-{\bf k}_1,{\bf k}_2;{\bf q}_1,-{\bf q}))^{\ast}.
\label{eq:5d}
\end{equation}

Now we consider the problem of deriving the probability for
elastic scattering of normal quark excitations off soft transverse
gluon excitations. In this case instead of replacements
(\ref{eq:5u}) it should be used the following those in
Eq.\,(\ref{eq:5y})
\[
I^{\mu_1\mu_1^{\prime}}(k_1)\rightarrow
P^{\mu_1\mu_1^{\prime}}(k_1)I^t(k_1),\quad
I^{\mu_2\mu_2^{\prime}}(k_2)\rightarrow
P^{\mu_2\mu_2^{\prime}}(k_2)I^t(k_2),
\]
\[
\Upsilon_{\beta^{\prime}\!\beta}(q_1)\rightarrow
(h_{+}({\hat{\bf q}}_1))_{\beta^{\prime}\!\beta}
\Upsilon^{+}({\bf q}_1)\delta(q_1^0-\omega^{+}_{{\bf q}_1}) +
(h_{-}({\hat{\bf q}}_1))_{\beta^{\prime}\!\beta}
\bar{\Upsilon}^{+}(-{\bf q}_1)\delta(q_1^0+\omega^{+}_{{\bf q}_1}),
\]
\[
\,^{\ast}\!S_{\alpha\alpha^{\prime}}(-q)\rightarrow
- \,(h_{+}({\hat{\bf q}}))_{\alpha\alpha^{\prime}}
(\,^{\ast}\!\Delta_{+}(q))^{\ast},
\]
and instead of (\ref{eq:5o}) it should be used the expansions
\begin{equation}
(h_{+}({\hat{\bf q}}))_{\alpha\alpha^{\prime}}=
\sum\limits_{\lambda=\pm}u_{\alpha}(\hat{\bf q},\lambda)
\bar{u}_{\alpha^{\prime}}(\hat{\bf q},\lambda),\quad
(h_{+}({\hat{\bf q}}_1))_{\beta^{\prime}\!\beta}=
\sum\limits_{\lambda_1=\pm}u_{\beta^{\prime}}(\hat{\bf q}_1,\lambda_1)
\bar{u}_{\beta}(\hat{\bf q}_1,\lambda_1).
\label{eq:5dd}
\end{equation}
Furthermore the transverse projectors
$P^{\mu_i\mu_i^{\prime}}(k_i),\,i=1,2$ will be also presented in
the form of expansion in eigenvectors of transverse polarization
(Kalashnikov, Klimov \cite{kalashnikov})
\begin{equation}
P^{\mu_i\mu_i^{\prime}}(k_i)=\sum\limits_{\xi_i=1,\,2}
\left(\frac{\epsilon^{\,\mu_i}(k_i,\xi_i)}
{\sqrt{\epsilon^2(k_i,\xi_i)}}\right)\!
\left(\frac{\epsilon^{\,\mu_i^{\prime}}(k_i,\xi_i)}
{\sqrt{\epsilon^2(k_i,\xi_i)}}\right),
\label{eq:5ddd}
\end{equation}
\[
\epsilon^{\,\mu_i}(k_i,\xi_i)\equiv {\rm e}^{(\xi_i)\mu_i}k_i^2
- k_i^{\mu_i}({\rm e}^{(\xi_i)}\cdot k_i),
\quad (k_i\cdot \epsilon(k_i,\xi_i))=0,\quad i=1,2,
\]
where ${\rm e}^{(\xi_i)\mu_i}$ are some four-vectors such that
$\epsilon^{\,\mu_i}(k_i,\xi_i)$ are linearly independent.

The reasoning similar to that used in deriving the probability of
plasmino-plasmon scattering (\ref{eq:5d}) results in the following expression
for desired probability
\begin{equation}
\delta^{ji} {\it w}_{qg\rightarrow qg}^{(+t;\,+t)}
({\bf q}, {\bf k}_1; {\bf q}_1, {\bf k}_2)\!=\!\!\!\!\!\!
\sum\limits_{\xi_1\!,\,\xi_2=1,\,2\,}
\sum\limits_{\lambda,\,\lambda_1=\,\pm}\!\!\!\!
{\rm T}^{a_1a_2,\,ii_1}_{\xi_1\xi_2,\,\lambda\lambda_1}
(-{\bf k}_1,{\bf k}_2;{\bf q}_1,-{\bf q})
({\rm T}^{a_1a_2,\,ji_1}_{\xi_1\xi_2,\,\lambda\lambda_1}
(-{\bf k}_1,{\bf k}_2;{\bf q}_1,-{\bf q}))^{\ast}\!,
\label{eq:5dddd}
\end{equation}
where the matrix element for elastic scattering of soft normal quark by soft
transverse gluon is defined now by the following expression:
\begin{equation}
{\rm T}^{a_1a_2,\,ii_1}_{\xi_1\xi_2,\,\lambda\lambda_1}
(-{\bf k}_1,{\bf k}_2;{\bf q}_1,-{\bf q})
= g^2
\left(\frac{{\rm Z}_{+}({\bf q})}{2}\right)^{\!1/2}\!
\left(\frac{{\rm Z}_{+}({\bf q}_1)}{2}\right)^{\!1/2}\!
\left(\frac{{\rm Z}_t({\bf k}_1)}{2\omega_{{\bf k}_1}^t}\right)^{\!1/2}\!
\left(\frac{{\rm Z}_t({\bf k}_2)}{2\omega_{{\bf k}_2}^t}\right)^{\!1/2}
\label{eq:5ddddd}
\end{equation}
\[
\times\,
\left(\frac{\epsilon^{\,\mu_1}(k_1,\xi_1)}
{\sqrt{\epsilon^2(k_1,\xi_1)}}\right)\!
\left(\frac{\epsilon^{\,\mu_2}(k_2,\xi_2)}
{\sqrt{\epsilon^2(k_2,\xi_2)}}\right)\!
\Bigl[\,\bar{u}_{\beta}(\hat{\bf q}_1,\lambda_1)
\!\,^{\ast}\tilde{\bar{\Gamma}}^{(Q)a_1a_2,\,ii_1}_{\mu_1\mu_2,\,\alpha\beta}\!
(-k_1,k_2;q_1,-q)u_{\alpha}(\hat{\bf q},\lambda)\Bigr]_{\,{\rm on-shell}}.
\]
Here, ${\rm Z}_{+}({\bf q})$ and ${\rm Z}_t({\bf p})$ are residues
of the HTL-resummed quark and gluon propagators at soft normal quark and
soft transverse gluon poles, respectively.

The calculation of probabilities for other types of elastic scattering
processes ${\it w}_{qg\rightarrow qg}^{\!(-\,t;\,-\,t)}$ and
${\it w}_{qg\rightarrow qg}^{\!(+l;\,+l)}$ is quite obvious.
In both cases it is reduced to corresponding
replacements of the gluon wave functions
in Eqs.\,(\ref{eq:5s}), (\ref{eq:5d}) and (\ref{eq:5dddd}), (\ref{eq:5ddddd})
\[
\left(\frac{{\rm Z}_l({\bf k}_i)}{2\omega_{{\bf k}_i}^l}\right)^{\!1/2}\!
\Biggl(\frac{\bar{u}^{\mu_i}(k_i)}{\sqrt{\bar{u}^2(k_i)}}\Biggr)
\Longleftrightarrow
\left(\frac{{\rm Z}_t({\bf k}_i)}{2\omega_{{\bf k}_i}^t}\right)^{\!1/2}\!
\left(\frac{\epsilon^{\,\mu_i}(k_i,\xi_i)}
{\sqrt{\epsilon^2(k_i,\xi_i)}}\right),
\]
$i=1,2$ and proper choice of mass-shell conditions on the right-hand
side of Eqs.\,(\ref{eq:5s}) and (\ref{eq:5ddddd}).
The factors ${\rm Z}_{t,\,l}^{1/2}({\bf k}_i)$ provide renormalization
of transverse and longitudinal gluon wave functions by thermal effects.

The probabilities for more complicated scattering processes, which
result in change of a type of one or both quasiparticles in a
final state (for example, the probabilities ${\it
w}_{qg\rightarrow qg}^{\!(-\,l;\,-\,t)},\, {\it w}_{qg\rightarrow
qg}^{\!(-\,l;\,+l)}$ and so on) can be obtained by similar
replacements of gluon and quark wave functions. However, according
to remark at the end of section 2 in this case it should be
performed further research on the existence of solutions for the
system of energy-momentum conservation laws.

\begin{figure}[hbtp]
\begin{center}
\includegraphics*[scale=0.5]{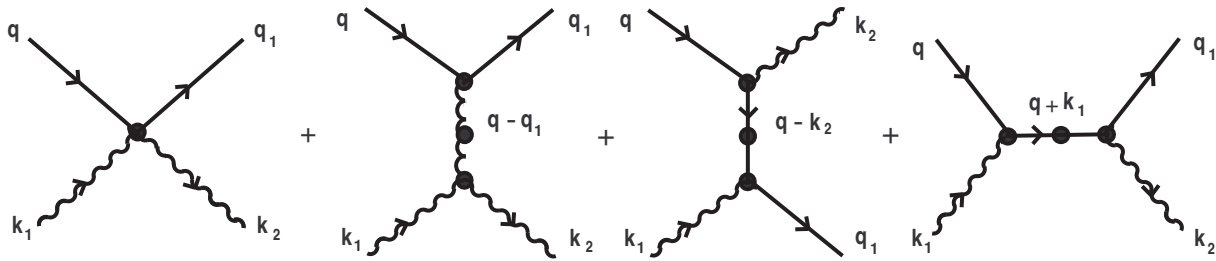}
\end{center}
\caption{\small The matrix element for soft-quark\,--\,\,soft-gluon scattering.
The straight and wave lines denote soft-quark and soft-gluon excitations,
correspondingly. The blob stands for HTL resummation.}
\label{fig1}
\end{figure}

Matrix elements (\ref{eq:5s}) and (\ref{eq:5ddddd}) owing to the
structure of effective amplitude (B.3) have a simple diagrammatic
representation drawn in Fig.\,\ref{fig1}.
The first graph in Fig.\,\ref{fig1} represents a direct
interaction of soft-quark modes with soft-gluon modes induced by HTL-amplitude
$\delta\Gamma_4^{(Q)}$ (Eq.\,(A.2)). The remaining terms
are connected with soft-quark\,--\,\,soft-gluon interaction induced by
two-quark-gluon HTL-resummed vertex and three-gluon HTL-resummed vertex
with intermediate virtual oscillations representing $s$\,- and
$t$\,-\,channel contributions. Matrix elements (\ref{eq:5s}),
(\ref{eq:5ddddd}) also contain $u$\,-\,channel contribution, which it is
not drawn in Fig.\,\ref{fig1}.

Let us consider in more detail a structure of matrix element
(\ref{eq:5s}). For this purpose in the initial
definition of effective amplitude (B.3) we proceed from the product of color
matrices $t^{a_1} t^{a_2}$ and $t^{a_2} t^{a_1}$ to their symmetric
$({\cal S})$ and anti-symmetric $({\cal A})$ combinations
\begin{equation}
t^{a_1}t^{a_2}\rightarrow \frac{1}{2}\,\{t^{a_1},t^{a_2}\}+
\frac{1}{2}\,[\,t^{a_1},t^{a_2}],\quad
t^{a_2}t^{a_1}\rightarrow \frac{1}{2}\,\{t^{a_1},t^{a_2}\}-
\frac{1}{2}\,[\,t^{a_1},t^{a_2}],
\label{eq:5f}
\end{equation}
where $\{\,,\}$ denotes anticommutator in a color space.

The first term only in the effective amplitude (B.3) is needed some comment.
Making use of an explicit definition of HTL-induced vertex function (A.2), we
rewrite the vertex in the following form (we suppress spinor indices and color
indices in the fundamental representation):
\begin{equation}
\delta \Gamma^{(Q)\,a_1a_2}_{\mu_1\mu_2}\!(k_1,k_2;q_1,-q)\!=\!
\frac{1}{2}\,\{t^{a_1},t^{a_2}\}
\delta \Gamma^{(Q;\,{\cal S})}_{\mu_1\mu_2}(k_1,k_2;q_1,-q) +
\frac{1}{2}\,[\,t^{a_1},t^{a_2}]
\delta \Gamma^{(Q;\,{\cal A})}_{\mu_1\mu_2}(k_1,k_2;q_1,-q),
\label{eq:5g}
\end{equation}
where
\[
\delta\Gamma^{(Q;\,{\cal S\!,\,A})}_{\mu_1\mu_2}(k_1,k_2;q_1,-q) =
\omega_0^2\!\!\int\!\frac{{\rm d} \Omega}{4 \pi}\,
\frac{v_{\mu_1}v_{\mu_2} \not\!v}{(v\cdot q_1 + i \epsilon)
(v\cdot q + i \epsilon)}
\hspace{3cm}
\]
\[
\hspace{3cm}
\times
\bigg(\frac{1}{v\cdot (q_1+k_1) + i \epsilon}
\pm \frac{1}{v\cdot (q_1+k_2) + i \epsilon} \bigg).
\]
The functions $\delta\Gamma^{(Q;\,{\cal S\!,\,A})}_{\mu_1\mu_2}$ possess
evident properties
\begin{equation}
\delta\Gamma^{(Q;\,{\cal S\!,\,A})}_{\mu_1\mu_2}(k_1,k_2;q_1,-q) =\pm\,
\delta\Gamma^{(Q;\,{\cal S\!,\,A})}_{\mu_1\mu_2}(k_2,k_1;q_1,-q),
\label{eq:5h}
\end{equation}
\[
\delta\Gamma^{(Q;\,{\cal S\!,\,A})}_{\mu_1\mu_2}(-k_1,-k_2;-q_1,q) =
\gamma^0(\delta\Gamma^{(Q;\,{\cal S\!,\,A})}_{\mu_1\mu_2}
(k_1,k_2;q_1,-q))^{\dagger}\gamma^0.
\]
Taking into account color decomposition (\ref{eq:5g}) and change (\ref{eq:5f}),
we can present the matrix element ${\rm T}^{a_1a_2,\,ii_1}_{\lambda\lambda_1}$ 
as follows:
\begin{equation}
{\rm T}^{a_1a_2,\,ii_1}_{\lambda\lambda_1}
(-{\bf k}_1,{\bf k}_2;{\bf q}_1,-{\bf q})=
\frac{1}{2}\,\{t^{a_2},t^{a_1}\}^{i_1i}
{\cal C}_{\cal S}^{-1\,}
{\rm T}^{({\cal S})}_{\lambda\lambda_1}
(-{\bf k}_1,{\bf k}_2;{\bf q}_1,-{\bf q})
\label{eq:5j}
\end{equation}
\[
\hspace{3.5cm}
+\,\frac{1}{2}\,[\,t^{a_2},t^{a_1}]^{i_1i}
\,{\cal C}_{\cal A}^{-1}
{\rm T}^{({\cal A})}_{\lambda\lambda_1}(-{\bf k}_1,{\bf k}_2;{\bf q}_1,-{\bf q}),
\]
where
\[
{\cal C}_{\cal S\!,\,A}\equiv
\left\{\frac{1}{2}\,C_F\biggl(C_F\mp\frac{1}{2N_c}\biggr)\right\}^{\!1/2},
\]
\begin{equation}
{\rm T}^{({\cal S\!,\,A})}_{\lambda\lambda_1}
(-{\bf k}_1,{\bf k}_2;{\bf q}_1,-{\bf q})
\equiv g^2
\,{\cal C}_{\cal S\!,\,A}\!
\left(\frac{{\rm Z}_{-}({\bf q})}{2}\right)^{\!1/2}\!\!
\left(\frac{{\rm Z}_{-}({\bf q}_1)}{2}\right)^{\!1/2}\!\!
\left(\frac{{\rm Z}_l({\bf k}_1)}{2\omega_{{\bf k}_1}^l}\right)^{\!1/2}\!\!
\left(\frac{{\rm Z}_l({\bf k}_2)}{2\omega_{{\bf k}_2}^l}\right)^{\!1/2}
\label{eq:5k}
\end{equation}
\[
\times\,
\Biggl(\frac{\bar{u}^{\mu_1}(k_1)}{\sqrt{\bar{u}^2(k_1)}}\Biggr)
\Biggl(\frac{\bar{u}^{\mu_1}(k_2)}{\sqrt{\bar{u}^2(k_2)}}\Biggr)
\Bigl[\,\bar{v}(\hat{\bf q}_1,\lambda_1)
{\cal T}^{({\cal S\!,\,A})}_{\mu_1\mu_2}(-{\bf k}_1,{\bf k}_2;{\bf q}_1,-{\bf q})
v(\hat{\bf q},\lambda)\Bigr]_{\,{\rm on-shell}},
\]
\begin{equation}
{\cal T}^{({\cal S})}_{\mu_1\mu_2}(-{\bf k}_1,{\bf k}_2;{\bf q}_1,-{\bf q})
\equiv
\delta\Gamma^{(Q;\,{\cal S})}_{\mu_1\mu_2}(k_1,-k_2;-q_1,q)
\label{eq:5kk}
\end{equation}
\[
-\,^{\ast}\Gamma^{(Q)}_{\mu_2}(k_2;-q_1-k_2,q_1)
\,^{\ast}\!S(-q_1-k_2)
\,^{\ast}\Gamma^{(Q)}_{\mu_1}(-k_1;-q,q+k_1)
\]
\[
-\,^{\ast}\Gamma^{(Q)}_{\mu_1}(-k_1;-q_1+k_1,q_1)
\,^{\ast}\!S(-q_1+k_1)
\,^{\ast}\Gamma^{(Q)}_{\mu_2}(k_2;-q,q-k_2),
\]
and
\[
{\cal T}^{({\cal A})}_{\mu_1\mu_2}(-{\bf k}_1,{\bf k}_2;{\bf q}_1,-{\bf q})
\equiv
\delta\Gamma^{(Q;\,{\cal A})}_{\mu_1\mu_2}(k_1,-k_2;-q_1,q)
\]
\[
+\,2\,^{\ast}\Gamma^{(Q)}_{\nu}(q-q_1;-q,q_1)
\,^{\ast}{\cal D}^{\nu\nu^{\prime}}(k_1-k_2)
\,^\ast\Gamma_{\nu^{\prime}\mu_1\mu_2}(k_1-k_2,-k_1,k_2)
\]
\[
-\,^{\ast}\Gamma^{(Q)}_{\mu_2}(k_2;-q_1-k_2,q_1)
\,^{\ast}\!S(-q_1-k_2)
\,^{\ast}\Gamma^{(Q)}_{\mu_1}(-k_1;-q,q+k_1)
\hspace{0.1cm}
\]
\[
+\,^{\ast}\Gamma^{(Q)}_{\mu_1}(-k_1;-q_1+k_1,q_1)
\,^{\ast}\!S(-q_1+k_1)
\,^{\ast}\Gamma^{(Q)}_{\mu_2}(k_2;-q,q-k_2).
\]
The functions ${\rm T}^{({\cal S\!,\,A})}_{\lambda\lambda_1}$ possess the
properties following from their definitions
\begin{equation}
{\rm T}^{({\cal S\!,\,A})}_{\lambda\lambda_1}(-{\bf k}_1,{\bf k}_2;{\bf q}_1,-{\bf q})
=\pm
{\rm T}^{({\cal S\!,\,A})}_{\lambda\lambda_1}({\bf k}_2,-{\bf k}_1;{\bf q}_1,-{\bf q}).
\label{eq:5l}
\end{equation}
The advantage of the choice of new color basis (\ref{eq:5f}) is defined by the
following properties:
\[
(\{t^{a_1},t^{a_2}\}\{t^{a_2},t^{a_1}\})^{ji}=
2C_F\biggl(C_F-\frac{1}{2N_c}\biggr)\delta^{ji},
\]
\begin{equation}
([\,t^{a_1},t^{a_2}][\,t^{a_2},t^{a_1}])^{ji}=
2C_F\biggl(C_F+\frac{1}{2N_c}\biggr)\delta^{ji},
\label{eq:5z}
\end{equation}
\[
(\{t^{a_1},t^{a_2}\}[\,t^{a_2},t^{a_1}])^{ji}=0,
\]
i.e., crossed terms vanish. Making use of the color algebra (\ref{eq:5z}), we
obtain instead of Eq.\,(\ref{eq:5d})
\begin{equation}
{\it w}_{qg\rightarrow qg}^{(-\,l;\,-\,l)}
({\bf q}, {\bf k}_1; {\bf q}_1, {\bf k}_2)=
{\it w}_{qg\rightarrow qg}^{({\cal S})}
({\bf q}, {\bf k}_1; {\bf q}_1, {\bf k}_2) +
{\it w}_{qg\rightarrow qg}^{({\cal A})}
({\bf q}, {\bf k}_1; {\bf q}_1, {\bf k}_2),
\label{eq:5x}
\end{equation}
where
\[
{\it w}_{qg\rightarrow qg}^{({\cal S\!,\,A})}
({\bf q}, {\bf k}_1; {\bf q}_1, {\bf k}_2)=
\sum\limits_{\lambda\!,\,\lambda_1=\pm}\!
\Bigl|\,
{\rm T}^{({\cal S\!,\,A})}_{\lambda\lambda_1}(-{\bf k}_1,{\bf k}_2;{\bf q}_1,-{\bf q})
\Bigr|^2_{\,{\rm on-shell}}.
\]
Completely similar decomposition is true and for probability
${\it w}_{qg\rightarrow qg}^{(+t;\,+t)}$ (Eq.\,(\ref{eq:5dddd}))
and also for probabilities ${\it w}_{qg\rightarrow qg}^{(+l;\,+l)}$, and
${\it w}_{qg\rightarrow qg}^{(-t;\,-t)}$.

The second term in integrand of Eq.\,(\ref{eq:5i})
defines the probability for plasmino-antiplas\-mino annihilation into two
plasmons $w_{q\bar{q}\rightarrow gg}^{(--;\,l\,l)}$. The probability is
obtained from (\ref{eq:5x}) by replacement
${\rm T}^{({\cal S\!,\,A})}_{\lambda\lambda_1}(-k_1, k_2; q_1, -q) \rightarrow
{\rm T}^{({\cal S\!,\,A})}_{\lambda\lambda_1}(k_1, k_2; -q_1, -q)$
(without change of a sign of vector ${\bf q}_1$ in spinor
$\bar{v}(\hat{\bf q}_1,\lambda_1)$). Accordingly the probability for
annihilation of soft normal quark-antiquark pair into two soft transverse
gluons $w_{q\bar{q}\rightarrow gg}^{(++;\,t\,t)}$ is obtained by
replacement ${\rm T}^{({\cal S\!,\,A})}_{\xi_1\xi_2,\,\lambda\lambda_1}
(-{\bf k}_1,{\bf k}_2;{\bf q}_1,-{\bf q}) \rightarrow
{\rm T}^{({\cal S\!,\,A})}_{\xi_1\xi_2,\,\lambda\lambda_1}
({\bf k}_1,{\bf k}_2;-{\bf q}_1,-{\bf q})$ also
without change of a sign of vector ${\bf q}_1$ in spinor
$\bar{u}(\hat{\bf q}_1,\lambda_1)$.

From Eq.\,(\ref{eq:5x}) we see that the processes of elastic
scattering and annihilation proceed through two physical
independent channels determined by a parity of state of two gluon
system with respect to permutation of external soft-gluon legs
(Eq.\,(\ref{eq:5l})). The symbols $({\cal S, A})$ belong to states of two
gluons being in even and odd states correspondingly (in the c.m.s.
of gluons).

We now return to initial equation (\ref{eq:5q}). Let us transform the left-hand
side of this equation similar to Ref.\,\cite{markov_PRD}. We result in
\[
-\,\delta^{ji}\delta(q^0-\omega_{\bf q}^{(f)})
\left(
\frac{\partial n_{\bf q}^{(f)}}{\partial t} +
{\bf v}_{\bf q}^{(f)}\cdot\frac{\partial n_{\bf q}^{(f)}}{\partial{\bf x}}
\right).
\]
Taking into account above-mentioned, finally we write out the expression for
kinetic equation defining the change in the colorless soft quark number
densities caused by `spontaneous' processes of soft-quark\,--\,soft-gluon
elastic scattering and annihilation of soft quark-antiquark pair
\begin{equation}
\left(
\frac{\partial n_{\bf q}^{(f)}}{\partial t} +
{\bf v}_{\bf q}^{(f)}\cdot\frac{\partial n_{\bf q}^{(f)}}{\partial{\bf x}}
\right)^{\!{\rm sp}}\!\!\!=
\!\!\!\sum\limits_{f_1=\pm}\sum\limits_{\,b_1\!,\,b_2=t,\,l}\Biggl\{
2\!\int\!d{\cal T}_{qg\rightarrow qg}^{(fb_1;\,f_1b_2)}
\, {\it w}_{qg\rightarrow qg}^{(fb_1;\,f_1b_2)}
({\bf q}, {\bf k}_1; {\bf q}_1, {\bf k}_2)
n_{{\bf q}_1}^{(f_1)}
N_{{\bf k}_1}^{(b_1)}N_{{\bf k}_2}^{(b_2)}
\label{eq:5c}
\end{equation}
\[
\hspace{4.9cm}
+\!\int\!d{\cal T}_{q\bar{q}\rightarrow gg}^{(ff_1;\,b_1b_2)}\,
{\it w}_{q\bar{q}\rightarrow gg}^{(ff_1;\,b_1b_2)}
({\bf q}, {\bf q}_1; {\bf k}_1, {\bf k}_2)
\Bigl(1-{\bar n}_{{\bf q}_1}^{(f_1)}\Bigl)
N_{{\bf k}_1}^{(b_1)}N_{{\bf k}_2}^{(b_2)}\!\Biggr\}.
\]
This equation follows from (\ref{eq:2w}), (\ref{eq:2y}) in the limit of
a small intensity of soft-quark mode $n^{(f)}_{\bf q} \rightarrow 0$ and
making use of the fact that the boson occupation numbers
$N^{(b_i)}_{{\bf k}_i}$ are much more than one,
$1 + N^{(b_i)}_{{\bf k}_i}\simeq N^{(b_i)}_{{\bf k}_i},\,i=1,2$.

\section{\bf Probabilities of soft-quark\,--\,soft-(anti)quark elastic 
scattering}
\setcounter{equation}{0}

Now we consider the problem of determination of the probabilities for the
scattering processes of soft-quark excitations off soft-quark and
soft-antiquark excitations. For this purpose we return to term (\ref{eq:5w})
and examine the following replacement:
\[
\bar{\psi}^i_{\alpha}(-q)\rightarrow
\tilde{\bar{\eta}}^{(0,3)i}_{\gamma}
(\bar{\psi}^{(0)},\bar{\psi}^{(0)},\psi^{(0)})(-q)
\,^{\ast}\!S_{\gamma\alpha}(-q),
\quad
\psi^{j^{\prime}}_{\beta}(q_1)\rightarrow \psi^{(0)j^{\prime}}_{\beta}(q_1),
\]
\[
A^{a_1\mu_1}(k_1)\rightarrow
-\,^{\ast}{\cal D}^{\mu_1\mu_1^{\prime}}(k_1)
\tilde{j}^{\Psi(0,2)a_1}_{\mu_1^{\prime}}(\bar{\psi}^{(0)},\psi^{(0)})(k_1),
\]
where the effective source $\tilde{\bar{\eta}}^{(0,3)i}_{\gamma}$ is defined by
Eq.\,(\ref{eq:3d}). This substitution results in sixth-order correlator with
respect to free quark fields $\psi^{(0)}$ and $\bar{\psi}^{(0)}$.
Furthermore, we decouple the sixth-order correlation function in terms of pair
correlators $\langle\bar{\psi}^{(0)}\psi^{(0)}\rangle$. One can show that just
two terms of this decoupling reproduce finally necessary factor in integrand:
$\delta(q+q_1-q_2-q_3)$. Taking into account these terms and the definition of
the effective vertex
$^{\ast}\tilde{\Gamma}^{ii_1i_2i_3}_{\alpha\alpha_1\alpha_2\alpha_3}$,
Eq.\,(\ref{eq:4a}), we get instead of (\ref{eq:5w})
\begin{equation}
-\,\frac{ig^4}{2!}\,^{\ast}\!S_{\alpha\alpha^{\prime}}(-q)\!
\int\! dq_1 dq_2 dq_3\,\delta(q+q_1-q_2-q_3)\,
\!\,^{\ast}\tilde{\bar{\Gamma}}^{ii_1i_2i_3}_{\alpha\alpha_1\alpha_2\alpha_3}
(q,q_1,-q_2,-q_3)
\label{eq:6q}
\end{equation}
\[
\times\,
\!\,^{\ast}\tilde{\Gamma}^{ji_1i_2i_3}_{\alpha^{\prime}\alpha_1^{\prime}
\alpha_2^{\prime}\alpha_3^{\prime}}(q,q_1,-q_2,-q_3)
\Upsilon_{\alpha_1\alpha_1^{\prime}}(q_1)
\Upsilon_{\alpha_2^{\prime}\alpha_2}(q_2)
\Upsilon_{\alpha_3^{\prime}\alpha_3}(q_3).
\]

At first we will obtain an expression for probability of elastic scattering of
plasmino off plasmino ${\it w}_{qq\rightarrow qq}^{\!(--;--)}
({\bf q}, {\bf q}_1; {\bf q}_2, {\bf q}_3)$. For this purpose we set
\[
\,^{\ast}\!S_{\alpha\alpha^{\prime}}(-q)\rightarrow
- \,(h_{-}({\hat{\bf q}}))_{\alpha\alpha^{\prime}}
(\,^{\ast}\!\Delta_{-}(q))^{\ast},
\]
\[
\Upsilon_{\!\alpha_1\alpha_1^{\prime}}(q_1)\rightarrow
(h_{-}({\hat{\bf q}}_1))_{\alpha_1\alpha_1^{\prime}}
\hat{\Upsilon}_{{\bf q}_1}^{-}\delta(q_1^0-\omega^{-}_{{\bf q}_1}) +
(h_{+}({\hat{\bf q}}_1))_{\alpha_1\alpha_1^{\prime}}
\hat{\bar{\Upsilon}}_{-{\bf q}_1}^{-}\delta(q_1^0+\omega^{-}_{{\bf q}_1})
\hspace{1.3cm}
\]
\[
\hspace{0.95cm}
\Upsilon_{\alpha_i^{\prime}\alpha_i}(q_i)\,\rightarrow
(h_{-}({\hat{\bf q}}_i))_{\alpha^{\prime}_i\alpha_i}
\Upsilon_{{\bf q}_i}^{-}\delta(q_i^0-\omega^{-}_{{\bf q}_i}) +
(h_{+}({\hat{\bf q}}_i))_{\alpha^{\prime}_i\alpha_i}
\bar{\Upsilon}_{-{\bf q}_i}^{-}\delta(q_i^0+\omega^{-}_{{\bf q}_i}),
\quad i=2,\,3.
\hspace{0.8cm}
\]
Let us especially note that the order of convolution of Dirac indices for
soft-quark spectral density $\Upsilon_{\!\alpha_1\alpha_1^{\prime}}(q_1)$
with the effective amplitudes 
$\!\,^{\ast}\tilde{\bar{\Gamma}}^{ii_1i_2i_3}_{\alpha\alpha_1\alpha_2\alpha_3}$
and 
$\!\,^{\ast}\tilde{\Gamma}^{ji_1i_2i_3}_{\alpha^{\prime}\alpha_1^{\prime}
\alpha_2^{\prime}\alpha_3^{\prime}}$ in Eq.\,(\ref{eq:6q}) is the
reverse of convolution of Dirac indices of soft-quark
spectral densities $\Upsilon_{\alpha_2^{\prime}\alpha_2}(q_2)$ and
$\Upsilon_{\alpha_3^{\prime}\alpha_3}(q_3)$ (the order of convolution
of the last functions coincides with that of convolution of the function 
$\Upsilon_{\!\alpha_1\alpha_1^{\prime}}(q_1)$ with the effective amplitudes
$\!\,^{\ast}\tilde{\bar{\Gamma}}^{(Q)a_1a_2,\,il}_{\mu_1\mu_2,\,\alpha\beta}$
and 
$\,\!\,^{\ast}\tilde{\Gamma}^{(Q)a_1a_2,\,jl}_{\mu_1^{\prime}\mu_2^{\prime},
\,\alpha^{\prime}\beta^{\prime}}$ in Eq.\,(\ref{eq:5y})). To account for this 
nontrivial fact in the expansion of the function 
$\Upsilon_{\!\alpha_1\alpha_1^{\prime}}(q_1)$ in terms of the spinor projectors
in comparison with similar expansion (\ref{eq:5u}) we introduce new functions
$\hat{\Upsilon}_{{\bf q}_1}^{-}$ and $\hat{\bar{\Upsilon}}_{-{\bf q}_1}^{-}$
(with hat above). In the product $\Upsilon_{\alpha_1\alpha_1^{\prime}}(q_1)
\Upsilon_{\alpha_2^{\prime}\alpha_2}(q_2)
\Upsilon_{\alpha_3^{\prime}\alpha_3}(q_3)$ we keep only the terms
preserved a form of $\delta$-function argument in integrand of
Eq.\,(\ref{eq:6q}) (up to a trivial permutations of 4-momenta $q_1$, $q_2$
and $q_3$). There exist three such terms:
\begin{equation}
(h_{-}({\hat{\bf q}}_1))_{\alpha_1\alpha_1^{\prime}}
\hat{\Upsilon}_{{\bf q}_1}^{-}\delta(q_1^0-\omega^{-}_{{\bf q}_1})
(h_{-}({\hat{\bf q}}_2))_{\alpha_2^{\prime}\alpha_2}
\Upsilon_{{\bf q}_2}^{-}\delta(q_2^0-\omega^{-}_{{\bf q}_2})
(h_{-}({\hat{\bf q}}_3))_{\alpha_3^{\prime}\alpha_3}
\Upsilon_{{\bf q}_3}^{-}\delta(q_3^0-\omega^{-}_{{\bf q}_3})
\label{eq:6w}
\end{equation}
\[
+\,
(h_{+}({\hat{\bf q}}_1))_{\alpha_1\alpha_1^{\prime}}
\hat{\bar{\Upsilon}}_{-{\bf q}_1}^{-}\delta(q_1^0+\omega^{-}_{{\bf q}_1})
(h_{+}({\hat{\bf q}}_2))_{\alpha_2^{\prime}\alpha_2}
\bar{\Upsilon}_{-{\bf q}_2}^{-}\delta(q_2^0+\omega^{-}_{{\bf q}_2})
(h_{-}({\hat{\bf q}}_3))_{\alpha_3^{\prime}\alpha_3}
\Upsilon_{{\bf q}_3}^{-}\delta(q_3^0-\omega^{-}_{{\bf q}_3})
\]
\[
+\,
(h_{+}({\hat{\bf q}}_1))_{\alpha_1\alpha_1^{\prime}}
\hat{\bar{\Upsilon}}_{-{\bf q}_1}^{-}\delta(q_1^0+\omega^{-}_{{\bf q}_1})
(h_{-}({\hat{\bf q}}_2))_{\alpha_2^{\prime}\alpha_2}
\Upsilon_{{\bf q}_2}^{-}\delta(q_2^0-\omega^{-}_{{\bf q}_2})
(h_{+}({\hat{\bf q}}_3))_{\alpha_3^{\prime}\alpha_3}
\bar{\Upsilon}_{-{\bf q}_3}^{-}\delta(q_3^0+\omega^{-}_{{\bf q}_3}).
\]

\begin{figure}[hbtp]
\begin{center}
\includegraphics[width=0.7\textwidth]{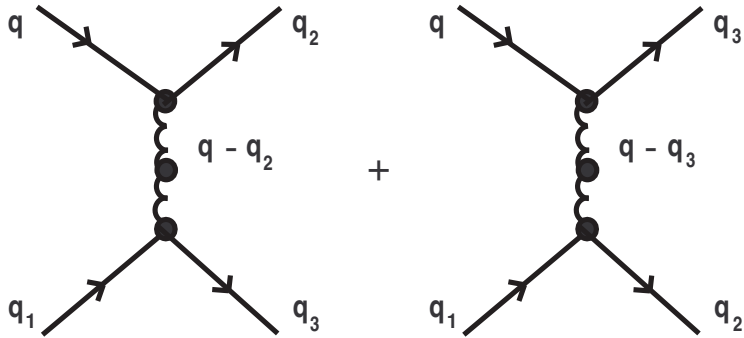}
\end{center}
\caption{\small M$\!\not{\!{\rm o}}$ller-like elastic
soft-quark by soft-quark scattering.}
\label{fig2}
\end{figure}

Let us consider a contribution of the expression in the first line.
We substitute this term into Eq.\,(\ref{eq:6q}) and perform integration
in $dq^0_1 dq^0_2 dq^0_3$. Further we expand spinor projectors
$h_{-}({\hat{\bf q}})$ in terms of eigenspinors (Eq.\,(\ref{eq:5o})) and turn
to the (anti)plasmino number densities according to rules (\ref{eq:5a})
supplemented by
\begin{equation}
1 - n_{\mathbf q}^{-} = (2\pi)^3\,
2{\rm Z}_{-}^{-1}({\bf q}) \hat{\Upsilon}_{\bf q}^{-},
\quad
\bar{n}_{\bf q}^{-} = (2\pi)^3\,
2{\rm Z}_{-}^{-1}({\bf q}) \hat{\bar{\Upsilon}}_{\bf q}^{-}.
\label{eq:6ww}
\end{equation}
The second term on the right-hand side of Eq.\,(\ref{eq:5q}) gives a
contribution complex conjugate to that of (\ref{eq:5w}). The sum of these two
terms will differ from (\ref{eq:6q}) by a simple replacement
\[
\frac{i}{2}\,(\,^{\ast}\!\Delta_{-}(q))^{\ast}\rightarrow
\frac{i}{2}\,\Bigl[\,(\,^{\ast}\!\Delta_{-}(q))^{\ast} -
\,^{\ast}\!\Delta_{-}(q)\Bigr]
\simeq
\pi{\rm Z}_{-}({\bf q})\delta(q^0-\omega^{-}_{\bf q}).
\]
Now we introduce the following matrix element of elastic plasmino-plasmino
scattering, setting
\begin{equation}
{\rm T}^{ii_1i_2i_3}_{\lambda\lambda_1\lambda_2\lambda_3}
({\bf q},{\bf q}_1;-{\bf q}_2,-{\bf q}_3)
\equiv g^2
\left(\frac{{\rm Z}_{-}({\bf q})}{2}\right)^{\!1/2}\!
\prod_{i=1}^{3}
\left(\frac{{\rm Z}_{-}({\bf q}_i)}{2}\right)^{\!1/2}
\label{eq:6e}
\end{equation}
\[
\times\,
v_{\alpha}(\hat{\bf q},\lambda)v_{\alpha_1}(\hat{\bf q}_1,\lambda_1)
\bar{v}_{\alpha_2}(\hat{\bf q}_2,\lambda_2)
\bar{v}_{\alpha_3}(\hat{\bf q}_3,\lambda_3)
\!\,^{\ast}\tilde{\bar{\Gamma}}^{ii_1i_2i_3}_{\alpha\alpha_1\alpha_2\alpha_3}
(q,q_1,-q_2,-q_3).
\]
The probability of plasmino-plasmino scattering is defined as
\begin{equation}
\delta^{ji} {\it w}_{qq\rightarrow qq}^{\!(--;--)}
({\bf q}, {\bf q}_1; {\bf q}_2, {\bf q}_3)=\!\!\!
\sum\limits_{\lambda\!,\,\lambda_1,\,\ldots\,=\pm}\!\!
{\rm T}^{ii_1i_2i_3}_{\lambda\lambda_1\lambda_2\lambda_3}
({\bf q},{\bf q}_1;-{\bf q}_2,-{\bf q}_3)
({\rm T}^{ii_1i_2i_3}_{\lambda\lambda_1\lambda_2\lambda_3}
({\bf q},{\bf q}_1;-{\bf q}_2,-{\bf q}_3))^{\ast}.
\label{eq:6r}
\end{equation}
The diagrammatic interpretation of the different terms of matrix element
(\ref{eq:6e}) is presented in Fig.\,\ref{fig2}.

The probability of elastic scattering of plasmino off soft normal quark
excitation ${\it w}_{qq\rightarrow qq}^{\!(-\,+;-\,+)}$ is obtained from
Eqs.\,(\ref{eq:6r}) and (\ref{eq:6e}) by simple replacement of quark wave
functions
\[
\left(\!\frac{{\rm Z}_{-}({\bf q}_1)}{2}\!\right)^{\!\!1/2}\!\!\!\!
v_{\alpha_1}(\hat{\bf q}_1,\lambda_1)\!
\rightarrow\!\!
\left(\!\frac{{\rm Z}_{+}({\bf q}_1)}{2}\!\right)^{\!\!1/2}\!\!\!\!
u_{\alpha_1}(\hat{\bf q}_1,\lambda_1),
\left(\!\frac{{\rm Z}_{-}({\bf q}_3)}{2}\!\right)^{\!\!1/2}\!\!\!\!
\bar{v}_{\alpha_3}(\hat{\bf q}_3,\lambda_3)\!
\rightarrow\!\!
\left(\!\frac{{\rm Z}_{+}({\bf q}_3)}{2}\!\right)^{\!\!1/2}\!\!\!\!
\bar{u}_{\alpha_3}(\hat{\bf q}_3,\lambda_3),
\]
and corresponding choice of mass-shell conditions.
For deriving the probability of mutually elastic scattering of
soft normal quark excitations 
${\it w}_{qq\rightarrow qq}^{\!(++;++)}$ it should be also added the 
following replacements to the above-stated those:
\[
\left(\frac{{\rm Z}_{-}({\bf q})}{2}\right)^{\!\!1/2}\!\!\!
v_{\alpha}(\hat{\bf q},\lambda)\!
\rightarrow\!
\left(\frac{{\rm Z}_{+}({\bf q})}{2}\right)^{\!\!1/2}\!\!\!
u_{\alpha}(\hat{\bf q},\lambda),
\,
\left(\frac{{\rm Z}_{-}({\bf q}_2)}{2}\right)^{\!\!1/2}\!\!\!
\bar{v}_{\alpha_2}(\hat{\bf q}_2,\lambda_2)\!
\rightarrow\!
\left(\frac{{\rm Z}_{+}({\bf q}_2)}{2}\right)^{\!\!1/2}\!\!\!
\bar{u}_{\alpha_2}(\hat{\bf q}_2,\lambda_2).
\]
In the remaining part of this section we shall restrict our attention to
study of the scattering processes including only plasmino and
antiplasmino. Results for the scattering processes with soft normal quark 
modes follows from previous by above-mentioned replacements of quark wave 
functions.

For more detailed analysis of structure of probability
(\ref{eq:6r}) we regroup the terms in the effective amplitudes
$\!\,^{\ast}\tilde{\bar{\Gamma}}^{ii_1i_2i_3}_{\alpha\alpha_1\alpha_2\alpha_3}
(q,q_1,-q_2,-q_3)$ and
$\!\,^{\ast}\tilde{\Gamma}^{ji_1i_2i_3}_{\alpha^{\prime}\alpha_1^{\prime}
\alpha_2^{\prime}\alpha_3^{\prime}}(q,q_1,-q_2,-q_3)$. For this purpose we
replace initial color factors by `orthogonal' ones
\[
(t^a)^{ii_2}(t^a)^{i_1i_3}\rightarrow\frac{1}{2}\,
\Bigl\{(t^a)^{ii_2}(t^a)^{i_1i_3} + (t^a)^{ii_3}(t^a)^{i_1i_2}\Bigr\}
+
\frac{1}{2}\,\Bigl\{(t^a)^{ii_2}(t^a)^{i_1i_3} -
(t^a)^{ii_3}(t^a)^{i_1i_2}\Bigr\},
\]
\[
(t^a)^{ii_3}(t^a)^{i_1i_2}\rightarrow\frac{1}{2}\,
\Bigl\{(t^a)^{ii_2}(t^a)^{i_1i_3} + (t^a)^{ii_3}(t^a)^{i_1i_2}\Bigr\}
-
\frac{1}{2}\,\Bigl\{(t^a)^{ii_2}(t^a)^{i_1i_3} -
(t^a)^{ii_3}(t^a)^{i_1i_2}\Bigr\}.
\]
Thus we can represent matrix element (\ref{eq:6e}) in the following form:
\begin{equation}
{\rm T}^{ii_1i_2i_3}_{\lambda\lambda_1\lambda_2\lambda_3}
({\bf q},{\bf q}_1;-{\bf q}_2,-{\bf q}_3)=
\label{eq:6rr}
\end{equation}
\[
= \frac{1}{2}\,
\Bigl\{(t^a)^{ii_2}(t^a)^{i_1i_3} + (t^a)^{ii_3}(t^a)^{i_1i_2}\Bigr\}
\,\check{\cal C}_{\cal A}^{-1}\,
{\rm T}^{({\cal A})}_{\lambda\lambda_1\lambda_2\lambda_3}
({\bf q},{\bf q}_1;-{\bf q}_2,-{\bf q}_3)
\]
\[
\hspace{0.2cm}
+\,\frac{1}{2}\,
\Bigl\{(t^a)^{ii_2}(t^a)^{i_1i_3} - (t^a)^{ii_3}(t^a)^{i_1i_2}\Bigr\}
\,\check{\cal C}_{\cal S}^{-1}\,
{\rm T}^{({\cal S})}_{\lambda\lambda_1\lambda_2\lambda_3}
({\bf q},{\bf q}_1;-{\bf q}_2,-{\bf q}_3),
\]
where
\[
\check{\cal C}_{\cal S,\,A}\equiv
\left\{\frac{1}{4}\,C_F\biggl(1\pm\frac{1}{N_c}\biggr)\right\}^{\!1/2},
\]
\begin{equation}
{\rm T}^{({\cal S,\,A})}_{\lambda\lambda_1\lambda_2\lambda_3}({\bf q},{\bf q}_1;-{\bf q}_2,-{\bf q}_3)
\equiv g^2 \,\check{\cal C}_{\cal S,\,A}\!
\left(\frac{{\rm Z}_{-}({\bf q})}{2}\right)^{\!1/2}\!
\prod_{i=1}^{3}
\left(\frac{{\rm Z}_{-}({\bf q}_i)}{2}\right)^{\!1/2}
\label{eq:6t}
\end{equation}
\[
\times\,
v_{\alpha}(\hat{\bf q},\lambda)v_{\alpha_1}(\hat{\bf q}_1,\lambda_1)
\bar{v}_{\alpha_2}(\hat{\bf q}_2,\lambda_2)\bar{v}_{\alpha_3}(\hat{\bf q}_3,\lambda_3)
\]
\[
\times\,
\Bigl[
\,^{\ast}\Gamma^{(Q)\nu}_{\alpha_2\alpha}(-q+q_2;-q_2,q)
\,^{\ast}{\cal D}_{\nu\nu^{\prime}}(-q+q_2)
\,^{\ast}\Gamma^{(G)\nu^{\prime}}_{\alpha_3\alpha_1}(q_1-q_3;-q_1,q_3)
\]
\[
\hspace{1.8cm}
\pm
\,^{\ast}\Gamma^{(Q)\nu}_{\alpha_3\alpha}(-q+q_3;-q_3,q)
\,^{\ast}{\cal D}_{\nu\nu^{\prime}}(-q+q_3)
\,^{\ast}\Gamma^{(G)\nu^{\prime}}_{\alpha_2\alpha_1}(q_1-q_2;-q_1,q_2)
\Bigr]_{\,{\rm on-shell}}.
\]
The functions ${\rm T}^{({\cal S,\,A})}_{\lambda\lambda_1\lambda_2\lambda_3}$
possess properties
\[
{\rm T}^{({\cal S,\,A})}_{\lambda\lambda_1\lambda_2\lambda_3}
({\bf q},{\bf q}_1;-{\bf q}_2,-{\bf q}_3)= \pm\,
{\rm T}^{({\cal S,\,A})}_{\lambda\lambda_1\lambda_3\lambda_2}
({\bf q},{\bf q}_1;-{\bf q}_3,-{\bf q}_2).
\]
Taking into account a color algebra
\[
\Bigl\{(t^a)^{i_2i}(t^a)^{i_3i_1} \pm (t^a)^{i_2i_1}(t^a)^{i_3i}\Bigr\}
\Bigl\{(t^b)^{ji_2}(t^b)^{i_1i_3} \pm (t^b)^{ji_3}(t^b)^{i_1i_2}\Bigr\}
\]
\[
=2\,\Bigl\{(t^at^b)^{ji}{\rm tr}(t^at^b)\pm
(t^at^bt^at^b)^{ji}\Bigr\} = \delta^{ji}C_F\biggl(1\mp\frac{1}{N_c}\biggr),
\]
\[
\Bigl\{(t^a)^{i_2i}(t^a)^{i_3i_1} \pm (t^a)^{i_2i_1}(t^a)^{i_3i}\Bigr\}
\Bigl\{(t^b)^{ji_2}(t^b)^{i_1i_3} \mp (t^b)^{ji_3}(t^b)^{i_1i_2}\Bigr\}=0,
\]
we can cast probability (\ref{eq:6r}) into a sum of two independent part
\begin{equation}
{\it w}_{qq\rightarrow qq}^{\!(--;--)}
({\bf q}, {\bf q}_1; {\bf q}_2, {\bf q}_3)=
{\it w}_{qq\rightarrow qq}^{({\cal S})}
({\bf q}, {\bf q}_1; {\bf q}_2, {\bf q}_3) +
{\it w}_{qq\rightarrow qq}^{({\cal A})}
({\bf q}, {\bf q}_1; {\bf q}_1, {\bf q}_2),
\label{eq:6y}
\end{equation}
where
\begin{equation}
{\it w}_{qq\rightarrow qq}^{({\cal S,\,A})}
({\bf q}, {\bf q}_1; {\bf q}_2, {\bf q}_3)=
\sum\limits_{\lambda\!,\,\lambda_1,\,\dots\,=\pm}\!
\Bigl|\,
{\rm T}^{({\cal S,\,A})}_{\lambda\lambda_1\lambda_2\lambda_3}({\bf q},{\bf q}_1;-{\bf q}_2,-{\bf q}_3)
\Bigr|^2_{\,{\rm on-shell}}.
\label{eq:6u}
\end{equation}
Thus, as in the case of elastic scattering of soft-quark excitations off 
soft-gluon excitations (see previous section), the elastic scattering process 
of soft-quark excitations off each other proceed through two 
physical independent channels determined by parity of final state of 
soft-quark quasiparticles system. We pay attention to the fact that in the 
color decomposition of matrix element (\ref{eq:6rr}) function 
${\rm T}^{({\cal A})}_{\lambda\lambda_1\lambda_2\lambda_3}$ stands with 
{\it symmetric} combination of color matrices, and 
${\rm T}^{({\cal S})}_{\lambda\lambda_1\lambda_2\lambda_3}$ stands with 
{\it anti-symmetric} combination.

\begin{figure}[hbtp]
\begin{center}
\includegraphics[width=0.95\textwidth]{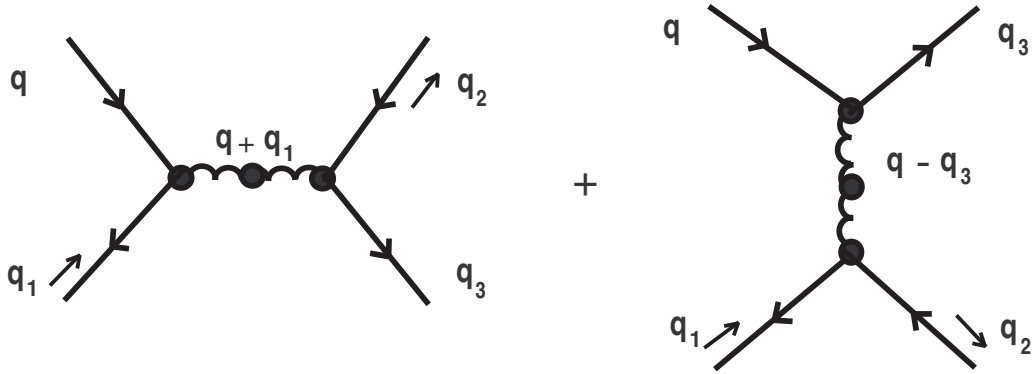}
\end{center}
\caption{\small Bhabha-like elastic scattering of soft-quark off 
soft-antiquark excitations.}
\label{fig3}
\end{figure}

The second term in Eq.\,(\ref{eq:6w}) defines the process of
elastic scattering of plasmino off antiplasmino. Reasoning similar
to previous one results in expression for the probability
${\it w}_{q\bar{q}\rightarrow q{\bar q}}^{\!(--;--)}
({\bf q}, {\bf q}_1; {\bf q}_2,
{\bf q}_3)$ analogous to (\ref{eq:6y}), (\ref{eq:6u}) with
replacement\footnote{Besides, in the kinetic equation it is
necessary to replace $(1-n^{-}_{{\bf q}_1})n^{-}_{{\bf q}_2}
n^{-}_{{\bf q}_3}$ by $(1-\bar{n}^{-}_{{\bf
q}_1})\bar{n}^{-}_{{\bf q}_2}n^{-}_{{\bf q}_3}$.} of the helical
amplitudes ${\rm T}^{({\cal S,\,A})}_{\lambda\lambda_1\lambda_2\lambda_3}$ by
\[
\bar{\rm T}^{({\cal S,\,A})}_{\lambda\lambda_1\lambda_2\lambda_3}
({\bf q},{\bf q}_1;-{\bf q}_2,-{\bf q}_3)
\equiv g^2 \,\check{\cal C}_{\cal S,\,A}\!
\left(\frac{{\rm Z}_{-}({\bf q})}{2}\right)^{\!1/2}\!
\prod_{i=1}^{3}
\left(\frac{{\rm Z}_{-}({\bf q}_i)}{2}\right)^{\!1/2}
\]
\[
\times\,
v_{\alpha}(\hat{\bf q},\lambda)v_{\alpha_2}(\hat{\bf q}_2,\lambda_2)
\bar{v}_{\alpha_1}(\hat{\bf q}_1,\lambda_1)\bar{v}_{\alpha_3}(\hat{\bf q}_3,\lambda_3)
\]
\[
\times\,
\Bigl[
\,^{\ast}\Gamma^{(Q)\nu}_{\alpha_1\alpha}(-q-q_1;q_1,q)
\,^{\ast}{\cal D}_{\nu\nu^{\prime}}(-q-q_1)
\,^{\ast}\Gamma^{(G)\nu^{\prime}}_{\alpha_3\alpha_2}(-q_2-q_3;q_2,q_3)
\]
\[
\hspace{1.8cm}
\pm
\,^{\ast}\Gamma^{(Q)\nu}_{\alpha_3\alpha}(-q+q_3;-q_3,q)
\,^{\ast}{\cal D}_{\nu\nu^{\prime}}(-q+q_3)
\,^{\ast}\Gamma^{(G)\nu^{\prime}}_{\alpha_1\alpha_2}(q_1-q_2;q_2,-q_1)
\Bigr]_{\,{\rm on-shell}}.
\]
The diagrammatic interpretation of the different terms of matrix element
in this case is presented in Fig.\,\ref{fig3}.

Finally, the third term in Eq.\,(\ref{eq:6w}) also defines the process of
elastic scattering of plasmino off antiplasmino. The matrix element of this
process $\hat{\bar{\rm T}}^{({\cal S,\,A})}_{\lambda\lambda_1\lambda_2\lambda_3}$ is associated with
the preceding matrix element by relation
\[
\hat{\bar{\rm T}}^{({\cal S,\,A})}_{\lambda\lambda_1\lambda_2\lambda_3}
({\bf q},{\bf q}_1;-{\bf q}_2,-{\bf q}_3) =\pm\,
\bar{\rm T}^{({\cal S,\,A})}_{\lambda\lambda_1\lambda_3\lambda_2}
({\bf q},{\bf q}_1;-{\bf q}_3,-{\bf q}_2).
\]
The whole contribution associated with the third term in
(\ref{eq:6w}) can be resulted in previous one by a simple
replacement $q_2 \rightleftharpoons q_3$.

Taking into account the aforesaid we write out the final expression for
`spontaneous' part of collision terms caused by the processes of 
soft-quark\,--\,soft-(anti)quark elastic scattering, which should be added 
to the right-hand side of Eq.\,(\ref{eq:5c})
\begin{equation}
\sum\limits_{f_1\!,\,f_2\!,\,f_3=\pm}\Biggl\{
\int\!\!d{\cal T}_{qq\rightarrow qq}^{(ff_1;\,f_2f_3)}
\, {\it w}_{qq\rightarrow qq}^{(ff_1;\,f_2f_3)}
({\bf q}, {\bf q}_1; {\bf q}_2, {\bf q}_3)
\Bigl(1-n_{{\bf q}_1}^{(f_1)}\Bigr)
n_{{\bf q}_2}^{(f_2)}n_{{\bf q}_3}^{(f_3)}
\label{eq:6i}
\end{equation}
\[
\hspace{1.5cm}
+\,2\!\int\!\!d{\cal T}_{q\bar{q}\rightarrow q\bar{q}}^{(ff_1;\,f_2f_3)} \,
{\it w}_{q\bar{q}\rightarrow q\bar{q}}^{(ff_1;\,f_2f_3)}
({\bf q}, {\bf q}_1; {\bf q}_2, {\bf q}_3)
\Bigl(1-\bar{n}_{{\bf q}_1}^{(f_1)}\Bigr)
\bar{n}_{{\bf q}_2}^{(f_2)}n_{{\bf q}_3}^{(f_3)}\!\Biggr\}.
\]

\section{\bf Elastic scattering of soft-gluon excitations off
soft-quark excitations}
\setcounter{equation}{0}

Here, we consider derivation of the kinetic equation describing a
change of soft gluon number densities $N_{\bf k}^{t,\,l}$ caused by their
interaction with soft-quark excitations.
In this case the initial equation is \cite{markov_PRD}
\begin{equation}
\frac{\partial}{\partial k_{\lambda}}[\,k^{2}g^{\mu \nu} - (1 + \xi^{-1})
k^{\mu}k^{\nu} + \delta \Pi^{{\rm H}\,\mu\nu}(k)]
\frac{\partial I_{\mu\nu}^{ab}(k,x)}{\partial x^{\lambda}} =
\label{eq:7q}
\end{equation}
\[
=-\,\frac{ig}{2}\,\!\int\!dk^{\prime} dk_1 dk_2\Bigl\{
(T^b)^{cd} \,^{\ast}\Gamma^{\mu\nu\lambda}(k,-k_1,-k_2)
\langle A^{\ast a}_{\mu}(k^{\prime})A^{c}_{\nu}(k_1)
A^{d}_{\lambda}(k_2)\rangle\,\delta(k - k_1 - k_2)
\]
\[
+\,(T^a)^{cd} (\,^{\ast}\Gamma^{\mu\nu\lambda}(k^{\prime},-k_1,-k_2))^{\ast}
\langle A^{b}_{\mu}(k)A^{\ast c}_{\nu}(k_1)
A^{\ast d}_{\lambda}(k_2)\rangle\,\delta(k^{\prime} - k_1 - k_2)
\Bigr\}
\]
\[
+\,ig\!\int\!dk^{\prime} dq_1 dq_2\,\Bigl\{
(t^b)^{ij}\,^{\ast}\Gamma^{(G)\mu}_{\alpha\beta}(k;q_1,-q_2)\,
\langle A^{\ast a}_{\mu}(k^{\prime})\bar{\psi}^i_{\alpha}(-q_1)
\psi^j_{\beta}(q_2)\rangle \,\delta(k + q_1 - q_2)
\]
\[
-\,
(t^a)^{ij}\,(\gamma^0[\,^{\ast}\Gamma^{(G)\mu}(k^{\prime};q_1,-q_2)]^{\dagger}
\gamma^0)_{\alpha\beta}\,
\langle A^{b}_{\mu}(k)\bar{\psi}^i_{\alpha}(-q_2)
\psi^j_{\beta}(q_1)\rangle \,\delta(k^{\prime} + q_1 - q_2)
\Bigr\}
\]
\[
+\,ig^2\!\int\!dk^{\prime} dk_1 dq_1 dq_2\,\Bigl\{
\delta{\Gamma}^{(G)bc,\,ij}_{\mu\nu,\,\alpha\beta}(k,-k_1;q_1,-q_2)\,
\langle A^{\ast a\mu}(k^{\prime})A^{c\nu}(k_1)
\bar{\psi}^i_{\alpha}(-q_1)\psi^j_{\beta}(q_2)\rangle
\]
\[
\times\,\delta(k + q_1 -k_1 - q_2)
\]
\[
-\,
(\gamma^0[\delta{\Gamma}^{(G)ac}_{\mu\nu}
(k^{\prime},-k_1;q_1,-q_2)]^{\dagger}\gamma^0)^{ij}_{\alpha\beta}\,
\langle A^{b\mu}(k)A^{\ast c\nu}(k_1)
\bar{\psi}^i_{\alpha}(-q_2)\psi^j_{\beta}(q_1)\rangle
\,\delta(k^{\prime} + q_1 -k_1 - q_2)\Bigr\}.
\]
In order to obtain the desired equation we must take into account only
the substitutions
that lead to the correlation function depending on two pairs of free-quark
fields $\psi^{(0)}\bar{\psi}^{(0)},\,\psi^{(0)}\bar{\psi}^{(0)}$ and two
free-gluon fields $A^{(0)},\,A^{\ast(0)}$. We consider the first
term on the right-hand side of Eq.\,(\ref{eq:7q}) containing the third-order
correlator of the interacting fields
\begin{equation}
-\,\frac{ig}{2}\,(T^b)^{cd}\!\int\!dk^{\prime} dk_1 dk_2
\,^{\ast}\Gamma^{\mu\nu\lambda}(k,-k_1,-k_2)
\langle A^{\ast a}_{\mu}(k^{\prime})A^{c}_{\nu}(k_1)
A^{d}_{\lambda}(k_2)\rangle\,\delta(k - k_1 - k_2).
\label{eq:7w}
\end{equation}
The replacement resulting in relevant correlation function has the following
form:
\[
A^{\ast a\mu}(k^{\prime})\rightarrow
-\,(\,^{\ast}{\cal D}^{\mu\mu^{\prime}}(k^{\prime}))^{\ast}
\tilde{j}^{\,\ast\Psi(1,2)a}_{\mu^{\prime}}(A^{\ast(0)},\bar{\psi}^{(0)},
\psi^{(0)})(k^{\prime}),
\]
\[
A^{c\nu}(k_1)\rightarrow
-\,^{\ast}{\cal D}^{\nu\nu^{\prime}}(k_1)
\tilde{j}^{\,\Psi(0,2)a}_{\nu^{\prime}}(\bar{\psi}^{(0)},\psi^{(0)})(k_1),
\quad
A^{d\lambda}(k_2)\rightarrow A^{(0)d\lambda}(k_2).
\]
Here and henceforth, we consider only the replacements that contain gluon
propagators $\,^{\ast}{\cal D}^{\mu\mu^{\prime}}(k)$ and
$\,^{\ast}{\cal D}^{\mu\mu^{\prime}}(k^{\prime})$. These propagators give
later on the factors proportional to
\[
\delta({\rm Re}\,^{\ast}\!\Delta^{\!-1\,t,\,l}(k^0,{\bf k}))=
\frac{1}{2\omega_{\bf k}^{t,\,l}}\,{\rm Z}_{t,\,l}({\bf k})\,[
\delta(\omega-\omega_{\bf k}^{t,\,l}) +
\delta(\omega+\omega_{\bf k}^{t,\,l})].
\]
The factors take into account an existence of soft transverse gluon modes and
plasmons with wave vector ${\bf k}$ and energy $\omega_{\bf k}^{t,\,l}$
(see remark after Eq.\,(\ref{eq:5q})).

For this substitution the following product of correlation functions of free
fields arises:
\[
\langle\bar{\psi}^{(0)i_1}_{\alpha_1}(-q_1)
\psi^{(0)i_2}_{\alpha_2}(q_2)\bar{\psi}^{(0)i_3}_{\alpha_3}(-q_3)
\psi^{(0)i_4}_{\alpha_4}(q_4)\rangle
\langle A^{\ast(0)a\mu}(k_1)A^{(0)b\nu}(k_2)\rangle
\]
\[
\simeq\Bigl\{
\delta^{i_1i_2}\Upsilon_{\alpha_2\alpha_1}(q_2)\delta(q_2-q_1)
\delta^{i_3i_4}\Upsilon_{\alpha_4\alpha_3}(q_4)\delta(q_4-q_3)
\]
\[
-\,\delta^{i_1i_4}\Upsilon_{\alpha_4\alpha_1}(q_4)\delta(q_4-q_1)
\delta^{i_3i_2}\Upsilon_{\alpha_2\alpha_3}(q_2)\delta(q_2-q_3)\Bigr\}\,
\delta^{ab}I^{\mu\nu}(k_1)\delta(k_1-k_2).
\]
On the right-hand side of the last expression only the second term in braces
results in collision terms with required conservation laws of energy and
momentum, Eq.\,(\ref{eq:2o}). Performing the calculations similar to those of
sections 5 and 6, we obtain instead of (\ref{eq:7w})
\begin{equation}
-\frac{ig^4}{2}\,[t^{b},t^{c}]^{ij}\,^{\ast}{\cal D}^{\mu\mu^{\prime}}\!(-k)\!
\!\int\!\!dk_1dq_1dq_2
\,^\ast\Gamma_{\mu\nu\lambda}(k,-k+k_1,-k_1)
\,^{\ast}{\cal D}^{\nu\nu^{\prime}}\!(k-k_1)
\label{eq:7e}
\end{equation}
\[
\times
\,^{\ast}\Gamma^{(G)}_{\nu^{\prime},\,\alpha^{\prime}\beta^{\prime}}
(k-k_1;q_1,-q_2)
\,^{\ast}\tilde{\bar{\Gamma}}^{(G)ac,\,ij}_{\mu^{\prime}\lambda^{\prime},
\,\alpha\beta}(k,-k_1;q_1,-q_2)
\]
\[
\times\,
I^{\lambda^{\prime}\lambda}(k_1)
\Upsilon_{\alpha\alpha^{\prime}}(q_1)
\Upsilon_{\beta^{\prime}\beta}(q_2)\,\delta(k+q_1-k_1-q_2).
\]

For term (\ref{eq:7w}) there exists another relevant substitution
\[
A^{\ast a\mu}(k^{\prime})\rightarrow
-\,(\,^{\ast}{\cal D}^{\mu\mu^{\prime}}(k^{\prime}))^{\ast}
\tilde{j}^{\,\ast\Psi(1,2)a}_{\mu^{\prime}}(A^{\ast(0)},\bar{\psi}^{(0)},
\psi^{(0)})(k^{\prime}),\quad
A^{c\nu}(k_1)\rightarrow A^{(0)c\nu}(k_1),
\]
\[
A^{d\lambda}(k_2)\rightarrow
-\,^{\ast}{\cal D}^{\lambda\lambda^{\prime}}(k_2)
\tilde{j}^{\,\Psi(0,2)d}_{\lambda^{\prime}}(\bar{\psi}^{(0)},\psi^{(0)})(k_2).
\]
The expression obtained for this replacement by a simple change of integration
variable can be resulted in (\ref{eq:7e}).

Furthermore, we consider the third term on the right-hand side of basic
equation (\ref{eq:7q}) also containing the third order correlation function of
interacting fields
\begin{equation}
ig\,(t^b)^{ij}\!\!\int\!dk^{\prime} dq_1 dq_2
\,^{\ast}\Gamma^{(G)\mu}_{\alpha\beta}(k;q_1,-q_2)\,
\langle A^{\ast a}_{\mu}(k^{\prime})\bar{\psi}^i_{\alpha}(-q_1)
\psi^j_{\beta}(q_2)\rangle \,\delta(k + q_1 - q_2).
\label{eq:7r}
\end{equation}
For this term there exist two substitutions leading to sixth-order correlation
function of free fields with conservation laws (\ref{eq:2o}):
\[
A^{\ast a\mu}(k^{\prime})\rightarrow
-\,(\,^{\ast}{\cal D}^{\mu\mu^{\prime}}(k^{\prime}))^{\ast}
\tilde{j}^{\,\ast\Psi(1,2)a}_{\mu^{\prime}}(A^{\ast(0)},\bar{\psi}^{(0)},
\psi^{(0)})(k^{\prime}),\quad
\bar{\psi}^i_{\alpha}(-q_1)\rightarrow\bar{\psi}^{(0)i}_{\alpha}(-q_1),
\]
\[
\psi^{j}_{\beta}(q_2)\rightarrow
-\,^{\ast}\!S_{\beta\beta^{\prime}}(q_2)
\tilde{\eta}^{(1,1)j}_{\beta^{\prime}}
(A^{(0)},\psi^{(0)})(q_2),
\]
and
\[
A^{\ast a\mu}(k^{\prime})\rightarrow
-\,(\,^{\ast}{\cal D}^{\mu\mu^{\prime}}(k^{\prime}))^{\ast}
\tilde{j}^{\,\ast\Psi(1,2)a}_{\mu^{\prime}}(A^{\ast(0)},\bar{\psi}^{(0)},
\psi^{(0)})(k^{\prime}),
\]
\[
\bar{\psi}^i_{\alpha}(-q_1)\rightarrow
\tilde{\bar{\eta}}^{(1,1)i}_{\alpha^{\prime}}
(A^{\ast(0)},\bar{\psi}^{(0)})\,^{\ast}\!S_{\alpha^{\prime}\alpha}(-q_1),
\quad
\psi^j_{\beta}(q_2)\rightarrow\psi^{(0)j}_{\beta}(q_2).
\]
These replacements after simple algebraic transformations lead term
(\ref{eq:7r}) to the following expression:
\begin{equation}
-ig^4\,^{\ast}{\cal D}^{\mu\mu^{\prime}}(-k)\!
\!\int\!\!dk_1dq_1dq_2\,
\label{eq:7t}
\end{equation}
\[
\times\Bigl\{(t^{b}t^{c})^{ij}
\,^{\ast}\Gamma^{(G)}_{\mu,\,\alpha^{\prime}\gamma}(k;q_1,-k-q_1)
\,^{\ast}S_{\gamma\gamma^{\prime}}(k+q_1)
\,^{\ast}\Gamma^{(Q)}_{\lambda,\,\gamma^{\prime}\beta^{\prime}}
(k_1;q_2,-k_1-q_2)
\hspace{0.7cm}
\]
\[
+\,(t^{c}t^{b})^{ij}
\,^{\ast}\Gamma^{(Q)}_{\lambda,\,\alpha^{\prime}\gamma}(k_1;-q_1,q_1-k_1)
\,^{\ast}S_{\gamma\gamma^{\prime}}(k-q_2)
\,^{\ast}\Gamma^{(G)}_{\mu,\,\gamma^{\prime}\beta^{\prime}}(k;-k+q_2,-q_2)
\Bigr\}
\]
\[
\times\,\!\,^{\ast}
\tilde{\bar{\Gamma}}^{(G)ac,\,ij}_{\mu^{\prime}\lambda^{\prime},\,\alpha\beta}
(k,-k_1;q_1,-q_2)I^{\lambda^{\prime}\lambda}(k_1)
\Upsilon_{\alpha\alpha^{\prime}}(q_1)
\Upsilon_{\beta^{\prime}\beta}(q_2)\,
\delta(k+q_1-k_1-q_2).
\]

Finally, we will consider the terms containing fourth-order correlation
functions on the right-hand side of Eq.\,(\ref{eq:7q}). Here, there exists
a unique relevant substitution
\[
A^{\ast a\mu}(k^{\prime})\rightarrow
-\,(\,^{\ast}{\cal D}^{\mu\mu^{\prime}}(k^{\prime}))^{\ast}
\tilde{j}^{\,\ast\Psi(1,2)a}_{\mu^{\prime}}(A^{\ast(0)},\bar{\psi}^{(0)},
\psi^{(0)})(k^{\prime}),\quad
A^{c\nu}(k_1)\rightarrow A^{(0)c\nu}(k_1),
\]
\[
\bar{\psi}^i_{\alpha}(-q_1)\rightarrow\bar{\psi}^{(0)i}_{\alpha}(-q_1),
\quad
\psi^j_{\beta}(q_2)\rightarrow\psi^{(0)j}_{\beta}(q_2).
\]
This replacement results the first term containing the fourth-order correlation
function of interacting fields in the following form:
\begin{equation}
-ig^4(\,^{\ast}{\cal D}^{\mu\mu^{\prime}}\!(k))^{\ast}
\!\int\!dk_1dq_1dq_2\,
\delta{\Gamma}^{(G)bc,\,ij}_{\mu\lambda,\,\alpha^{\prime}\beta^{\prime}}
(k,-k_1;q_1,-q_2)\,\!\,^{\ast}
\tilde{\bar{\Gamma}}^{(G)ac,\,ij}_{\mu^{\prime}\lambda^{\prime},\,\alpha\beta}
(k,-k_1;q_1,-q_2)
\label{eq:7y}
\end{equation}
\[
\times\,
I^{\lambda^{\prime}\lambda}(k_1)
\Upsilon_{\alpha\alpha^{\prime}}(q_1)
\Upsilon_{\beta^{\prime}\beta}(q_2)\,\delta(k+q_1-k_1-q_2).
\]
Now we add together expressions obtained (\ref{eq:7e}), (\ref{eq:7t}) and
(\ref{eq:7y}). Taking into account the definition of effective vertex
(\ref{eq:4i}), we find that the sum of the first, third and fifth terms on 
the right-hand side of Eq.\,(\ref{eq:7q}) can be
presented in the form similar to Eq.\,(\ref{eq:5y})
\begin{equation}
-ig^4\,^{\ast}{\cal D}^{\mu\mu^{\prime}}\!(-k)
\!\int\!dk_1dq_1dq_2\,
\!\,^{\ast}\tilde{\bar{\Gamma}}^{(G)ac,\,ij}_{\mu^{\prime}\lambda^{\prime},
\,\alpha\beta}(k,-k_1;q_1,-q_2)\,
\!\,^{\ast}\tilde{\Gamma}^{(G)bc,\,ij}_{\mu\lambda,
\,\alpha^{\prime}\beta^{\prime}}
(k,-k_1;q_1,-q_2)
\label{eq:7u}
\end{equation}
\[
\times\,
I^{\lambda^{\prime}\lambda}(k_1)
\Upsilon_{\alpha\alpha^{\prime}}(q_1)
\Upsilon_{\beta^{\prime}\beta}(q_2)\,\delta(k+q_1-k_1-q_2).
\]
Further, we transform expression (\ref{eq:7u}) within the scheme
suggested in section 5. Let us obtain at first the probability of
plasmon-plasmino elastic scattering. For this purpose at the
beginning we extract a purely plasmon part from gluon spectral
density $I^{\lambda^{\prime}\lambda}(k_1)$ setting
\[
I^{\lambda^{\prime}\lambda}(k_1)\rightarrow
\frac{\bar{u}^{\lambda^{\prime}}(k_1)\bar{u}^{\lambda}(k_1)}
{\bar{u}^2(k_1)}\,\Bigl\{
I_{{\bf k}_1}^l \delta (k^0_1 - \omega_{{\bf k}_1}^l)
+ I_{-{\bf k}_1}^l \delta (k^0_1 + \omega_{{\bf k}_1}^l)\Bigr\},
\]
\[
(\,^{\ast}{\cal D}^{\mu\mu^{\prime}}\!(k))^{\ast}\rightarrow
-\,\frac{\bar{u}^{\mu}(k)\bar{u}^{\mu^{\prime}}(k)}
{\bar{u}^2(k)}\,(\,^{\ast}\!\Delta^l(k))^{\ast}.
\]
By virtue of $\delta$-functions we can perform integration with respect to
$dk^0_1$.
In term containing $I^l_{-{\bf k}_1}$ we make a replacement of integration
variable: ${\bf k}_1 \rightarrow - {\bf k}_1$
($\omega_{{\bf k}_1}^l\rightarrow \omega_{{\bf k}_1}^l$).
Further, we extract the plasmino part of the functions
$\Upsilon_{\alpha\alpha^{\prime}}(q_1)$ and
$\Upsilon_{\beta^{\prime}\beta}(q_2)$
\[
\Upsilon_{\alpha\alpha^{\prime}}(q_1)\rightarrow
(h_{-}({\hat{\bf q}}_1))_{\alpha\alpha^{\prime}}
\hat{\Upsilon}_{{\bf q}_1}^{-}\delta(q_1^0-\omega^{-}_{{\bf q}_1}) +
(h_{+}({\hat{\bf q}}_1))_{\alpha\alpha^{\prime}}
\hat{\bar{\Upsilon}}_{-{\bf q}_1}^{-}\delta(q_1^0+\omega^{-}_{{\bf q}_1}),
\]
\[
\hspace{0.07cm}
\Upsilon_{\beta^{\prime}\beta}(q_2)\rightarrow
(h_{-}({\hat{\bf q}}_2))_{\beta^{\prime}\beta}
\Upsilon_{{\bf q}_2}^{-}\delta(q_2^0-\omega^{-}_{{\bf q}_2}) +
(h_{+}({\hat{\bf q}}_2))_{\beta^{\prime}\beta}
\bar{\Upsilon}_{-{\bf q}_2}^{-}\delta(q_2^0+\omega^{-}_{{\bf q}_2}).
\]
In the product
$\Upsilon_{\alpha\alpha^{\prime}}(q_1)\Upsilon_{\beta^{\prime}\beta}(q_2)$
for collision term containing conservation law $\delta(k+q_1-k_1-q_2)$
(Eq.\,(\ref{eq:7u})) it is necessary to keep only two terms of a `direct'
production, i.e.,
\begin{equation}
\Upsilon_{\alpha\alpha^{\prime}}(q_1)\Upsilon_{\beta^{\prime}\beta}(q_2)
\rightarrow
(h_{-}({\hat{\bf q}}_1))_{\alpha\alpha^{\prime}}
\hat{\Upsilon}_{{\bf q}_1}^{-}\delta(q_1^0-\omega^{-}_{{\bf q}_1})
(h_{-}({\hat{\bf q}}_2))_{\beta^{\prime}\beta}
\Upsilon_{{\bf q}_2}^{-}\delta(q_2^0-\omega^{-}_{{\bf q}_2})
\label{eq:7i}
\end{equation}
\[
\hspace{2.8cm}
+\,
(h_{+}({\hat{\bf q}}_1))_{\alpha\alpha^{\prime}}
\hat{\bar{\Upsilon}}_{-{\bf q}_1}^{-}\delta(q_1^0+\omega^{-}_{{\bf q}_1})
(h_{+}({\hat{\bf q}}_2))_{\beta^{\prime}\beta}
\bar{\Upsilon}_{-{\bf q}_2}^{-}\delta(q_2^0+\omega^{-}_{{\bf q}_2})
\]
and in the term containing conservation law $\delta(k+q_1+k_1-q_2)$ it is
necessary to keep only `crossed' term
\begin{equation}
\Upsilon_{\alpha\alpha^{\prime}}(q_1)\Upsilon_{\beta^{\prime}\beta}(q_2)
\rightarrow
(h_{+}({\hat{\bf q}}_1))_{\alpha\alpha^{\prime}}
\hat{\bar{\Upsilon}}_{-{\bf q}_1}^{-}
\delta(q_1^0+\omega^{-}_{{\bf q}_1})
(h_{-}({\hat{\bf q}}_2))_{\beta^{\prime}\beta}
\Upsilon_{{\bf q}_2}^{-}\delta(q_2^0-\omega^{-}_{{\bf q}_2}).
\label{eq:7o}
\end{equation}
The first term on the right-hand side of Eq.\,(\ref{eq:7i}) is associated with
the scattering process of plasmon off plasmino and the second one is associated
with the scattering of plasmon off antiplasmino.  Finally, contribution
(\ref{eq:7o}) defines the process of creation of plasmino-antiplasmino pair
by two plasmon fusion.

Let us consider for the sake of definiteness the first term on the right-hand
side of Eq.\,(\ref{eq:7i}). We substitute this term into
Eq.\,(\ref{eq:7u}) and perform integration in $dq^0_1dq^0_1$.
For spinor projectors $h_{-}({\hat{\bf q}}_1)$ and
$h_{-}({\hat{\bf q}}_2)$ we use representation (\ref{eq:5o}). The
spectral densities $I_{{\bf k}_1}^l,\,\hat{\Upsilon}^{-}_{{\bf q}_1}$, and
$\Upsilon^{-}_{{\bf q}_2}$ are replaced by plasmon
and (anti)plasmino number densities according to rules
(\ref{eq:5a}) and (\ref{eq:6ww}). The remaining conjugate terms on the
right-hand side of Eq.\,(\ref{eq:7q}) lead to the expression
similar to (\ref{eq:7u}) with a unique replacement
$(\,^{\ast}{\cal D}^{\mu \mu^{\prime}}(k))^{\ast} \rightarrow \,^{\ast}
{\cal D}^{\mu \mu^{\prime}}(k)$. Thus in the case of plasmon-plasmino
interaction for a weak-absorption medium when
${\rm Im} ^{\ast}\Delta^{-1\,l}(k) \rightarrow 0$, this results in simple
replacement $(\,^{\ast}\Delta^l(k))^{\ast}$ by
\[
(\,^{\ast}\!\Delta^l(k))^{\ast}-\,^{\ast}\!\Delta^l(k)=
-2i\,{\rm Im}\,^{\ast}\!\Delta^l(k)\simeq2\pi i\,{\rm sign}(k^0)
\left(\frac{{\rm Z}_l({\bf k})}{2\omega_{{\bf k}}^l}\right)
\delta (k^0 - \omega_{{\bf k}}^l).
\]
Here, on the right-hand side we keep only positively-frequency part.

Taking into account above-mentioned, we introduce the following matrix element
of plasmon-plasmino scattering:
\begin{equation}
{\rm M}^{aa_1\!,\,i_1i_2}_{\lambda\lambda_1}({\bf k},-{\bf k}_1;{\bf q}_1,-{\bf q}_2)
\equiv g^2
\left(\frac{{\rm Z}_l({\bf k})}{2\omega_{{\bf k}}^l}\right)^{\!1/2}\!
\left(\frac{{\rm Z}_l({\bf k}_1)}{2\omega_{{\bf k}_1}^l}\right)^{\!1/2}\!
\left(\frac{{\rm Z}_{-}({\bf q}_1)}{2}\right)^{\!1/2}\!
\left(\frac{{\rm Z}_{-}({\bf q}_2)}{2}\right)^{\!1/2}\!
\label{eq:7p}
\end{equation}
\[
\times\,
\Biggl(\frac{\bar{u}^{\mu}(k)}{\sqrt{\bar{u}^2(k)}}\Biggr)
\Biggl(\frac{\bar{u}^{\mu_1}(k_1)}{\sqrt{\bar{u}^1(k_1)}}\Biggr)
\Bigl[\,\bar{v}_{\beta}(\hat{\bf q}_2,\lambda_2)\,
\!\,^{\ast}\tilde{\bar{\Gamma}}^{(G)aa_1\!,\,i_1i_2}_{\mu\mu_1,\,\alpha\beta}\!
(k,-k_1;q_1,-q_2) v_{\alpha}(\hat{\bf q}_1,\lambda_1)\Bigr]_{\,{\rm on-shell}}.
\]
The probability of Compton-like plasmon-plasmino scattering in this case
is defined as
\begin{equation}
\delta^{ab} {\it w}_{gq\rightarrow gq}^{\!(l-;\,l-)}
({\bf k}, {\bf q}_1; {\bf k}_1, {\bf q}_2)=\!\!\!
\sum\limits_{\lambda_1\!,\,\lambda_2=\pm}\!\!
{\rm M}^{aa_1\!,\,i_1i_2}_{\lambda_1\lambda_2}
({\bf k},-{\bf k}_1;{\bf q}_1,-{\bf q}_2)
({\rm M}^{ba_1\!,\,i_1i_2}_{\lambda_1\lambda_2}
({\bf k},-{\bf k}_1;{\bf q}_1,-{\bf q}_2))^{\ast}.
\label{eq:7a}
\end{equation}
The different terms of matrix element (\ref{eq:7p}) have a similar diagrammatic
interpretation as is depicted in Fig.\,\ref{fig1} with relevant replacements
of momenta and HTL-resummed vertices only. The probability for the elastic
scattering of soft transverse gluon excitation off soft normal quark
excitation is defined by the following expression:
\begin{equation}
\delta^{ab}{\it w}_{gq\rightarrow gq}^{\!(t+;\,t+)}
({\bf k}, {\bf q}_1; {\bf k}_1, {\bf q}_2)\!=\!\!\!\!\!
\sum\limits_{\xi,\,\xi_1=1,2}
\sum\limits_{\lambda_1,\,\lambda_2=\pm}\!\!\!
{\rm M}^{aa_1\!,\,i_1i_2}_{\xi\xi_1\!,\,\lambda_1\lambda_2}
({\bf k},-{\bf k}_1;{\bf q}_1,-{\bf q}_2)
({\rm M}^{ba_1\!,\,i_1i_2}_{\xi\xi_1\!,\,\lambda_1\lambda_2}
({\bf k},-{\bf k}_1;{\bf q}_1,-{\bf q}_2))^{\ast},
\label{eq:7aa}
\end{equation}
where now the matrix element is
\begin{equation}
{\rm M}^{aa_1\!,\,i_1i_2}_{\xi\xi_1\!,\,\lambda_1\lambda_2}
({\bf k},-{\bf k}_1;{\bf q}_1,-{\bf q}_2)
\equiv g^2
\left(\frac{{\rm Z}_t({\bf k})}{2\omega_{{\bf k}}^t}\right)^{\!1/2}\!
\left(\frac{{\rm Z}_t({\bf k}_1)}{2\omega_{{\bf k}_1}^t}\right)^{\!1/2}\!
\left(\frac{{\rm Z}_{+}({\bf q}_1)}{2}\right)^{\!1/2}\!
\left(\frac{{\rm Z}_{+}({\bf q}_2)}{2}\right)^{\!1/2}\!
\label{eq:7aaa}
\end{equation}
\[
\times\!
\left(\frac{\epsilon^{\,\mu}(k,\xi)}
{\sqrt{\epsilon^2(k,\xi)}}\right)\!\!
\left(\frac{\epsilon^{\,\mu_1}(k_1,\xi_1)}
{\sqrt{\epsilon^2(k_1,\xi_1)}}\right)\!
\Bigl[\,\bar{u}_{\beta}(\hat{\bf q}_2,\lambda_2)\,
\!\,^{\ast}\tilde{\bar{\Gamma}}^{(G)aa_1\!,\,i_1i_2}_{\mu\mu_1,\,\alpha\beta}
\!(k,-k_1;q_1,-q_2)
u_{\alpha}(\hat{\bf q}_1,\lambda_1)\Bigr]_{\,{\rm on-shell}}.
\]
The probabilities for other types of the elastic scattering
${\it w}_{gq\rightarrow gq}^{\!(t-;\,t-)}$,
${\it w}_{gq\rightarrow gq}^{\!(l+;\,l+)}$
are obtained from previous by the appropriate replacements of gluon and quark
wave functions and choice of mass-shell conditions on the
right-hand sides of Eqs.\,(\ref{eq:7p}) and (\ref{eq:7aaa}).

As in section 5 we regroup the terms in the effective amplitudes
$\!\,^{\ast}\tilde{\bar{\Gamma}}^{(G)aa_1,\,i_1 i_2}_{\mu \mu_1,
\,\alpha\beta}$ and
$\!\,^{\ast}\tilde{\Gamma}^{(G)aa_1,\,i_1 i_2}_{\mu \mu_1,\,\alpha\beta}$
according to transition (\ref{eq:5f}). We present the HTL-induced vertex
$\delta \Gamma^{(G)a a_1}_{\mu\mu_1}$ in the form similar to (\ref{eq:5g}),
where now by virtue of initial definition
(A.1) instead of $\delta \Gamma^{(Q;\,{\cal S,\,A})}_{\mu\mu_1}$ we imply
\[
\delta\Gamma^{(G;\,{\cal S,\,A})}_{\mu\mu_1}(k,-k_1;q_1,-q_2) =
\omega_0^2\!\int\!\frac{{\rm d} \Omega}{4 \pi}\,
\frac{v_{\mu}v_{\mu_1} \not\!v}{(v\cdot q_1 - i \epsilon)
(v\cdot q_2 + i \epsilon)}
\hspace{3.5cm}
\]
\[
\hspace{5.5cm}
\times
\bigg(\frac{1}{v\cdot (q_1+k) + i \epsilon}
\pm \frac{1}{v\cdot (q_1-k_1) - i \epsilon} \bigg).
\]
Taking into account this expression, we present matrix element (\ref{eq:7p})
in the following form:
\begin{equation}
{\rm M}^{aa_1\!,\,i_1i_2}_{\lambda_1\lambda_2}
({\bf k},-{\bf k}_1;{\bf q}_1,-{\bf q}_2)=
\frac{1}{2}\,\{t^{a_1},t^{a}\}^{i_2i_1}
\breve{\cal C}_{\cal S}^{-1\,}
{\rm M}^{({\cal S})}_{\lambda_1\lambda_2}
({\bf k},-{\bf k}_1;{\bf q}_1,-{\bf q}_2)
\label{eq:7s}
\end{equation}
\[
\hspace{3.65cm}
+\,\frac{1}{2}\,[\,t^{a_1},t^{a}]^{i_2i_1}
\,\breve{\cal C}_{\cal A}^{-1}
{\rm M}^{({\cal A})}_{\lambda_1\lambda_2}
({\bf k},-{\bf k}_1;{\bf q}_1,-{\bf q}_2),
\]
where
\[
\breve{\cal C}_{\cal S,\,A}\equiv
\left\{\frac{1}{2}\,T_F\biggl(C_F\mp\frac{1}{2N_c}\biggr)\right\}^{\!1/2},
\]
\begin{equation}
{\rm M}^{({\cal S,\,A})}_{\lambda_1\lambda_2}({\bf k},-{\bf k}_1;{\bf q}_1,-{\bf q}_2)
\equiv g^2
\,\breve{\cal C}_{\cal S,\,A}\!
\left(\frac{{\rm Z}_l({\bf k})}{2\omega_{{\bf k}}^l}\right)^{\!1/2}\!\!
\left(\frac{{\rm Z}_l({\bf k}_1)}{2\omega_{{\bf k}_1}^l}\right)^{\!1/2}\!\!
\left(\frac{{\rm Z}_{-}({\bf q}_1)}{2}\right)^{\!1/2}\!\!
\left(\frac{{\rm Z}_{-}({\bf q}_2)}{2}\right)^{\!1/2}\!\!
\label{eq:7d}
\end{equation}
\[
\times\,
\Biggl(\frac{\bar{u}^{\mu}(k)}{\sqrt{\bar{u}^2(k)}}\Biggr)
\Biggl(\frac{\bar{u}^{\mu_1}(k_1)}{\sqrt{\bar{u}^2(k_1)}}\Biggr)
\Bigl[\,\bar{v}(\hat{\bf q}_2,\lambda_2)
{\cal M}^{({\cal S,\,A})}_{\mu\mu_1}({\bf k},-{\bf k}_1;{\bf q}_1,-{\bf q}_2)
v(\hat{\bf q}_1,\lambda_1)\Bigr]_{\,{\rm on-shell}}.
\]
\noindent The functions ${\cal M}^{({\cal S,\,A})}_{\mu\mu_1}$ are defined by
the following expressions:
\[
{\cal M}^{({\cal S})}_{\mu\mu_1}({\bf k},-{\bf k}_1;{\bf q}_1,-{\bf q}_2)
\equiv
\delta\Gamma^{(G;\,{\cal S})}_{\mu\mu_1}(-k,k_1;-q_1,q_2)
\]
\[
-\,^{\ast}\Gamma^{(Q)}_{\mu_1}(-k_1;-q_2,k_1+q_2)
\,^{\ast}\!S(-k-q_1)
\,^{\ast}\Gamma^{(G)}_{\mu}(-k;k+q_1,-q_1)
\]
\[
+\,^{\ast}\Gamma^{(G)}_{\mu}(-k;k-q_2,q_2)
\,^{\ast}\!S(-k_1+q_1)
\,^{\ast}\Gamma^{(Q)}_{\mu_1}(-k_1;q_1,-q_1+k_1)
\]
and
\[
{\cal M}^{({\cal A})}_{\mu\mu_1}({\bf k},-{\bf k}_1;{\bf q}_1,-{\bf q}_2)
\equiv
\delta\Gamma^{(G;\,{\cal A})}_{\mu\mu_1}(-k,k_1;-q_1,q_2)
\]
\[
+\,2\,^{\ast}\Gamma^{(G)}_{\nu}(q_1-q_2;-q_1,q_2)
\,^{\ast}{\cal D}^{\nu\nu^{\prime}}(-k+k_1)
\,^\ast\Gamma_{\mu\mu_1\nu^{\prime}}(-k,k_1,k-k_1)
\]
\[
-\,^{\ast}\Gamma^{(Q)}_{\mu_1}(-k_1;-q_2,k_1+q_2)
\,^{\ast}\!S(-k-q_1)
\,^{\ast}\Gamma^{(G)}_{\mu}(-k;k+q_1,-q_1)
\]
\[
-\,^{\ast}\Gamma^{(G)}_{\mu}(-k;k-q_2,q_2)
\,^{\ast}\!S(-k_1+q_1)
\,^{\ast}\Gamma^{(Q)}_{\mu_1}(-k_1;q_1,-q_1+k_1).
\]
Here, $T_F$ is index of the fundamental representation.

By virtue of representation (\ref{eq:7s}) probability (\ref{eq:7a}) can
be introduced in the form of a sum of two independent parts similar to
(\ref{eq:5x})
\begin{equation}
{\it w}_{gq\rightarrow gq}^{\!(l-;\,l-)}
({\bf k}, {\bf q}_1; {\bf k}_1, {\bf q}_2)=
{\it w}_{gq\rightarrow gq}^{({\cal S})}
({\bf k}, {\bf q}_1; {\bf k}_1, {\bf q}_2)+
{\it w}_{gq\rightarrow gq}^{({\cal A})}
({\bf k}, {\bf q}_1; {\bf k}_1, {\bf q}_2),
\label{eq:7f}
\end{equation}
where
\[
{\it w}_{gq\rightarrow gq}^{({\cal S,\,A})}
({\bf k}, {\bf q}_1; {\bf k}_1, {\bf q}_2)=
\sum\limits_{\lambda_1\!,\,\lambda_2=\pm}\!
\Bigl|\,
{\rm M}^{({\cal S,\,A})}_{\lambda_1\lambda_2}
({\bf k},-{\bf k}_1;{\bf q}_1,-{\bf q}_2)
\Bigr|^2_{\,{\rm on-shell}}.
\]
In deriving this expression a color algebra was used
\[
{\rm tr}(\{t^{a_1},t^{a}\}\{t^{b},t^{a_1}\})=
2\delta^{ab}T_F\biggl(C_F-\frac{1}{2N_c}\biggr),
\]
\[
{\rm tr}([\,t^{a_1},t^{a}][\,t^{b},t^{a_1}])=
2\delta^{ab}T_F\biggl(C_F+\frac{1}{2N_c}\biggr),
\]
\[
{\rm tr}(\{t^{a_1},t^{a}\}[\,t^{b},t^{a_1}])=0.
\]
The decomposition (\ref{eq:7f}) is true also for probability
(\ref{eq:7aa}) and for other probabilities of elastic scattering:
${\it w}_{gq\rightarrow gq}^{\!(t-;\,t-)}$ and
${\it w}_{gq\rightarrow gq}^{\!(l+;\,l+)}$.

\begin{figure}[hbtp]
\begin{center}
\includegraphics*[height=3.65cm]{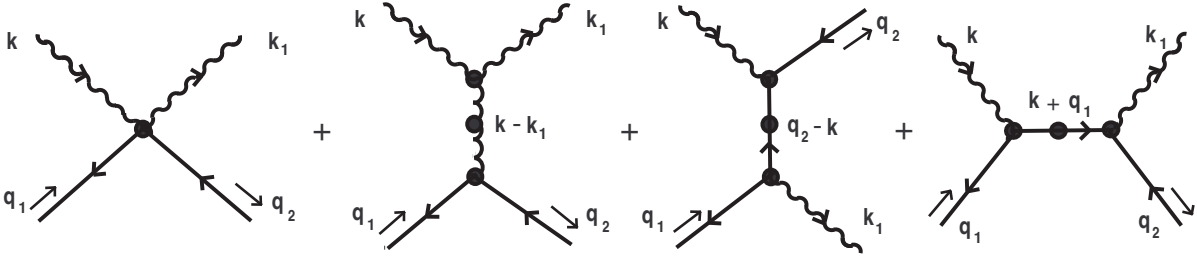}
\end{center}
\caption{\small Compton scattering of soft-gluon off soft-antiquark 
excitations.}
\label{fig4}
\end{figure}

Now we consider the second term on the right-hand side of Eq.\,(\ref{eq:7i}).
As was mentioned above this term defines the process of plasmon scattering off
antiplasmino. The reasoning similar to previous one results in scattering
probability
${\it w}_{g\bar{q}\rightarrow g\bar{q}}^{\!(l-;\,l-)}
({\bf k}, {\bf k}_1; {\bf q}_1, {\bf q}_2)$, as was defined by
Eq.\,(\ref{eq:7f}), where now
\[
{\it w}_{g\bar{q}\rightarrow g\bar{q}}^{({\cal S,\,A})}
({\bf k}, {\bf q}_1; {\bf k}_1, {\bf q}_2)=
\sum\limits_{\lambda_1\!,\,\lambda_2=\pm}\!
\Bigl|\,
\bar{\rm M}^{({\cal S,\,A})}_{\lambda_1\lambda_2}({\bf k},-{\bf k}_1;{\bf q}_1,-{\bf q}_2)
\Bigr|^2_{\,{\rm on-shell}}\,,
\]
and
\begin{equation}
\bar{\rm M}^{({\cal S,\,A})}_{\lambda_1\lambda_2}
({\bf k},-{\bf k}_1;{\bf q}_1,-{\bf q}_2)
\equiv g^2
\,\breve{\cal C}_{\cal S,\,A}\!
\left(\frac{{\rm Z}_l({\bf k})}{2\omega_{{\bf k}}^l}\right)^{\!1/2}\!\!
\left(\frac{{\rm Z}_l({\bf k}_1)}{2\omega_{{\bf k}_1}^l}\right)^{\!1/2}\!\!
\left(\frac{{\rm Z}_{-}({\bf q}_1)}{2}\right)^{\!1/2}\!\!
\left(\frac{{\rm Z}_{-}({\bf q}_2)}{2}\right)^{\!1/2}\!\!
\label{eq:7g}
\end{equation}
\[
\times\,
\Biggl(\frac{\bar{u}^{\mu}(k)}{\sqrt{\bar{u}^2(k)}}\Biggr)
\Biggl(\frac{\bar{u}^{\mu_1}(k_1)}{\sqrt{\bar{u}^2(k_1)}}\Biggr)
\Bigl[\,\bar{v}(\hat{\bf q}_1,\lambda_1)
{\cal M}^{({\cal S,\,A})}_{\mu\mu_1}
({\bf k},-{\bf k}_1;-{\bf q}_2,{\bf q}_1)
v(\hat{\bf q}_2,\lambda_2)\Bigr]_{\,{\rm on-shell}}\,.
\]
In the kinetic equation it is necessary to replace
$n^{-}_{{\bf q}_1}$, $n^{-}_{{\bf q}_2}$ by
$\bar{n}^{-}_{{\bf q}_1}$, $\bar{n}^{-}_{{\bf q}_2}$.
The integration measure here
remains invariable, as it was defined on the first line of
Eq.\,(\ref{eq:2o}). The diagrammatic interpretation of the different
terms entering into matrix element (\ref{eq:7g}) is presented in
Fig.\,\ref{fig4}.

\begin{figure}[hbtp]
\begin{center}
\includegraphics*[height=3.7cm]{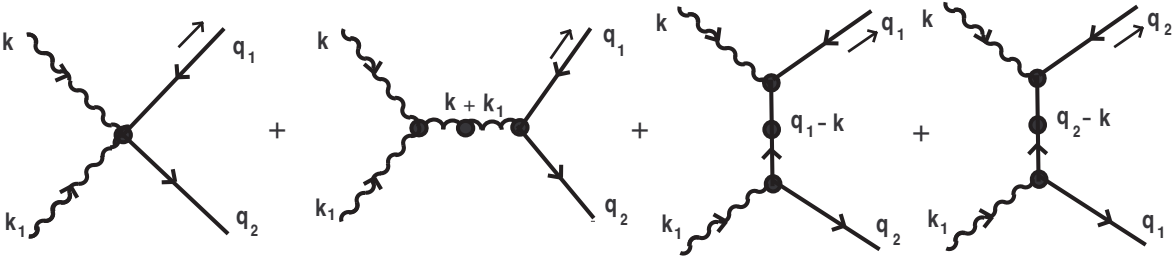}
\end{center}
\caption{\small Production of soft quark-antiquark pair.}
\label{fig5}
\end{figure}

Finally we will consider the contribution of term (\ref{eq:7o}) determining
a pair production $gg\rightarrow q\bar{q}$. The probability of this
process ${\it w}_{gg\rightarrow q\bar{q}}^{\!(l\,l;\,--)}
({\bf k}, {\bf k}_1; {\bf q}_1, {\bf q}_2)$ is defined by an expression, which
is similar to Eqs.\,(\ref{eq:7f}), (\ref{eq:7d}) with a unique
difference\footnote{In the kinetic equation we replace
$1-n^{-}_{{\bf q}_1}$ by $\bar{n}^{-}_{{\bf q}_1}$ and
$1+N_{{\bf k}_1}^{l}$ by $N_{{\bf k}_1}^{l}$. The integration
measure is defined by the second equation in (\ref{eq:2o}).} that
functions ${\cal M}^{({\cal S,\,A})}_{\mu\mu_1}
({\bf k},-{\bf k}_1;{\bf q}_1,-{\bf q}_2)$ in (\ref{eq:7d}) should
be replaced by ${\cal M}^{({\cal S,\,A})}_{\mu\mu_1}
({\bf k},{\bf k}_1;-{\bf q}_1,-{\bf q}_2)$. The diagrammatic interpretation
of the different terms entering into the matrix element of pair production
is presented in Fig.\,\ref{fig5}.

Now we return to initial equation (\ref{eq:7q}). We transform the left-hand
side similar to \cite{markov3}. As result we have
\[
-\,\delta^{ab}\,\delta(k^0-\omega_{\bf k}^{(b)})\!
\left(
\frac{\partial N_{\bf k}^{(b)}}{\partial t} +
{\bf v}_{\bf k}^{(b)}\cdot\frac{\partial N_{\bf k}^{(b)}}{\partial {\bf x}}
\right).
\]
Taking into account above-mentioned, we write out the final expression for the
kinetic equation defining the change of the colorless soft-gluon number
densities caused by `spontaneous' processes of
soft-gluon\,--\,soft-quark elastic scattering and
pair production by fusion of two soft-gluon excitations
\[
\left(\!
\frac{\partial N_{\bf k}^{(b)}}{\partial t} +
{\bf v}_{\bf k}^{(b)}\!\cdot\!\frac{\partial N_{\bf k}^{(b)}}{\partial{\bf x}}
\!\right)^{\!{\rm sp}}\!\!\!\!=
\!\!\!\sum\limits_{b_1=t,\,l}\sum\limits_{\,f_1,\,f_2=\pm}\!\Biggl\{
\int\!\!d{\cal T}_{gq\rightarrow gq}^{(bf_1;\,b_1f_2)}
{\it w}_{gq\rightarrow gq}^{(bf_1;\,b_1f_2)}
({\bf k}, {\bf q}_1; {\bf k}_1, {\bf q}_2)N_{{\bf k}_1}^{(b_1)}
\Bigl(\!1-n_{{\bf q}_1}^{(f_1)}\!\Bigr)
n_{{\bf q}_2}^{(f_1)}
\]
\[
\hspace{5.5cm}
+\!\int\!d{\cal T}_{g\bar{q}\rightarrow g\bar{q}}^{(bf_1;\,b_1f_2)}\,
{\it w}_{g\bar{q}\rightarrow g\bar{q}}^{(bf_1;\,b_1f_2)}
({\bf k}, {\bf q}_1; {\bf k}_1, {\bf q}_2)N_{{\bf k}_1}^{(b_1)}
\Bigl(1-\bar{n}_{{\bf q}_1}^{(f_1)}\Bigr)
\bar{n}_{{\bf q}_2}^{(f_2)}
\]
\vspace{0.1cm}
\[
\hspace{4.8cm}
+\!\int\!d{\cal T}_{gg\rightarrow q\bar{q}}^{(b\,b_1;\,f_1f_2)}\,
{\it w}_{gg\rightarrow q\bar{q}}^{(b\,b_1;\,f_1f_2)}
({\bf k}, {\bf k}_1; {\bf q}_1, {\bf q}_2)N_{{\bf k}_1}^{(b_1)}
\bar{n}_{{\bf q}_1}^{(f_1)}n_{{\bf q}_2}^{(f_2)}\Biggr\}.
\]
This equation follows from (\ref{eq:2e}), (\ref{eq:2i}) in the limit of
a small intensity of soft-gluon excitations
$N^{(b)}_{\bf k} \rightarrow 0$ and making use of
the condition $1 + N^{(b_1)}_{{\bf k}_1}\simeq N^{(b_1)}_{{\bf k}_1}$.

\section{\bf Generalized formula of emitted radiant power in QGP}
\setcounter{equation}{0}

In previous sections we have explicitly shown how one can define
the probabilities of the simplest processes of nonlinear
interaction between the soft collective modes Fermi and Bose
statistics in QGP. This has been done by immediate extracting all
possible contributions to matrix elements of these processes.
However the use of such a direct approach in the general case of
the processes including an arbitrary number of soft fermion and
boson excitations becomes ineffective because of awkwardness and
complexity of computations. In this case the method developed by
Tsytovich \cite{gailitis, tsytovich} in the theory of usual plasma
known as the {\it correspondence principle} is more convenient for
deriving an explicit form of the scattering probabilities. We have
already used this approach for deriving the general expression for
the probability of $2n+2$ -colorless soft-gluon decay processes,
$n=1,2, ...$ \cite{markov1}. Here, we don't set ourselves a
purpose to obtain similar general expressions for the
probabilities of the processes including an arbitrary (even)
number of soft boson and  fermion excitations. We
concentrate our attention on deriving the most general expression
for the emitted radiant power, which would take into account an
existence of fermionic sector of plasma excitations in QGP. Having
at hand this expression and following standard scheme stated in
section 4 of the paper \cite{markov1} with employing general
expressions for effective currents and sources (\ref{eq:4t}), it
is not difficult to obtain the general expressions for the
probabilities in this more complicated case. Thus the problem
stated is reduced to calculation of explicit form of the effective
amplitudes in integrand of effective currents and sources
(\ref{eq:4t}). Calculating algorithm of the effective amplitudes
was proposed in section 4 of the present paper.

For derivation of the probabilities of soft-gluon decay processes
in \cite{markov1} we have used the following formula for the emitted
radiant power:
\[
{\cal I} = \lim\limits_{\tau,\,V\rightarrow\infty}
\frac{1}{\tau V}\!\int\limits_{-\tau/2}^{\tau/2}\!\int\limits_{V}
d{\bf x}dt\,\langle {\bf E}^{a} ({\bf x},t)\cdot
{\bf j}^{\,a} ({\bf x},t) \rangle,
\]
where $V$ is a spatial volume of integration. Averaging over the time is
made for elimination of oscillating terms in ${\cal I}$. Thus the emitted
radiant power is defined as a work per time done by radiation field with color
current (in this case ${\bf j}^{\,a}({\bf x},t)\equiv{\bf j}^{Aa}({\bf x},t))$
creating this field. The chromoelectric field
${\bf E}^{a}({\bf x},t)$ is associated with ${\bf j}^{\,a}({\bf x},t)$ by the
field equation.

We propose that generalization of the above-written formula for emitted
radiant power taking into account an existence of soft fermion excitations in
the system is
\[
{\cal I}={\cal I}_{\,\cal B}+{\cal I}_{\cal F},
\]
where
\begin{equation}
{\cal I}_{\,\cal B}= \!\lim\limits_{\tau,\,V\rightarrow\infty}
\frac{1}{\tau V}\!\int\limits_{-\tau/2}^{\tau/2}\!\int\limits_{V}
\!d{\bf x}dt\,\langle {\bf E}^{a} ({\bf x},t)\cdot
{\bf j}^{Aa} ({\bf x},t) \rangle
+\!
\lim\limits_{\tau,\,V\rightarrow\infty}
\frac{1}{\tau V}\!\int\limits_{-\tau/2}^{\tau/2}\!\int\limits_{V}
\!d{\bf x}dt\,\langle {\bf E}^{a} ({\bf x},t)\cdot
{\bf j}^{\Psi a} ({\bf x},t) \rangle
\label{eq:8q}
\end{equation}
\[
=\!
\lim\limits_{\tau,\,V\rightarrow\infty}
\frac{(2\pi)^4}{\tau V}
\int\!d{\bf k}dk^0\,\langle {\bf E}^a ({\bf k},k^0)\cdot
{\bf j}^{Aa}({\bf k},k^0) \rangle +
\lim\limits_{\tau,\,V\rightarrow\infty}
\frac{(2\pi)^4}{\tau V}
\int\!d{\bf k}dk^0\,\langle {\bf E}^a ({\bf k},k^0)\cdot
{\bf j}^{\Psi a}({\bf k},k^0) \rangle
\]
and
\begin{equation}
{\cal I}_{\cal F}= -\,\frac{1}{2}\,
\!\lim\limits_{\tau,\,V\rightarrow\infty} \frac{1}{\tau
V}\!\int\limits_{-\tau/2}^{\tau/2}\!\int\limits_{V}\!d{\bf x}dt
\Bigr\{\langle (\bar{\psi}_{\alpha}({\bf
x},t)\overleftarrow{D}_0^{\dagger})^i\, \eta_{\alpha}^{\,i}
({\bf x},t)\rangle + \langle \bar{\eta}_{\alpha}^{\,i}({\bf x},t)\,
(\overrightarrow{D}_0\psi_{\alpha}({\bf x},t))^{i}\rangle\Bigl\}
\label{eq:8w}
\end{equation}
\[
= -\,\frac{i}{2}\,
\lim\limits_{\tau,\,V\rightarrow\infty}
\frac{(2\pi)^4}{\tau V}
\int\!d{\bf q}dq^0\,q^0 \Bigr\{
\langle\bar{\psi}_{\alpha}^{\,i}(-q)\eta_{\alpha}^{i}(q)\rangle -
\langle\bar{\eta}_{\alpha}^{\,i}(-q)\psi_{\alpha}^{i}(q)\rangle
\Bigl\}
\]
\[
\hspace{1.3cm} +\,\frac{ig}{2}\,
\lim\limits_{\tau,\,V\rightarrow\infty} \frac{(2\pi)^4}{\tau V}
\int\!dqdq^{\prime}dk \Bigr\{(t^a)^{ji}
\langle\bar{\psi}_{\alpha}^{\,j}(-q)A_0^a(k)
\eta_{\alpha}^{\,i}(q^{\prime})\rangle
\,\delta(q-q^{\prime}-k)
\]
\[
\hspace{6.5cm}
-\,(t^a)^{ij}
\langle\bar{\eta}_{\alpha}^{\,i}(-q^{\prime})
A_0^a(k)\psi_{\alpha}^j(q)\rangle
\,\delta(q-q^{\prime}+k)
\Bigl\}.
\]
Here, on the first line\footnote{The authors are grateful to
reviewer pointed to the necessity of entering covariant derivative
$D_0$ instead of usual time derivative $\partial/\partial t$ in
Eq.\,(\ref{eq:8w}) that is necessary for maintenance of gauge symmetry.
The terms in the emitted radiant power ${\cal I}_{\cal F}$ including
$A_0$ should be taken into account in the consideration of more
complicated scattering processes than the processes presented in
this work. For this reason in the subsequent discission these
terms will be omitted (or one can assume that we work in
$A_0$-gauge). For recovery of Lorentz symmetry of
Eq.\,(\ref{eq:8w}) instead of $D_0$ it should be used
$(u\cdot D)$, where $u$ is four-velocity of
plasma.} of Eq.\,(\ref{eq:8w}) we make use of covariant derivative
$D_0\equiv\partial/\partial t + igA_0^a({\bf x},t)t^a$. For
Fourier transformation we have used a relation
\[
\lim\limits_{\tau\rightarrow\infty}\!\int\limits_{-\tau/2}^{\tau/2}
\!{\rm e}^{i\omega t}
dt=2\pi\delta(\omega).
\]
The appearance of the second term on the right-hand side of Eq.\,(\ref{eq:8q})
containing the induced current ${\bf j}^{\Psi a}({\bf x},t)$ is rather evident.
The appearance of contribution (\ref{eq:8w}) is a principal new feature, and
therefore our further attention will be concentrated on a proof of the fact
that expression (\ref{eq:8w}) correctly reproduces the scattering probabilities
obtained in sections 5 and 6 within the framework of the generalized
correspondence principle.

According to the correspondence principle for determination of the scattering
probabilities of the processes of a type
\[
qg \rightarrow qg, \quad qq \rightarrow qq,
\quad q\bar{q} \rightarrow q\bar{q},\;\;
\mbox{etc}
\]
the expression (\ref{eq:8w}) should be compared with the
expression determining a change of energy of soft fermionic plasma
excitations caused by the spontaneous processes of soft-quark and soft-gluon
waves emission only
\begin{equation}
\left( \frac{{\rm d}{\cal E}}{{\rm d} t} \right)^{\!{\rm sp}} =
\sum\limits_{f=\pm}\frac{\rm d}{{\rm d}t}\,\Biggl\{
\int\!\frac{d{\bf q}}{(2 \pi)^3} \;
\omega^{(f)}_{\bf q} \,n^{(f)}_{\bf q}
+
\int\!\frac{d{\bf q}}{(2 \pi)^3} \;
\omega^{(f)}_{\bf q}\,\bar{n}^{(f)}_{\bf q}
\!\Biggr\}
\label{eq:8e}
\end{equation}
\[
=\!\!\!\sum\limits_{f,\,f_1=\pm}\,\sum\limits_{\,b_1,\,b_2=t,\,l}\Biggl\{
2\!\int\!\frac{d{\bf q}}{(2 \pi)^3}
\!\int\!d{\cal T}_{qg\rightarrow qg}^{(fb_1;\,f_1b_2)}
\omega^{(f)}_{\bf q}\, {\it w}_{qg\rightarrow qg}^{(fb_1;\,f_1b_2)}
({\bf q}, {\bf q}_1; {\bf k}_1, {\bf k}_2)\,n_{{\bf q}_1}^{(f_1)}
N_{{\bf k}_1}^{(b_1)}N_{{\bf k}_2}^{(b_2)}
\hspace{0.5cm}
\]
\[
\hspace{3.1cm}
+\int\!\frac{d{\bf q}}{(2 \pi)^3}
\!\int\!d{\cal T}_{q\bar{q}\rightarrow gg}^{(ff_1;\,b_1b_2)}
\omega^{(f)}_{\bf q}\,{\it w}_{q\bar{q}\rightarrow gg}^{(ff_1;\,b_1b_2)}
({\bf q}, {\bf q}_1; {\bf k}_1, {\bf k}_2)\,
\Bigl(1-\bar{n}_{{\bf q}_1}^{(f_1)}\Bigr)
N_{{\bf k}_1}^{(b_1)}N_{{\bf k}_2}^{(b_2)}
\]
\[
\hspace{1.8cm}
+\,\!\!\!\sum\limits_{\!f,\,f_1,...=\pm}
\Biggl\{\int\!\frac{d{\bf q}}{(2 \pi)^3}
\!\int\!\!d{\cal T}_{qq\rightarrow qq}^{(ff_1;\,f_2f_3)} \,
\omega^{(f)}_{\bf q}\,{\it w}_{qq\rightarrow qq}^{(ff_1;\,f_2f_3)}
({\bf q}, {\bf q}_1; {\bf q}_2, {\bf q}_3)\,
\Bigl(1-n_{{\bf q}_1}^{(f_1)}\Bigr)
n_{{\bf q}_2}^{(f_2)}n_{{\bf q}_3}^{(f_3)}\!\Biggr\}
\]
\[
\hspace{3cm}
+\,2\int\!\frac{d{\bf q}}{(2 \pi)^3}
\!\int\!\!d{\cal T}_{q\bar{q}\rightarrow q\bar{q}}^{(ff_1;\,f_2f_3)}\,
\omega^{(f)}_{\bf q}\,{\it w}_{q\bar{q}\rightarrow q\bar{q}}^{(ff_1;\,f_2f_3)}
({\bf q}, {\bf q}_1; {\bf q}_2, {\bf q}_3)\,
\Bigl(1-\bar{n}_{{\bf q}_1}^{(f_1)}\Bigr)
\bar{n}_{{\bf q}_2}^{(f_2)}n_{{\bf q}_3}^{(f_3)}\!\Biggr\}
\]
\[
+\,\biggl(n^{(f_i)}_{{\bf q}_i}\rightleftharpoons\bar{n}^{(f_i)}_{{\bf q}_i},
\;i=1,2,3\biggr).
\]
In deriving the right-hand side of (\ref{eq:8e}) we have used equations
(\ref{eq:5c}) and (\ref{eq:6i}). To proof (\ref{eq:8w}) it is sufficient
restored some scattering probability on the right-hand side of
Eq.\,(\ref{eq:8e}). For definiteness we consider the probability of
elastic scattering of plasmino off plasmon entering into the first term.

At first step we transform Eq.\,(\ref{eq:8w}) to more suitable form. The
Fourier components of quark fields $\psi_{\alpha}^i(q)$,
$\bar{\psi}_{\alpha}^i(-q)$ are associated with induced sources
$\eta_{\alpha}^i(q)$ and $\bar{\eta}_{\alpha}^i(-q)$ by the Dirac equation
\[
\psi_{\alpha}^i(q)= -\,^{\ast}\!S_{\alpha\beta}(q)\,\eta_{\beta}^i(q),\quad
\bar{\psi}_{\alpha}^i(-q)=\bar{\eta}_{\beta}^i(-q)
\,^{\ast}\!S_{\beta\alpha}(-q).
\]
Substituting the last expressions into Eq.\,(\ref{eq:8w}), we obtain
instead of (\ref{eq:8w})
\begin{equation}
{\cal I}_{\cal F}= -\,\frac{i}{2}\,
\lim\limits_{\tau,\,V\rightarrow\infty}
\frac{(2\pi)^4}{\tau V}
\int\!d{\bf q}dq^0\,q^0
\langle\bar{\eta}_{\alpha}^i(-q)\eta_{\beta}^i(q)\rangle\,
\bigr\{\,^{\ast}\!S_{\alpha\beta}(-q)+\,^{\ast}\!S_{\alpha\beta}(q)\bigl\}.
\label{eq:8r}
\end{equation}
By virtue of the properties of the quark scalar propagators
$^{\ast}\!\Delta_{\pm}(-q)= (^{\ast}\!\Delta_{\mp}(q))^{\ast}$ the sum in braces
in integrand of (\ref{eq:8r}) can be written in the form
\[
2i\,{\rm Im}\,(\!\,^{\ast}\!\Delta_{+}(q))(h_{+}(\hat{\bf q}))_{\alpha\beta}
+
2i\,{\rm Im}\,(\!\,^{\ast}\!\Delta_{-}(q))(h_{-}(\hat{\bf q}))_{\alpha\beta}.
\]
Furthermore, we express the projection operators 
$h_{\pm}(\hat{\bf q})$ in terms of simultaneous eigenspinors of chirality and
helicity
\[
(h_{+}({\hat{\bf q}}))_{\alpha\beta}=
\sum\limits_{\lambda=\pm}u_{\alpha}(\hat{\bf q},\lambda)
\bar{u}_{\beta}(\hat{\bf q},\lambda),
\quad
(h_{-}({\hat{\bf q}}))_{\alpha\beta}=
\sum\limits_{\lambda=\pm}v_{\alpha}(\hat{\bf q},\lambda)
\bar{v}_{\beta}(\hat{\bf q},\lambda),
\]
and approximate imaginary parts of the scalar propagators for
a weak-absorption medium by the following expressions:
\[
{\rm Im}\,^{\ast}\!\Delta_{\pm}(q)\simeq
\mp\,\pi\,{\rm Z}_{\pm}({\bf q})\,
\delta (q^0 - \omega_{\bf q}^{\pm})
\pm\,\pi\,{\rm Z}_{\mp}({\bf q})\,
\delta (q^0 + \omega_{\bf q}^{\mp}).
\]
It is not difficult to see that the term containing function
$\delta(q^0+\omega^{\mp}_{\bf q})$ after substitution into (\ref{eq:8r}),
performing integration in $dq_0$ and replacement of variable
${\bf q}\rightarrow - {\bf q}$ results in contribution to emitted radiant
power ${\cal I}_{\cal F}$ exactly equal the
contribution of term with $\delta(q^0 - \omega^{\pm}_{\bf q})$. Taking into
account above-mentioned, we define the final expression for the emitted radiant
power ${\cal I}_{\cal F}$, which is more convenient for
concrete applications
\begin{equation}
{\cal I}_{\cal F}= 2\pi\!
\lim\limits_{\tau,\,V\rightarrow\infty}
\frac{(2\pi)^4}{\tau V}\sum_{\lambda=\pm}
\int\!d{\bf q}\,
\Bigl\{\omega_{\bf q}^{+}\,{\rm Z}_{+}({\bf q})\,
\langle|\,\bar{\eta}_{\alpha}^i(-q)u_{\alpha}
(\hat{\bf q},\lambda)|^{\,2}_{\,q^0=\omega_{\bf q}^{+}}\rangle
\label{eq:8t}
\end{equation}
\[
\hspace{5.7cm}
+\,
\omega_{\bf q}^{-}\,{\rm Z}_{-}({\bf q})\,
\langle|\,\bar{\eta}_{\alpha}^i(-q)
v_{\alpha}(\hat{\bf q},\lambda)|^{\,2}_{\,q^0=\omega_{\bf q}^{-}}
\rangle\Bigr\}.
\]

For derivation of the probability for the elastic scattering of plasmino
off plasmon in the last term on the right-hand side of Eq.\,(\ref{eq:8t})
we perform the following replacement:
\[
\bar{\eta}_{\alpha}^i(-q)\rightarrow
\tilde{\bar{\eta}}^{(2,1)i}_{\alpha}
(A^{\ast(0)},A^{\ast(0)},\bar{\psi}^{(0)})(-q),
\]
where the effective source $\tilde{\bar{\eta}}_{\alpha}^{(2,1)i}(-q)$ is defined by
equation (\ref{eq:3s}). As a result of such a replacement for `plasmino' part
of expression (\ref{eq:8t}) we have
\[
({\cal I}_{\cal F})_{\rm plasmino}= \pi g^4\!\!
\lim\limits_{\tau,\,V\rightarrow\infty}
\frac{(2\pi)^4}{\tau V}\sum_{\lambda=\pm}
\int\!d{\bf q}\,\,\omega_{\bf q}^{-}\,{\rm Z}_{-}({\bf q})\Bigl(
\!\,^{\ast}\tilde{\bar{\Gamma}}^{(Q)a_1a_2,\,ii_1}_{\mu_1\mu_2,\,\alpha\gamma}
(k_1,k_2;q_1,-q)v_{\alpha}(\hat{\bf q},\lambda)\Bigr)
\]
\[
\times
\Bigl(\!\,^{\ast}\tilde{\Gamma}^{(Q)a_1^{\prime}a_2^{\prime},\,ii_1^{\prime}
}_{\mu_1^{\prime}\mu_2^{\prime},\,\beta\gamma^{\prime}}
(k_1^{\prime},k_2^{\prime};q_1^{\prime},-q)
\bar{v}_{\beta}(\hat{\bf q},\lambda)\Bigr)\,
\langle
A^{\ast(0)a_1\mu_1}(k_1)A^{\ast(0)a_2\mu_2}(k_2)
A^{(0)a_1^{\prime}\mu_1^{\prime}}(k_1^{\prime})
A^{(0)a_2^{\prime}\mu_2^{\prime}}(k_2^{\prime})
\rangle
\]
\[
\times\langle\bar{\psi}^{(0)i_1}_{\gamma}(-q_1)
\psi^{(0)i_1^{\prime}}_{\gamma^{\prime}}(q_1^{\prime})
\rangle
\,\delta(q - q_1 - k_1 -k_2)\,
\delta(q - q_1^{\prime} - k_1^{\prime} -k_2^{\prime})\,
dq_1dk_1dk_2\,dq_1^{\prime}dk_1^{\prime}dk_2^{\prime}.
\]
Here, the product of correlation functions is transformed by formula
(\ref{eq:5e}). Performing integration in
$dq_1^{\prime}dk_1^{\prime}dk_2^{\prime}$, we obtain instead of the
last expression
\begin{equation}
({\cal I}_{\cal F})_{\rm plasmino}= \pi g^4\!\!
\lim\limits_{\tau,\,V\rightarrow\infty}
\frac{(2\pi)^4}{\tau V}\sum_{\lambda=\pm}
\int\!d{\bf q}\,\,\omega_{\bf q}^{-}\,{\rm Z}_{-}({\bf q})\,
I^{\mu_1\mu_1^{\prime}}(k_1)I^{\mu_2\mu_2^{\prime}}(k_2)
\Upsilon_{\gamma^{\prime}\!\gamma}(q_1)
\label{eq:8y}
\end{equation}
\[
\times\Bigl(\!
\,^{\ast}\tilde{\bar{\Gamma}}^{(Q)a_1a_2,\,ii_1}_{\mu_1\mu_2,\,\alpha\gamma}
(k_1,k_2;q_1,-q)v_{\alpha}(\hat{\bf q},\lambda)\Bigr)
\,\Bigl(\!\,^{\ast}\tilde{\Gamma}^{(Q)a_1a_2,\,ii_1}_
{\mu_1^{\prime}\mu_2^{\prime},\,\beta\gamma^{\prime}}
(k_1,k_2;q_1,-q)\bar{v}_{\beta}(\hat{\bf q},\lambda)\Bigr)
\]
\[
\times\,
[\delta(q - q_1 - k_1 -k_2)]^{\,2}
dq_1dk_1dk_2.
\]
By $\delta$-function squared we mean
\[
[\delta(q - q_1 - k_1 -k_2)]^2 =
\frac{1}{(2\pi)^4}\,\tau V\delta(q - q_1 - k_1 -k_2).
\]
In functions $I^{\mu_1\mu_1^{\prime}}(k_1)$,
$I^{\mu_2\mu_2^{\prime}}(k_2)$ and
$\Upsilon_{\gamma^{\prime}\!\gamma}(q_1)$ we extract plasmon and
plasmino part\footnote{In the product of the spectral densities
$I_{k_1}^lI_{k_2}^l$ considering quasiparticle approximation
(\ref{eq:5p}) we keep only the terms corresponding to the scattering
of plasmino off plasmon.} according to Eq.\,(\ref{eq:5u}). Further,
going to the plasmino and plasmon number densities
(Eq.\,(\ref{eq:5a})), we result equation (\ref{eq:8y}) in the final
form
\[
({\cal I}_{\cal F})_{\rm plasmino}=
2\!\int\!\frac{d{\bf q}}{(2 \pi)^3}
\frac{d{\bf q}_1}{(2\pi)^3}
\frac{d{\bf k}_1}{(2\pi)^3}\!
\frac{d{\bf k}_2}{(2 \pi)^3}\,
(2\pi)^4\delta({\bf q}+{\bf k}_1-{\bf q}_1-{\bf k}_2)
\,\delta(\omega_{\bf q}^{-}+\omega_{{\bf k}_1}^{l}-
\omega_{{\bf q}_1}^{-}-\omega_{{\bf k}_2}^{l})
\]
\[
\times\,\omega^{-}_{\bf q}\Biggl(\,
\sum\limits_{\lambda\!,\,\lambda_1=\,\pm}\!\!
{\rm T}^{a_1a_2,\,ii_1}_{\lambda\lambda_1}(-{\bf k}_1,{\bf k}_2;{\bf q}_1,-{\bf q})
\,({\rm T}^{a_1a_2,\,ii_1}_{\lambda\lambda_1}
(-{\bf k}_1,{\bf k}_2;{\bf q}_1,-{\bf q}))^{\ast}\Biggr)
\,n_{{\bf q}_1}^{-}N_{{\bf k}_1}^lN_{{\bf k}_2}^l,
\]
where the function ${\rm T}^{a_1a_2,\,ii_1}_{\lambda\lambda_1}$  is defined by
Eq.\,(\ref{eq:5s}). The expression obtained exactly coincides with
corresponding term on the right-hand side of Eq.\,(\ref{eq:8e}).

Thus suggested expression (\ref{eq:8w}) correctly reproduces the
scattering probabilities obtained by a different way. Formulae for
emitted radiant power (\ref{eq:8q}) and (\ref{eq:8w}) open a way
for more direct and economic deriving the scattering
probabilities of the other higher processes of nonlinear
interaction of soft fermion and boson excitations in QGP.

\section{\bf Linearized version of Boltzmann equation}
\setcounter{equation}{0}

In this section we briefly consider the linearization procedure of
self-consistent system of kinetic equations (\ref{eq:2w}) and
(\ref{eq:2e}). We examine plasmino kinetic equation (\ref{eq:2w}) with
generalized rates (\ref{eq:2t}) and (\ref{eq:2y}), where we will consider
only the contribution with $f_1=-$ and $b_1=b_2=l$. We assume that the
off-equilibrium fluctuations are perturbative small and present the number
densities of colorless (anti)plasminos and plasmons in the form
\[
n_{\bf q}^{-}=n_{eq}^{-}({\bf q})+\delta n_{\bf q}^{-},\quad
\bar{n}_{\bf q}^{-}=\bar{n}_{eq}^{-}({\bf q})+\delta\bar{n}_{\bf q}^{-},\quad
N_{\bf k}^l=N_{eq}^{l}({\bf k})+\delta N_{\bf k}^l,
\]
where $n_{eq}^{-}({\bf q})=\bar{n}_{eq}^{-}({\bf q})=
(e^{\,\omega^{-}_{\bf q}/T} + 1)^{-1}$ and
$N_{eq}^{l}({\bf k})=(e^{\,\omega^{\,l}_{\bf k}/T} - 1)^{-1}$.
Parametrizating  off-equilibrium fluctuations of the occupation numbers
$\delta n_{\bf q}^{-}$, $\delta\bar{n}_{\bf q}^{-}$ and $\delta N^l_{\bf k}$
as follows
\[
\delta n_{\bf q}^{-}\equiv-\, \frac{dn^{-}_{eq}({\bf q})}
{d\omega^{-}_{\bf q}}\,w^{-}_{\bf q}=
\frac{1}{T}\,n_{eq}^{-}({\bf q})(1-n_{eq}^{-}({\bf q}))
w^{-}_{\bf q},
\]
\[
\delta \bar{n}_{\bf q}^{-}\equiv-\, \frac{dn^{-}_{eq}({\bf q})}
{d\omega^{-}_{\bf q}}\,\bar{w}^{-}_{\bf q}=
\frac{1}{T}\,n_{eq}^{-}({\bf q})(1-n_{eq}^{-}({\bf q}))
\bar{w}^{-}_{\bf q},
\]
\[
\hspace{0.2cm}
\delta N^l_{\bf k} \equiv -\, \frac{dN^l_{eq}({\bf k})}
{d\omega^l_{\bf k}}\,{\cal W}^l_{\bf k} =
\frac{1}{T}\,N^l_{eq}({\bf k})(1+N^l_{eq}({\bf k}))
{\cal W}^l_{\bf k},
\]
we derive from Eqs.\,(\ref{eq:2w}), (\ref{eq:2t}) and (\ref{eq:2y}),
after simple algebraic transformations the linearized Boltzmann equation for
the function $w_{\bf q}^{-}$
\begin{equation}
\frac{\partial w_{\bf q}^{-}}{\partial t} +
{\bf v}_{\bf q}^{-}\cdot\frac{\partial w_{\bf q}^{-}}{\partial {\bf x}} =
\label{eq:9q}
\end{equation}
\[
-\,2\!\int\!d{\cal T}_{qg\rightarrow qg} \,
\frac{n^{-}_{eq}({\bf q}_1)N^l_{eq}({\bf k}_2)(1+N^l_{eq}({\bf k}_1))}
{n^{-}_{eq}({\bf q})}\, {\it w}_{qg\rightarrow qg}
({\bf q}, {\bf k}_1; {\bf q}_1, {\bf k}_2)
\Bigl\{w^{-}_{\bf q}+{\cal W}^l_{{\bf k}_1}-w^{-}_{{\bf q}_1}-
{\cal W}^l_{{\bf k}_2}\Bigr\}
\]
\[
\hspace{0.2cm}
-\!\int\!d{\cal T}_{q\bar{q}\rightarrow gg} \,
\frac{(1-n^{-}_{eq}({\bf q}_1))N^l_{eq}({\bf k}_1)N^l_{eq}({\bf k}_2)}
{n^{-}_{eq}({\bf q})}\, {\it w}_{q\bar{q}\rightarrow gg}
({\bf q}, {\bf q}_1; {\bf k}_1, {\bf k}_2)
\Bigl\{w^{-}_{\bf q}+\bar{w}^{-}_{{\bf q}_1}-{\cal W}^l_{{\bf k}_1}-
{\cal W}^l_{{\bf k}_2}\Bigr\}
\]
\[
-\!\int\!d{\cal T}_{qq\rightarrow qq} \,
\frac{(1-n^{-}_{eq}({\bf q}_1))n^{-}_{eq}({\bf q}_2)n^{-}_{eq}({\bf q}_3)}
{n^{-}_{eq}({\bf q})}\, {\it w}_{qq\rightarrow qq}
({\bf q}, {\bf q}_1; {\bf k}_1, {\bf k}_2)\,
\{w^{-}_{\bf q}+w^{-}_{{\bf q}_1}- w^{-}_{{\bf q}_2}-
w^{-}_{{\bf q}_3}\}
\hspace{0.3cm}
\]
\[
-\,2\!\int\!d{\cal T}_{q\bar{q}\rightarrow q\bar{q}} \,
\frac{(1-n^{-}_{eq}({\bf q}_1))n^{-}_{eq}({\bf q}_2)n^{-}_{eq}({\bf q}_3)}
{n^{-}_{eq}({\bf q})}\, {\it w}_{q\bar{q}\rightarrow q\bar{q}}
({\bf q}, {\bf q}_1; {\bf k}_1, {\bf k}_2)\,
\{w^{-}_{\bf q}+\bar{w}^{-}_{{\bf q}_1}- \bar{w}^{-}_{{\bf q}_2}-
w^{-}_{{\bf q}_3}\}.
\hspace{1cm}
\]
In deriving Eq.\,(\ref{eq:9q}) we have used the identities
\[
(1-n^{-}_{eq}({\bf q}))n^{-}_{eq}({\bf q}_1)
(1+N^l_{eq}({\bf k}_1))N^l_{eq}({\bf k}_2)=
n^{-}_{eq}({\bf q})(1-n^{-}_{eq}({\bf q}_1))
N^l_{eq}({\bf k}_1)(1+N^l_{eq}({\bf k}_2)),
\]
\[
(1-n^{-}_{eq}({\bf q}))(1-\bar{n}^{-}_{eq}({\bf q}_1))
N^l_{eq}({\bf k}_1)N^l_{eq}({\bf k}_2)=
n^{-}_{eq}({\bf q})\bar{n}^{-}_{eq}({\bf q}_1)\,
(1+N^l_{eq}({\bf k}_1))(1+N^l_{eq}({\bf k}_2)),
\]
\[
(1-n^{-}_{eq}({\bf q}))(1-n^{-}_{eq}({\bf q}_1))\,
n^{-}_{eq}({\bf q}_2)n^{-}_{eq}({\bf q}_3)=
n^{-}_{eq}({\bf q})n^{-}_{eq}({\bf q}_1)\,
(1-n^{-}_{eq}({\bf q}_2))(1-n^{-}_{eq}({\bf q}_3)),
\hspace{0.1cm}
\]
which hold by virtue of corresponding conservation laws of energy.

In the remaining part of this section we make a little digression from 
the general line of our present work and consider the problem of the minimal
generalization of Boltzmann equation (\ref{eq:2w}) to the case of
the dynamics of color soft-quark and soft-gluon excitations. For
this  purpose we assume that a time-space dependent external
perturbation (e.g. external color current $j_{\mu}^{{\rm
ext}\,a}(x)$) starts acting on the system. In the presence of the
external color perturbation the soft gauge field develops an
expectation value $\langle A_{\mu}^{a}(x)\rangle\equiv{\cal
A}_{\mu}^a(x)\ne 0$ (but nevertheless\footnote{We can formally
suggest that in the system there exists also an external color
source $\eta_{\alpha}^{{\rm ext}\,i}(x)$ produced mean quark field
$\langle\psi^i_{\alpha}(x)\rangle \neq 0$ (see, for example
\cite{vega}). Here, we don't consider such a case for
simplification of the problem.} $\langle\psi^i_{\alpha}(x)\rangle=
\langle\bar{\psi}^i_{\alpha}(x)\rangle=0$) and the soft-quark and
soft-gluon number densities acquire a non-diagonal color
structure. In this case, we expect the time-space evolution of
$n^{-}_{\bf q} = (n^{-\,ij}_{\bf q})$ to be described by the
(matrix) plasmino Vlasov-Boltzmann equation instead of
(\ref{eq:2w})
\begin{equation}
\Bigl({\cal D}_t + {\bf v}_{\bf q}^{-}\cdot{\cal D}_{\bf x}\Bigr)n^{-}_{\bf q}
-\frac{1}{2}\,g\Bigl\{\Bigl({\bf {\cal E}}(x) + ({\bf v}_{\bf q}^{-}\times
{\bf {\cal B}}(x))\Bigr)_i,\nabla_{{\rm p}_i}n^{-}_{\bf q}\Bigr\} =
-\,{\rm C}\,[n^{-}_{\bf q},N_{\bf k}^l],
\label{eq:9w}
\end{equation}
where ${\cal D}_{\mu}$ is the covariant derivative acting as
\[
{\cal D}_{\mu}n^{-}_{\bf q} \equiv \partial_{\mu}n^{-}_{\bf q}
+ig[{\cal A}_{\mu}(x),n^{-}_{\bf q}], \quad
{\cal A}_{\mu}(x)=t^a{\cal A}_{\mu}^a(x).
\]
The functions ${\cal E}^i(x)$ and ${\cal B}^i(x)$ are mean
chromoelectric and chromomagnetic fields. The collision term ${\rm
C}\,[n^{-}_{\bf q},N_{\bf k}^l]$ has the following structure:
\begin{equation}
{\rm C}\,[n_{\bf q}^{-},N_{\bf p}^l]=\frac{1}{2}\,\Bigl\{
n^{-}_{\bf q},\Gamma_{\rm d}^{(-)}[n^{-}_{\bf q},N_{\bf k}^l]\Bigr\} -
\frac{1}{2}\,\Bigl\{( 1 - n^{-}_{\bf q}),
\Gamma_{\rm i}^{(-)}[n^{-}_{\bf q},N_{\bf k}^l]\Bigr\}-\ldots,
\label{eq:9e}
\end{equation}
where $\Gamma_{\rm d}^{(-)}[n^{-}_{\bf q},N_{\bf k}^l] =
(\Gamma_{\rm d}^{ii^{\prime}}[n^{-}_{\bf q},N_{\bf k}^l])$ and
$\Gamma_{\rm i}^{(-)}[n^{-}_{\bf q},N_{\bf k}^l]
=(\Gamma_{\rm i}^{ii^{\prime}}[n^{-}_{\bf q},N_{\bf k}^l])$ represent the
generalized decay and regenerating rates of color plasminos,
respectively. The dots on the right-hand side of
Eq.\,(\ref{eq:9e}) is refereed to possible contributions
containing commutators of $[\,{\rm Re}\,\Sigma_{R},n^{-}_{\bf q}]$
type, where $\Sigma_{R}$ is retarded soft-quark self-energy.

As above, we restrict our attention to the process of the elastic scattering of 
color plasmino off color plasmon. In this particular case the decay and 
regenerating rates have the following form:
\[
\Gamma_{\rm d}^{ii^{\prime}}[n^{-}_{\bf q},N_{\bf k}^l] =\!
\sum\limits_{\lambda,\,\lambda_1=\,\pm}\!
\int\!d{\cal T}_{qg\rightarrow qg}^{(-\,l;\,-\,l)} \,
{\rm T}^{\,a_1^{\prime}a_2^{\prime},\,i^{\prime}i_1^{\prime}}_{\lambda\lambda_1}
(-{\bf k}_1,{\bf k}_2;{\bf q}_1,-{\bf q})
\,({\rm T}^{\,a_1a_2,\,ii_1}_{\lambda\lambda_1}
(-{\bf k}_1,{\bf k}_2;{\bf q}_1,-{\bf q}))^{\ast}
\]
\begin{equation}
\hspace{0.5cm}
\times\,
(1 - n^{-}_{{\bf q}_1})^{i_1^{\prime}i_1}(N_{{\bf k}_1}^l)^{a_1a_1^{\prime}}
(1+N_{{\bf k}_2}^l)^{a_2^{\prime}a_2},
\label{eq:9r}
\end{equation}
\[
\Gamma_{\rm i}^{ii^{\prime}}[n^{-}_{\bf q},N_{\bf k}^l] =\!
\sum\limits_{\lambda,\,\lambda_1=\,\pm}\!
\int\!d{\cal T}_{qg\rightarrow qg}^{(-\,l;\,-\,l)} \,
{\rm T}^{\,a_1^{\prime}a_2^{\prime},\,i^{\prime}i_1^{\prime}}_{\lambda\lambda_1}
(-{\bf k}_1,{\bf k}_2;{\bf q}_1,-{\bf q})
\,({\rm T}^{\,a_1a_2,\,ii_1}_{\lambda\lambda_1}
(-{\bf k}_1,{\bf k}_2;{\bf q}_1,-{\bf q}))^{\ast}
\]
\[
\times\,
(n^{-}_{{\bf q}_1})^{i_1^{\prime}i_1}(1+N_{{\bf k}_1}^l)^{a_1^{\prime}a_1}
(N_{{\bf k}_2}^l)^{a_2a_2^{\prime}},
\]
where matrix element ${\rm T}^{\,a_1a_2,\,ii_1}_{\lambda\lambda_1}$ is defined by
Eq.\,(\ref{eq:5s}). We recall that this matrix element has a physical sensible
color decomposition:
\begin{equation}
{\rm T}^{a_1a_2\,ii_1}_{\lambda\lambda_1}(-{\bf k}_1,{\bf k}_2;{\bf q}_1,-{\bf q})=
\frac{1}{2}\,\{t^{a_2},t^{a_1}\}^{i_1i}
\,{\cal C}_{\cal S}^{-1\,}
{\rm T}^{({\cal S})}_{\lambda\lambda_1}(-{\bf k}_1,{\bf k}_2;{\bf q}_1,-{\bf q})
\label{eq:9t}
\end{equation}
\[
\hspace{4.6cm}
+\,\frac{1}{2}\,[\,t^{a_2},t^{a_1}]^{i_1i}
\,{\cal C}_{\cal A}^{-1}\,
{\rm T}^{({\cal A})}_{\lambda\lambda_1}(-{\bf k}_1,{\bf k}_2;{\bf q}_1,-{\bf q}).
\]
Here, helical amplitudes ${\rm T}^{({\cal S,\,A})}_{\lambda\lambda_1}$
are given by Eq.\,(\ref{eq:5k}) and
${\cal C}_{\cal S,\,A}\!\!=\!\!\{C_F(C_F\mp 1/(2N_c))/2\}^{\!1/2}\!$.

Furthermore, we consider the linearized version of
Vlasov-Boltzmann equation (\ref{eq:9w}). We write the number
densities of color plasminos and plasmons as follows:
\[
(n_{\bf q}^{-})^{ij}=n_{eq}^{-}({\bf k})\delta^{ij}
-\, \frac{dn^{-}_{eq}({\bf q})}
{d\omega^{-}_{\bf q}}\,w^{-\,a}_{\bf q}(t^a)^{ij},
\]
\[
\hspace{0.2cm}
(N_{\bf k}^l)^{ab}=N_{eq}^{l}({\bf k})\delta^{ab}
-\,\frac{dN^l_{eq}({\bf k})}
{d\omega^l_{\bf k}}\,{\cal W}^{\,lc}_{\bf k}(T^c)^{ab}.
\]
After some cumbersome algebraic transformations we derive the linearized
kinetic equation for color plasminos from
Eqs.\,(\ref{eq:9w})\,--\,(\ref{eq:9t})
\begin{equation}
\Bigl({\cal D}_t + {\bf v}_{\bf q}^{-}\cdot{\cal D}_{\bf x}\Bigr)
w^{-}_{\bf q}=
-\,g\,({\bf v}_{\bf q}^{-}\cdot{\bf {\cal E}}(x))
-\,\delta{\rm C}\,[w^{-}_{\bf q},{\cal W}_{\bf k}^{\,l}],
\label{eq:9y}
\end{equation}
where plasmino-plasmon linearized collision term $\delta{\rm C}$ has the form:
\begin{equation}
\delta{\rm C}\,[w^{-}_{\bf q},{\cal W}_{\bf k}^l]=
\sum\limits_{\lambda,\,\lambda_1=\pm}
\!\int\!d{\cal T}_{qg\rightarrow qg}^{(-\,l;\,-\,l)}\,
\frac{n^{-}_{eq}({\bf q}_1)N^l_{eq}({\bf k}_2)(1+N^l_{eq}({\bf k}_1))}
{n^{-}_{eq}({\bf q})}\,
\left[\Bigl\{|{\rm T}^{({\cal S})}_{\lambda\lambda_1}|^{\,2}
+ |{\rm T}^{({\cal A})}_{\lambda\lambda_1}|^{\,2}\Bigr\}\,w^{-}_{\bf q}\right.
\label{eq:9u}
\end{equation}
\[
+\left\{
\frac{1}{4}\left[\biggl(C_F-\frac{1}{N_c}\biggr)\frac{N_c}{2}
-\frac{1}{4}\,\right]{\cal C}_{\cal S}^{-2}\,
|{\rm T}^{({\cal S})}_{\lambda\lambda_1}|^{\,2}
\right.
+
\frac{1}{4}\left[\biggl(C_F+\frac{1}{N_c}\biggr)\frac{N_c}{2}
-\frac{1}{4}\,\right]{\cal C}_{\cal A}^{-2}\,
|{\rm T}^{({\cal A})}_{\lambda\lambda_1}|^{\,2}
\]
\[
\left.
+\,\frac{1}{2}\biggl(C_F+\frac{1}{2N_c}\biggr)\frac{N_c}{2}
\,({\cal C}_{\cal S}{\cal C}_{\cal A})^{-1}\,
{\rm Re}\,({\rm T}^{\,\ast({\cal S})}_{\lambda\lambda_1}
{\rm T}^{({\cal A})}_{\lambda\lambda_1})
\right\}{\cal W}_{{\bf k}_1}^{\,l}
\]
\[
-\left\{
\frac{1}{4}\left[\biggl(C_F-\frac{1}{N_c}\biggr)\frac{N_c}{2}
-\frac{1}{4}\,\right]{\cal C}_{\cal S}^{-2}\,
|{\rm T}^{({\cal S})}_{\lambda\lambda_1}|^{\,2}
\right.
+
\frac{1}{4}\left[\biggl(C_F+\frac{1}{N_c}\biggr)\frac{N_c}{2}
-\frac{1}{4}\,\right]{\cal C}_{\cal A}^{-2}\,
|{\rm T}^{({\cal A})}_{\lambda\lambda_1}|^{\,2}
\]
\[
\left.
-\,\frac{1}{2}\biggl(C_F+\frac{1}{2N_c}\biggr)\frac{N_c}{2}
\,({\cal C}_{\cal S}{\cal C}_{\cal A})^{-1}\,
{\rm Re}\,({\rm T}^{\,\ast({\cal S})}_{\lambda\lambda_1}{\rm T}^{({\cal A})}_{\lambda\lambda_1})
\right\}{\cal W}_{{\bf k}_2}^{\,l}
\]
\[
\left.
-\left\{\frac{1}{2}\,\biggl[\frac{1}{4N_c^2}+\biggl(\frac{1}{4N_c^2}+
\frac{1}{4}\biggr)\biggr]\,{\cal C}_{\cal S}^{-2}
|{\rm T}^{({\cal S})}_{\lambda\lambda_1}|^{\,2}
+\frac{1}{2}\,\biggl[\frac{1}{4N_c^2}-\biggl(\frac{1}{4N_c^2}+
\frac{1}{4}\biggr)\biggr]\,{\cal C}_{\cal A}^{-2}
|{\rm T}^{({\cal A})}_{\lambda\lambda_1}|^{\,2}\right\} w_{{\bf q}_1}^{-}
\right].
\]
Here, $w_{\bf q}^{-}\equiv w_{\bf q}^{-\,a}t^a$,
${\cal W}_{\bf k}^{\,l}\equiv{\cal W}_{\bf k}^{\,la}t^a$. In deriving
(\ref{eq:9u}) we use the following identities:
\[
(\{t^{a_1^{\prime}},t^{a_2}\}\{t^{a_2},t^{a_1}\})^{ii^{\prime}}
(T^b)^{a_1^{\prime}a_1}=
\left[\biggl(C_F-\frac{1}{N_c}\biggr)\frac{N_c}{2}-\frac{1}{4}\,\right]
(t^b)^{ii^{\prime}},
\]
\[
(\{t^{a_1^{\prime}},t^{a_2}\}[t^{a_2},t^{a_1}])^{ii^{\prime}}
(T^b)^{a_1^{\prime}a_1}=
\frac{1}{2}\biggl(C_F+\frac{1}{2N_c}\biggr)N_c\,(t^b)^{ii^{\prime}},
\hspace{0.3cm}
\]
\[
(\{t^{a_1},t^{a_2}\}t^b\{t^{a_2},t^{a_1}\})^{ii^{\prime}}=
2\biggl[\frac{1}{4N_c^2}+\biggl(\frac{1}{4N_c^2}+
\frac{1}{4}\biggr)\biggr](t^b)^{ii^{\prime}},
\]
etc. Since the equilibrium plasmino number density is proportional to
the identity, the commutator terms on the right-hand side of
Eq.\,(\ref{eq:9e}) vanish. Therefore if the system is in
conditions, when the off-equilibrium function
$\delta n^{-}_{\bf q}$ is perturbatively small, then linearized
Vlasov-Boltzmann equation (\ref{eq:9y}), (\ref{eq:9u}) in a certain sense
is exact. Finally by analogy with color current induced by color
plasmons (Eq.\,(9.12) in \cite{markov1}), we can define the
color current resulting from the color-plasmino number density
$j_{\mu}^{(-)}(x)=(j_{0}^{(-)}(x),\,{\bf j}^{(-)}(x))$, where now
\[
j_{0}^{(-)}(x)\!=\!-\,gt^a\!\!\int\!\!\frac{d{\bf q}}{(2\pi)^3}\biggr(
\frac{\partial n_{eq}({\bf q})}{\partial\omega_{\bf q}^{-}}\biggr)
(w_{\bf q}^{-\,a}-\bar{w}_{\bf q}^{-\,a}),\,
{\bf j}^{(-)}(x)\!=\!-\,gt^a\!\!\int\!\!\frac{d{\bf q}}{(2\pi)^3}\biggr(
\frac{\partial n_{eq}({\bf q})}{\partial{\bf q}}\biggr)
(w_{\bf q}^{-\,a}-\bar{w}_{\bf q}^{-\,a}).
\]

For closing kinetic equation (\ref{eq:9y}) (and similar equations for the
functions $\bar{w}_{\bf q}^{-\,a}$ and ${\cal W}_{\bf k}^{\,la}$ it remains
only to write out the mean field equation defining a change of mean gauge field
in a system in a self-consistent manner
\[
{\cal D}^{\nu}(x){\cal F}_{\mu\nu}(x) = j_{\mu}^{(-)}(x) + j_{\mu}^{\,(l)}(x)
+ j_{\mu}^{\,{\rm ext}}(x).
\]
Here, the color-plasmon induced current $j_{\mu}^{\,(l)}(x)$ is defined by
Eq.\,(9.12) in \cite{markov1} and $j_{\mu}^{\,{\rm ext}}$ is the external
current that plays a part of initial color perturbation.
It should be noted that an existence of a time-space dependence of background
gauge field in the system leads, generally speaking, to the replacement of
dispersion equations (\ref{eq:2q}) by
\[
{\rm Re}\,^{\ast}\!\Delta^{\!-1}_{-}(q^0\!,{\bf q};t,{\bf x})= 0,\quad
{\rm Re}\,^{\ast}\!\Delta^{\!-1\,l}(k^0\!,{\bf k};t,{\bf x}) = 0,
\]
slowly depending on $x = (t, {\bf x})$ that in turn results in
modification of the dispersion relations for plasminos and
plasmons: $\omega^{-}=\omega^{-}({\bf q},t,{\bf x})$,
$\omega^{\,l} = \omega^{\,l}({\bf k},t,{\bf x})$. Besides a color
background gauge field removes an degeneracy of spectrum of plasma
excitations, when the same dispersion relation corresponds to
various oscillations with  different color indices. The
consequence of the last property is the fact that the dispersion
relations for color plasminos and plasmons acquire a matrix
character: $\omega^{-}=(\omega^{-\,ij}),\, \omega^l =
(\omega^{l\,ab})$. In this case the left-hand side of
Eq.\,(\ref{eq:9w}) should be supplemented by term of the following
type:
\[
\frac{1}{2}\,\biggl\{ \frac{\partial\omega^{-}({\bf q},x)}{\partial{\bf x}},
\frac{\partial n_{\bf q}^{-}}{\partial{\bf q}}\biggr\},
\]
which can be considered as additional `force' term acting on the color
quasiparticles. We can assume nevertheless that in the weak-field limit
background gauge field weakly influences on dispersion properties in the
system and for the first approximation one can successfully neglected by it.

\section{\bf Conclusion}
\setcounter{equation}{0}

In this paper within the framework of the hard thermal loops effective
theory we have presented the formal scheme for deriving the system of the
semiclassical Boltzmann equations, which describes the space-time evolution
of the soft-quark and soft-gluon number densities due to their interactions
among themselves in hot non-Abeliam plasma. This type of the nonlinear
interactions forms in itself an important element in the general dynamics
of the processes occurring in quark-gluon plasma at least in a
weak coupling regime. For highly excited plasma
states, when typical time of relaxation of the hard thermal
particle distributions is commensurable with the typical time of
relaxation of soft oscillations or even significantly exceeds
it, along with the kinetic equations for hard thermal particles it should be
used the kinetic equations for soft modes of type
(\ref{eq:2w}), (\ref{eq:2e}). However, for completeness of the
description the right-hand sides of kinetic equations (\ref{eq:2w}),
(\ref{eq:2e}) should be added by collision terms taking into account the
scattering processes of soft plasma excitations on hard thermal particles.
For moderate level of plasma excitations the processes
of this type can even play more important role in the dynamics of the system
in comparison with the above-mentioned processes of self-interaction of soft
modes. In our forthcoming papers a formalism of regular calculation of
such collision terms will be presented to the full extent.

Furthermore, we would like to discuss briefly how the approach presented here
and in our early papers, is agreed with the more traditional calculations from
the Feynman graphs (like the original calculations of the damping rates for
soft quasiparticles within the framework of the imaginary-time formalism).

One of the strongest arguments of plausibility of this approach
is the fact that damping rate of a plasmino at
rest \cite{markov_PRD} calculated on its basis, exactly coincides with the
damping rate of the standing plasmino derived
by Kobes, Kunstater, and Mak \cite{kobes} in the framework of the
Braaten-Pisarski effective theory. The same can be said about the damping rate
of a plasmon computed within the framework of our approach \cite{markov4}.
Thus remaining in the context of the Blaizot-Iancu equations and deriving
from them the effective kinetic equations for the soft plasma modes,
we are able to reproduce the gauge-invariant damping rates of the soft
quasiparticles (to the leading order in the coupling), which corresponds
to the damping rates from the resummed perturbation theory
\cite{braaten3, kobes}.

Notice that in the work \cite{braaten4}, Braaten and Thoma have pointed to
an existence of correspondence of such a kind in calculation of the soft
contribution to energy loss of a heavy fermion in hot QED plasma. They have
shown that the result of calculation using the rigorous imaginary-time formalism and
the result employing the alternative field-theoretical calculation
(that consists in a simple replacement of photon propagator by
effective photon propagator) coincide among themselves. At the same time
they have noted that ``{\it...a rigorous justification must await the
development of resummation methods for the real-time formalism that are as
powerful as the resummation methods for the imaginary-time formalism}". This
in full measure can be related to our approach of deriving the effective
kinetic equations for the soft modes.

In recent years significant efforts have been undertaken to construction of
the hard thermal loop resummation technique in the non-equilibrium field
theory \cite{carrington, fueki} (see, also \cite{smilga}) allowing one
effectively to work not only with the single-particle
propagators, but also with three- and four-point functions. The methods and
approaches suggested in these works enable one at least in principle within
the framework of standard methods of real-time finite temperature field theory,
to construct a calculation scheme of kinetic equations for soft quasiparticles
in QGP. As a guide idea here, it can be used elegant, thoroughly considered
approach suggested by Blaizot and Iancu in deriving the Boltzmann equation
for hard transverse gluons in high-temperature plasma \cite{blaizot1}.
Their derivation relies on a gauge-covariant gradient expansion of
the Kadanoff-Baym equations for the gluon two-point function. The Boltzmann
equation has emerged as the quantum transport equation
to leading order in $g$ for the gauge-covariant fluctuation
$\delta\acute{G}$ of a hard gluon propagator.

Besides in the above-mentioned paper, the Kadanoff-Baym equations for the
off-equilibrium propagator of the soft gluon ${\cal D}_{\mu\nu}(X,Y)$,
which are formally identical to those for a hard gluon propagator
$G_{\mu\nu}(X,Y)$, are written out. These equations for
${\cal D}_{\mu\nu}(X,Y)$ are used in
\cite{blaizot1} only to deduce the relation between the off-equilibrium
gauge-covariant fluctuation $\acute{{\cal D}^{<}}(k,x)$ and the
gauge-covariant fluctuation of the leading-order soft gluon polarization
tensor $\acute{\delta\Pi^{<}}(k,x)$. The problem of self-interactions of
the soft fields is not considered here. However, in principal, there is nothing
to forbid the use of these equations for research of the soft-field dynamics
and construction of the relevant transport equations within the framework
of the scheme suggested by Blaisot and Iancu.
Here, by $\acute{\delta\Pi^{<}}(k,x)$ we mean the fluctuation of the
next-to-leading order of the soft-gluon self-energy involving three- and
four-gluon off-equilibrium vertices with soft external lines. The relevant
effects of self-interaction of the soft fields are encoded in these vertex
functions. It would allow us to derive the transport equations for the
soft gluons directly from the underlying quantum field theory and compare
them with the equation obtained in this paper in the context of the
semiclassical approximation. For deriving the kinetic equations for soft
fermion modes it should be used the Kadanoff-Baym equations for the quark
two-point function. The development of this approach is also needed to specify
the limits of validity of the semiclassical kinetic approach suggested in
present work to the research of the processes of nonlinear interaction of
the soft fields in hot QCD plasma.

Unfortunately on this way considerable technical complications
resulting from the doubling of degrees of freedom arise.
Although it is suggested a number of receptions \cite{carrington},
which to a certain extent simplify intermediate computations in
deriving convolutions of propagators and vertices, nevertheless a
practical application of these methods to solving specific problem
on deriving soft-quasiparticle kinetic equations is at present very
complicated and requires using the symbolic manipulation program.
In this sense our less strict approach gives more economic method
for calculation of desired kinetic equations, although it also is
very complex and lengthy. The ``memory" of initial rigorous theory
here, is a presence of different types of vertices
$^{\ast}\Gamma^{(G)}$ and $^{\ast}\Gamma^{(Q)}$ connected with
different chronological orders of incoming lines. However their
amount is small that gives a possibility to perform calculation
in visible form without resorting to a program for evaluating real
time Feynman amplitudes. It can be assumed that in rigorous
approach the final kinetic equations (at least their linearized
versions) will be practically the same, as the kinetic equations
obtained within the framework of our more phenomenological
approach.

\section*{\bf Acknowledgments}
This work was supported by the Russian Foundation for Basic Research
(project no 03-02-16797).

\newpage
\section*{\bf Appendix A}
\setcounter{equation}{0}

The explicit form of the HTL-induced vertices between quark pair and two
gluons is defined by the following expressions:
$$
\small{
\delta \Gamma^{(G)\,ab}_{\mu\nu}(k_1,k_2;q_1,q_2)\!=\!
-\,\omega_0^2\!\!\int\!\frac{{\rm d} \Omega}{4 \pi}\,
\frac{v_{\mu}v_{\nu} \not\!v}{(v\cdot q_1 - i \epsilon)
(v\cdot q_2 - i \epsilon)}
\bigg(\frac{t^at^b}{v\cdot (q_1+k_1) + i \epsilon}
+ \frac{t^bt^a}{v\cdot (q_1+k_2) - i \epsilon} \bigg),
}
\eqno{({\rm A}.1)}
$$
$$
\small{
\delta \Gamma^{(Q)\,ab}_{\mu\nu}(k_1,k_2;q_1,q_2)\!=\!
-\,\omega_0^2\!\!\int\!\frac{{\rm d} \Omega}{4 \pi}\,
\frac{v_{\mu}v_{\nu} \not\!v}{(v\cdot q_1 + i \epsilon)
(v\cdot q_2 - i \epsilon)}
\bigg(\frac{t^at^b}{v\cdot (q_1+k_1) + i \epsilon}
+ \frac{t^bt^a}{v\cdot (q_1+k_2) + i \epsilon} \bigg).
}
\eqno{({\rm A}.2)}
$$

Below we list the properties of HTL-resummed three-gluon,
two-quark\,--\,one-gluon and two-quark\,--\,two-gluon vertex
functions, which used in the text
$$
\left(\!\,^{\ast}\Gamma_{\mu\mu_1\mu_2}(-k_1-k_2,k_1,k_2)\right)^{\ast}=
-\,^{\ast}\Gamma_{\mu\mu_1\mu_2}(k_1+k_2,-k_1,-k_2)=
\,^{\ast}\Gamma_{\mu\mu_1\mu_2}(k_1+k_2,-k_2,-k_1),
$$
$$
\gamma^0(\!\,^{\ast}\Gamma^{(Q)}_{\mu}(k;q_1,q_2))^{\dagger}\gamma^0
= \,^{\ast}\Gamma^{(Q)}_{\mu}(k;q_2,q_1)
= \,^{\ast}\Gamma^{(Q)}_{\mu}(-k;-q_1,-q_2),
\eqno{({\rm A}.3)}
$$
$$
\hspace{1cm}\gamma^0
(\!\,^{\ast}\Gamma^{(G)}_{\mu}(k;q_1,q_2))^{\dagger}\gamma^0
= \,^{\ast}\Gamma^{(G)}_{\mu}(-k;-q_1,-q_2)
= \,^{\ast}\Gamma^{(G)}_{\mu}(-k;-q_2,-q_1),
$$
\vspace{0.2cm}
$$
\delta\Gamma^{(Q) ab}_{\mu \nu}(k_1,k_2;q_1,q_2)=
\delta\Gamma^{(Q) ab}_{\mu \nu}(-k_2,-k_1;-q_2,-q_1)=
\delta\Gamma^{(Q) ba}_{\mu \nu}(k_2,k_1;q_1,q_2),
\eqno{({\rm A}.4)}
$$
$$
\gamma^0
(\delta\Gamma^{(Q)ab}_{\mu \nu}(k_1,k_2;q_1,q_2))^{\dagger}\gamma^0
= -\,\delta\Gamma^{(Q)ab}_{\mu \nu}(k_1,k_2;q_2,q_1)
= -\,\delta\Gamma^{(Q)ba}_{\mu \nu}(-k_1,-k_2;-q_1,-q_2),
$$
\vspace{0.2cm}
$$
\delta\Gamma^{(G) ab}_{\mu \nu}(k_1,k_2;q_1,q_2)=
-\,\delta\Gamma^{(G) ba}_{\mu \nu}(k_1,k_2;q_2,q_1),
\eqno{({\rm A}.5)}
$$
$$
\gamma^0
(\delta\Gamma^{(G)ab}_{\mu \nu}(k_1,k_2;q_1,q_2))^{\dagger}
\gamma^0
= -\,\delta\Gamma^{(G)ba}_{\mu \nu}(-k_1,-k_2;-q_1,-q_2).
$$


\section*{\bf Appendix B}
\setcounter{equation}{0}

In this Appendix we give the explicit expressions for the coefficient functions
entering into integrands of the conjugate effective currents and sources.
For the conjugate effective current
$$
\tilde{j}^{\ast\Psi(1,2)a}_{\mu}(A^{\ast(0)},\bar{\psi}^{(0)},\psi^{(0)})\!=\!
-\,g^2\!\!\int\!\!\,^{\ast}
\tilde{\bar{\Gamma}}^{(G)aa_1,\,ij}_{\mu{\mu}_1,\,\alpha\beta}(k,-k_1;-q_1,-q_2)
A^{\ast(0)a_1{\mu}_1}(k_1)\bar{\psi}^{(0)j}_{\beta}(-q_2)
\psi^{(0)i}_{\alpha}(q_1)
$$
$$
\times\,\delta(k - k_1 - q_1 - q_2) dk_1 dq_1 dq_2
\eqno{({\rm B}.1)}
$$
we have
$$
\!\,^{\ast}\tilde{\bar{\Gamma}}^{(G)aa_1,\,ij}_{\mu{\mu}_1,\,\alpha\beta}
(k,-k_1;-q_1,-q_2)
=
\delta{\Gamma}^{(G)a_1a,\,ji}_{\mu{\mu}_1,\,\beta\alpha}(-k,k_1;q_1,q_2)
\eqno{({\rm B}.2)}
$$
$$
-\,
[t^{a_1},t^{a}]^{ji}
\,^\ast\Gamma_{\mu\nu\mu_1}(-k,k-k_1,k_1)
\,^{\ast}{\cal D}^{\nu\nu^{\prime}}(-k+k_1)
\,^{\ast}\Gamma^{(G)}_{\nu^{\prime},\,\beta\alpha}(-q_1-q_2;q_1,q_2)
\hspace{0.6cm}
$$
$$
-\,(t^{a_1}t^{a})^{ji}
\,^{\ast}\Gamma^{(Q)}_{\mu_1,\,\beta\gamma}(-k_1;-q_2,k_1+q_2)
\,^{\ast}S_{\gamma\gamma^{\prime}}(-k+q_1)
\,^{\ast}\Gamma^{(G)}_{\mu,\,\gamma^{\prime}\alpha}(-k;k-q_1,q_1)
\hspace{0.35cm}
$$
$$
+\,(t^{a}t^{a_1})^{ji}
\,^{\ast}\Gamma^{(G)}_{\mu,\,\beta\gamma}(-k;k-q_2,q_2)
\,^{\ast}S_{\gamma\gamma^{\prime}}(-k_1-q_1)
\,^{\ast}\Gamma^{(Q)}_{\mu_1,\,\gamma^{\prime}\alpha}(-k_1;-q_1,q_1+k_1).
$$
Note that this expression cannot be obtained by replacements
$a\rightleftharpoons a_1$, $\alpha\rightleftharpoons\beta$,
$k\rightleftharpoons -k$, $k_1\rightleftharpoons -k_1$, and
$q_{1,\,2}\rightleftharpoons -q_{1,\,2}$ from expression (\ref{eq:4i}) by
virtue of nontrivial (spinor) structure of two last terms. However, here
we can point to the following formula connecting these effective amplitudes
$$
\!\,^{\ast}\tilde{\Gamma}^{(G)aa_1,\,ij}_{\mu{\mu}_1,\,\alpha\beta}
(k,-k_1;-q_1,-q_2)=-\,
\!\,^{\ast}\tilde{\bar{\Gamma}}^{(G)aa_1,\,ji}_{\mu{\mu}_1,\,\beta\alpha}
(-k,k_1;q_2,q_1).
$$

For Dirac conjugate effective source
$\tilde{\bar{\eta}}^{(2,1)i}_{\alpha}$ (\ref{eq:3s}) we have
expression for conjugate effective amplitude in integrand
$$
\!\,^{\ast}\tilde{\bar{\Gamma}}^{(Q)a_1a_2,\,ii_1}_{\mu_1\mu_2,\,\alpha\beta}
(k_1,k_2;q_1,-q)=-\,\Bigl\{
\delta{\Gamma}^{(Q)a_2a_1,\,i_1i}_{\mu_1\mu_2,\,\beta\alpha}
(-k_1,-k_2;-q_1,q)
\eqno{({\rm B}.3)}
$$
$$
+\,[t^{a_2},t^{a_1}]^{i_1i}
\,^{\ast}\Gamma^{(Q)}_{\nu,\,\beta\alpha}(q-q_1;-q,q_1)
\,^{\ast}{\cal D}^{\nu\nu^{\prime}}(-k_1-k_2)
\,^\ast\Gamma_{\nu^{\prime}\mu_1\mu_2}(-k_1-k_2,k_1,k_2)
$$
$$
-\,(t^{a_2}t^{a_1})^{i_1i}
\,^{\ast}\Gamma^{(Q)}_{\mu_2,\,\beta\gamma}(k_2;-q_1-k_2,q_1)
\,^{\ast}\!S_{\gamma\gamma^{\prime}}(-q_1-k_2)
\,^{\ast}\Gamma^{(Q)}_{\mu_1,\,\gamma^{\prime}\alpha}(k_1;-q,q-k_1)
\hspace{0.1cm}
$$
$$
\hspace{0.25cm}
-\,(t^{a_1}t^{a_2})^{i_1i}
\,^{\ast}\Gamma^{(Q)}_{\mu_1,\,\beta\gamma}(k_1;-q_1-k_1,q_1)
\,^{\ast}\!S_{\gamma\gamma^{\prime}}(-q_1-k_1)
\,^{\ast}\Gamma^{(Q)}_{\mu_2,\,\gamma^{\prime}\alpha}(k_2;-q,q-k_2)
\Bigr\}.
$$
As in the case of effective amplitude (B.2) this effective one cannot be
obtained from expression (\ref{eq:4o}) by a simple replacement of indices
and momenta of the following type: $a_1\rightleftharpoons a_2$,
$i\rightleftharpoons i_1$, $k_1\rightleftharpoons -k_1$ etc.
Effective amplitude (B.3) possesses a property
$$
\!\,^{\ast}\tilde{\bar{\Gamma}}^{(Q)a_1a_2,\,ii_1}_{\mu_1\mu_2,\,\alpha\beta}
(k_1,k_2;q_1,-q)=
\!\,^{\ast}\tilde{\bar{\Gamma}}^{(Q)a_2a_1,\,ii_1}_{\mu_2\mu_1,\,\alpha\beta}
(k_2,k_1;q_1,-q).
\eqno{({\rm B}.4)}
$$

Finally we consider the conjugate effective source
$\tilde{\bar{\eta}}^{(0,3)\,i}_{\beta}$, Eq.\,(\ref{eq:3d}). By
direct calculation it is not difficult to show that the conjugate
effective amplitude
$\tilde{\bar{\Gamma}}^{ii_1i_2i_3}_{\alpha\alpha_1\alpha_2\alpha_3}$
is associated with
$\tilde{\Gamma}^{ii_1i_2i_3}_{\alpha\alpha_1\alpha_2\alpha_3}$ by
the following expression:
$$
\!\,^{\ast}\tilde{\bar{\Gamma}}^{ii_1i_2i_3}_{\alpha\alpha_1\alpha_2\alpha_3}
(q,q_1,-q_2,-q_3)=-\,
\!\,^{\ast}\tilde{\Gamma}^{i_2i_3ii_1}_{\alpha_2\alpha_3\alpha\alpha_1}
(-q,-q_1,q_2,q_3).
\eqno{({\rm B}.5)}
$$

\end{document}